\catcode`\@=11
\newif\if@fewtab\@fewtabtrue

{\count255=\time\divide\count255 by 60
\xdef\hourmin{\number\count255}
\multiply\count255 by-60\advance\count255 by\time
\xdef\hourmin{\hourmin:\ifnum\count255<10 0\fi\the\count255}}
\def\ps@draft{\let\@mkboth\@gobbletwo
    \def\@oddhead{}
    \def\@oddfoot
       {\hbox to 7 cm{$\scriptstyle Draft\ version:\ \draftdate$
       \hfil}\hskip -7cm\hfil\rm\thepage \hfil}
    \def\@evenhead{}\let\@evenfoot\@oddfoot}

\def\ceqno{\global\@fewtabfalse
    \ifcase\@eqcnt \def\@tempa{& & &}\or \def\@tempa{& &}
      \or \def\@tempa{&}
      \or\def\@tempa{}\fi\@tempa
{\rm(\theequation)}}

\def\aeqno#1{\global\@fewtabfalse
    \ifcase\@eqcnt \def\@tempa{& & &}\or \def\@tempa{& &}
      \or \def\@tempa{&}
      \or\def\@tempa{}\fi\@tempa
{\rm(\theequation,#1)}}

\def\label#1{\ifnum\draftcontrol=1
 \global\def\draftnote{$\scriptstyle #1$}\fi
 \@bsphack\if@filesw {\let\thepage\relax
   \def\protect{\noexpand\noexpand\noexpand}%
\xdef\@gtempa{\write\@auxout{\string
      \newlabel{#1}{{\@currentlabel}{\thepage}}}}}\@gtempa
   \if@nobreak \ifvmode\nobreak\fi\fi\fi
  \@esphack}

\def\alabel#1#2{\label{#1}\global\@fewtabfalse
    \ifcase\@eqcnt \def\@tempa{& & &}\or \def\@tempa{& &}
      \or \def\@tempa{&}
      \or\def\@tempa{}\fi\@tempa
{\hbox to 3cm{\phantom{\rm(\theequation,#2)}
\draftnote \hfil}\hskip -3cm {\rm(\theequation,#2)}}}

\def\clabel#1{\label{#1}\global\@fewtabfalse
    \ifcase\@eqcnt \def\@tempa{& & &}\or \def\@tempa{& &}
      \or \def\@tempa{&}
      \or\def\@tempa{}\fi\@tempa
{\hbox to 3cm{\phantom{\rm(\theequation)}
\draftnote \hfil}\hskip -3cm{\rm(\theequation)}}}

\def\eqnarray{\def\draftnote{{}}\global\@fewtabtrue
\stepcounter{equation}\let\@currentlabel=\theequation
\global\@eqnswtrue
\global\@eqcnt\z@\tabskip\@centering\let\\=\@eqncr
$$\halign to \displaywidth\bgroup\@eqnsel\hskip\@centering\@eqcnt\z@
  $\displaystyle\tabskip\z@{##}$&\global\@eqcnt\@ne
  \hskip 1\arraycolsep \hfil${##}$\hfil
  &\global\@eqcnt\tw@ \hskip 1\arraycolsep
$\displaystyle\tabskip\z@{##}$
\hfil  \tabskip\@centering&\global\@eqcnt\thr@@\llap{##}\tabskip\z@
\cr}

\def\endeqnarray{\@@eqncr\egroup
      \global\advance\c@equation\m@ne$$\global\@ignoretrue}

\def\@eqnnum{\hbox to 3cm{\phantom{\rm(\theequation)} \draftnote
                         \hfil}\hskip -3cm {\rm(\theequation)}}

\def\@@eqncr{\let\@tempa\relax
    \ifcase\@eqcnt \def\@tempa{& & &}\or \def\@tempa{& &}
      \or \def\@tempa{&}
      \or\def\@tempa{}
\fi\@tempa
\if@eqnsw
\if@fewtab\@eqnnum\fi
\stepcounter{equation}\fi\global
\@eqnswtrue\global\@eqcnt\z@\global\@fewtabtrue\cr}

\def\draftcite#1{\ifnum\draftcontrol=1#1\else{}\fi}

\def\@lbibitem[#1]#2{\item{}\hskip -3cm \hbox to 2cm
{\hfil$\scriptstyle\draftcite{#2}$}\hskip
1cm[\@biblabel{#1}]\if@filesw
     {\def\protect##1{\string ##1\space}\immediate
      \write\@auxout{\string\bibcite{#2}{#1}}}\fi\ignorespaces}

\def\@bibitem#1{\item\hskip -3cm \hbox to 2cm
{\hfil $\scriptstyle\draftcite{#1}$}\hskip 1cm
\if@filesw \immediate\write\@auxout
       {\string\bibcite{#1}{\the\value{\@listctr}}}\fi\ignorespaces}

 \def\nsection#1{\section{#1}\setcounter{equation}{0}}

     \def\nappendix#1{\vskip 1cm\no{\bf Appendix
         #1}\def\thesection{#1} \setcounter{equation}{0}}

\font\tendl=msbm10  scaled \magstep1
\font\sevendl=msbm7 scaled \magstep1
\font\fivedl=msbm5 scaled \magstep1
\font\tengl=eufm10  scaled \magstep1
\font\sevengl=eufm7 scaled \magstep1
\font\fivegl=eufm5 scaled \magstep1

\newfam\dlfam \def\dl{\fam\dlfam\tendl} 
\textfont\dlfam=\tendl \scriptfont\dlfam=\sevendl
\scriptscriptfont\dlfam=\fivedl
\newfam\glfam  
\textfont\glfam=\tengl \scriptfont\glfam=\sevengl
\scriptscriptfont\glfam=\fivegl

\def\draftdate{\number\month/\number\day/\number\year\ \ \ \hourmin }

\global\def\draftcontrol{0}
\catcode`\@=12
\def\tilde{\widetilde}
\def\hat{\widehat}

\documentstyle[11pt,epsf]{article}

\def\theequation{{\thesection.\arabic{equation}}}
\newcommand{\be}{\begin{eqnarray}}
\newcommand{\en}{\end{eqnarray}\vs 0.5 cm}
\newcommand{\non}{\nonumber}
\newcommand{\no}{\noindent}
\newcommand{\vs}{\vskip}

\newcommand{\NR}{{{\dl R}}}
\newcommand{\NC}{{{\dl C}}}
\newcommand{\NZ}{{{\dl Z}}}
\newcommand{\NH}{{{\dl H}}}
\newcommand{\Ng}{{{\bf g}}}
\newcommand{\Nt}{{{\bf t}}}
\newcommand{\qq}{\begin{eqnarray}}

\newcommand{\ee}{{\rm e}}

\newcommand{\qqq}{\end{eqnarray}}

\newcommand{\tr}{\hbox{tr}}

\newcommand{\CA}{{\cal A}}

\newcommand{\CC}{{\cal C}}
\newcommand{\CD}{{\cal D}}
\newcommand{\CE}{{\cal E}}

\newcommand{\CG}{{\cal G}}
\newcommand{\CH}{{\cal H}}

\newcommand{\CK}{{\cal K}}
\newcommand{\CL}{{\cal L}}

\newcommand{\CN}{{\cal N}}
\newcommand{\CO}{{\cal O}}

\newcommand{\CT}{{\cal T}}
\newcommand{\CU}{{\cal U}}

\newcommand{\CY}{{\cal Y}}
\newcommand{\CZ}{{\cal Z}}

\begin{document}

\title{WZW branes and gerbes}

\author{\ \\Krzysztof Gaw\c{e}dzki\footnote{membre du C.N.R.S.}\\
Nuno Reis\footnote{supported by the grant Praxis XXI/BD/18138/98
from FCT (Portugal)}\\ \\Laboratoire de Physique, ENS-Lyon,\\46, All\'ee
d'Italie, F-69364 Lyon, France}
\date{ }
\maketitle

\vskip 0.3cm
\begin{center}
\end{center}

\begin{abstract}
\vskip 0.3cm \noindent We reconsider the role that bundle gerbes
play in the formulation of the WZW model on closed and open
surfaces. In particular, we show how an analysis of bundle gerbes
on groups covered by $SU(N)$ permits to determine the spectrum of
symmetric branes in the boundary version of the WZW model with
such groups as the target. We also describe a simple relation
between the open string amplitudes in the WZW models based 
on simply connected groups and in their simple-current orbifolds.  

\end{abstract}
\vskip 2cm

\nsection{Introduction}

The WZW (Wess-Zumino-Witten) model  \cite{WZW}, a version of a
two-dimension sigma model with a group manifold $G$ as the target,
constitutes an important laboratory for conformal field theory
(CFT). It is a source of numerous rational models of CFT
\cite{Zoo} and a building block of certain string vacua
\cite{stv}. It is also closely connected to the topological
3-dimensional Chern-Simons gauge theory \cite{Jones}\cite{FFFS}.
It has been clear from the very start that the model involves
topological effects of a new type which are due to the presence of
the topological Wess-Zumino term in its action functional. The
topological intricacies of the model appear already at the
classical level as global obstructions in the definition of the
action functional. Those obstructions lead to the quantization of
the coupling constant (the level) of the model. The phenomenon is
similar to the Dirac quantization of the magnetic monopole charge
but in the loop space rather than in the physical space. In the
physical space, it involves closed 3-forms instead of magnetic
field 2-forms. Quite simple for simply connected groups $G$, this
effect becomes more subtle for non-simply connected ones leading
to more involved selection rules for the level and, possibly,
multiple (theta-)vacua of the quantum theory \cite{FGK}. \vskip
0.2cm

As is well known, a convenient mathematical framework for the Dirac
monopoles, their quantization and the Bohm-Aharonov effect is
provided by the theory of line bundles with hermitian connections.
Up to isomorphism, such bundles may be characterized by certain
sheaf cohomology classes. More exactly, they correspond to the
elements of the real version of the degree 2 Deligne cohomology
\cite{EV}\cite{Gajer}. It was realized in \cite{Gaw0} that the
Deligne cohomology in degree 3 provides a mathematical
language to treat the topological intricacies in the WZW model.
The theory is somewhat analogous to the degree 2 case when the
original space is replaced by its loop space. Indeed, a third
degree real Deligne class determines a (unique up to isomorphism)
hermitian line bundle with connection on the loop space
\cite{Gaw0}. The degree 3 theory appears, however, to be much
richer. In particular, one of the basic constructions in degree
2, that of the parallel transport along curves, becomes that of
the ``parallel transport'' around two-dimensional surfaces which
may have different topology. For closed surfaces one obtains the
$U(1)$-valued ``holonomies'' that enter the Feynman amplitudes of
classical field configurations in the WZW model. For surfaces with
boundary, the amplitudes take instead values in the product of
lines associated to the boundary loops. New phenomena appear when
the boundary components or their pieces are restricted to special
submanifolds (D-branes) over which the Deligne cohomology class
trivializes. The discussion in \cite{Gaw0} extends easily to that
case as was briefly evoked in \cite{ist}. This is precisely the
situation that one confronts when studying boundary conditions in
the WZW models that preserve (half of the) symmetries of the bulk
theory. One of the main points of this paper is to show how the
order 3 Deligne classes enter the classification of such
boundary conditions, i.e. of the WZW branes. \vskip 0.2cm

Although the whole discussion may be made using the cohomological
language, it is convenient to have at ones disposal geometric
objects whose isomorphism classes are characterized by the degree
3 real Deligne cohomology classes. This was recognized in
ref.\,\,\cite{Bryl} which proposed to use the theory of ``gerbes''
\cite{Gir} to provide for such objects. It seems, that the most
appropriate geometric notions are those of (hermitian) bundle
gerbes with connection defined in \cite{Murray} and of their
stable isomorphisms introduced in \cite{MurrS}. The bundle gerbes
with connection are simple geometric objects whose stable
isomorphism classes are exactly described by the order 3 real
Deligne cohomology. Their use allows to translate the
cohomological discussions of \cite{Gaw0} to a more geometric
language which is indeed useful when discussing the issues related
to branes. See also \cite{CMM}\cite{BCMMS}\cite{Hitchin} for the
discussions of gerbes in different, although related, contexts.
\vskip 0.2cm

The paper is organized as follows. In Sect.\,\,2 we recall the
essential points of \cite{Gaw0}, with some of the details
relegated to Sect.\,\,10.1, and discuss their translation to the
bundle-gerbe language, In Sect.\,\,3 we present an explicit
construction of gerbes over the $SU(N)$ groups. Sect.\,\,4 is
devoted to the case of non-simply connected groups covered by
$SU(N)$. In parenthetical Sect.\,\,5, we explain how to define
gerbes on discrete quotient spaces, an issue which was previously
discussed in the context of discrete torsion in \cite{Sharpe}. How
the construction from Sect.\,\,4 fits into this general scheme is
shown in Appendix B. In Sect.\,\,6, we describe the line bundles
with connection on the loop spaces induced by gerbes and relevant
for the geometric description of closed string amplitudes.
Sect.\,\,7 shows how those line bundles may be trivialized when
restricted to loop spaces of branes and how to describe the brane
structure in terms of gerbes. We also discuss the line bundles
induced by gerbes on the space of paths with ends on branes, the
open string counterpart of the loop space construction. In
Sect.\,\,8.1, we examine branes in the $SU(N)$ groups and in
Sect.\,\,8.2 the ones in the groups covered by $SU(N)$. In the
latter case, we obtain a completely explicit description of the
(symmetric) branes confirming the results based on studying
consistency of quantum amplitudes. In Sect.\,\,9, we evoke the
bearing that the geometric constructions discussed in this paper
have on the spectrum and the boundary partition functions of the
WZW models based on the groups covered by $SU(N)$. We identify a
general relation, that seems at least partially new in the context 
of the WZW theory, between the spaces of states for the boundary 
WZW models with non-simply 
connected groups and the ones for the models based on the covering 
groups. Sect.\,\,10 briefly indicates how to extend this
relation to general open string quantum amplitudes. In Section 11, 
we present a local description of the line bundles over the loop 
spaces and open path spaces induced by gerbes, discussed 
before in more abstract terms. Conclusions give a brief
summary of what was achieved in the paper and list some open
problems. More technical calculations referred to in the main text 
have been collected in Appendices. \vskip 0.2cm

\nsection{Topological action functionals and gerbes}

Let us start by recalling the some basic points of \cite{Gaw0}, changing
the notations to more up-to-date ones.

\subsection{Dirac monopoles and line bundles}

Suppose that $B$ is a (magnetic field) closed 2-form on a manifold
$M$. To describe a particle of unit charge moving in such a field
along a trajectory $\varphi(t)$, one has to add to the action
functional the coupling term $\int\varphi^* A$ where $A=d^{-1}B$
is the vector potential of ${B}$, i.e.\,\, a 1-form such that
$dA=B$. The problem arises when ${B}$ is not exact so that there
is no global $A$ (like for the magnetic field of a monopole).
Dirac's solution of the problem, when translated to a geometric
language, is to define the Feynman amplitudes
$\,\ee^{\,i\int\varphi^*d^{-1}{B}}\,$ of closed particle paths
$\varphi$ as holonomies in a hermitian line bundle $L$ with
(hermitian) connection $\nabla$ of curvature $curv(\nabla)={B}$,
provided such bundle exists. This is the case if the closed 2-form
${1\over2\pi}{B}$ is integral in the sense that its periods over
closed 2-cycles in $M$ are integers. \vskip 0.2cm

Let $(O_i)$ be a sufficiently fine open covering of $M$. We shall
use the standard notation $O_{ij}$, $O_{ijk}$ etc. for the
multiple intersections of the sets $O_i$. A choice of local
sections $\,s_i:O_i \rightarrow L\,$ of length 1 gives rise to the
local data $(g_{ij},{A}_i)$ for ${L}$ such that $s_j= g_{ij}s_i$
and $\nabla s_i={1\over i}{A}_i s_i$. They have the following
properties: \vskip 0.2cm

1. \ $g_{ij}=g_{ji}^{-1}:O_{ij}\rightarrow
U(1)\,$ and on $O_{ijk}$
\qq
g_{jk}\,g_{ik}^{-1}\,g_{ij}=1\,,
\qqq
\vskip 0.2cm

2. \ ${A}_i$ are real 1-forms on $O_i$ such that
\qq
d{A}_i={B},
\qqq

3. \ on $O_{ij}$
\qq
{A}_j-{A}_i=i\,g_{ij}^{-1}
dg_{ij}\,.
\qqq
\vskip 0.2cm

\noindent If $s'_i$ correspond to a different choice of local
sections so that $\,s'_i=f_i s_i\,$ then \qq g'_{ij}=g_{ij}f_j
f_i^{-1} \qquad{\rm and}\qquad{A}_i'={A}_i+i\,f_i^{-1} df_i\,.
\label{eq1} \qqq The local data also naturally restrict to finer
coverings. The two collections of local data are considered
equivalent if they are related by (\ref{eq1}) when restricted to a
sufficiently fine common covering. The equivalence classes
$w=[g_{ij},{A}_i]$ may be viewed as (real, degree 2) Deligne
(hyper-)cohomology classes \cite{Gaw0}. The class of local data
depends only on the bundle ${L}$ with connection and not on the
choice of its local sections. Besides, isomorphic bundles give
rise to the same Deligne class. \vskip 0.2cm

Denote by $W(M,{B})$ the set of equivalence classes $w$ of local
data corresponding to fixed ${B}$ and by $w({L})$ the class of local
data of the bundle ${L}$. In fact, the line bundle ${L}$ together
with its hermitian structure and connection may be reconstructed
from the local data $(g_{ij},{A}_i)$ up to isomorphism. One just
takes the disjoint union $\,\mathop{\sqcup}\limits_i(O_i\times\NC)
\equiv\mathop{\cup}\limits_i(O_i\times\{i\}\times\NC)\,$ of trivial
bundles and one divides it by the equivalence relation \qq
(x,i,g_{ij}z)\,\sim\,(x,j,z)\,. \label{iden} \qqq
The covariant derivative given by $\nabla=d+{1 \over i}{A}_i$
on $O_i$ defines a connection on the quotient bundle. It follows that
the elements of $W(M,{B})$ are in one-to-one correspondence with
the isomorphism classes of hermitian line bundles with
connections. In particular, $W(M,{B})$ in non-empty if and only if
the 2-form ${1\over2\pi}{B}$ is integral. In the latter case, the
cohomology group $H^1(M,U(1))$ acts on $W(M,{B})$ in a free,
transitive way, i.e. $W(M,B)$ is a $H^1(M,U(1))$-torsor. The action
sends $\,w=[g_{ij},{A}_i]\,$ to $\,uw=[u_{ij}g_{ij},{A}_i]$,
\,where $\,(u_{ij})\,$ is the ${\rm C}^{\hspace{-0.2cm}^\vee}$ech
cocycle representing $u\in H^1(M,U(1))$. The latter group may be
also viewed as that of characters of the fundamental group of $M$.
The line bundles corresponding to $w$ and to $uw$ have holonomies
differing by the corresponding character. In particular, the set
$W(M,0)$ of isomorphism classes of flat hermitian line bundles may
be identified with the group $H^1(M,U(1))$. The multiplication in
$H^1(M,U(1))$ corresponds to the tensor product of flat bundles.
\vskip 0.2cm

The holonomy $\CH(\varphi)$ in $L$ along a closed loop $\,\varphi:
\ell\rightarrow M\,$ may be expressed using the local data of
${L}$. One splits $\ell$ into small closed intervals $b$ with
common vertices $v$ in such a way that $b\subset O_{i_b}$ for some
$i_b$, choosing also for each vertex $v$ and index $i_v$ so that
$v\in O_{i_v}$. Then \qq \CH(\varphi)\ =\
\exp\Big[i\sum\limits_b\int\limits_b\varphi^*{A}_{i_b}
\Big]\prod\limits_{v\in b} g_{i_vi_b}(\varphi(v))\,, \label{holo}
\qqq where the product $\prod\limits_{v\in b}$ is taken with the
convention that the entry following it is inverted if $v$ is the
beginning of $\,b$. Since the holonomy depends only on the
isomorphism class of $\,{L}$, the right hand side depends only on
the class $w$ of the local data $(g_{ij},{A}_i)$. More generally,
for arbitrary curves $\,\varphi:\ell\rightarrow M$, the parallel
transport in ${L}$ defines an element in
${L}_{\varphi(v_-)}^{-1}\otimes{L}_{\varphi(v_+)}$, where $v_\pm$
are the ends of $\,\ell$, $\,{L}_m$ denotes the fiber of ${L}$
over $m\in M$, and ${L}^{-1}$ is the bundle dual to ${L}$. Using
local sections $s_{i_\pm}$ such that $\varphi(v_\pm) \in
O_{i_\pm}$, this element may be represented by a number that is
still given by the right hand side of (\ref{holo}). The value of
(\ref{holo}) changes now upon changing the indices $i_\pm$
assigned to the endpoints of $\,\ell$ according to the
identifications (\ref{iden}). It also changes when ones changes
the local data $(g_{ij},{A}_i)$ within the class $w$, but in the
way consistent with isomorphisms of the line bundles reconstructed
from such data.

\subsection{Topological actions in two-dimensional field theories}

In the two dimensional field theory, for example in the WZW model
(see \cite{Gaw0} for another example), one needs to make sense of
action functionals written formally as $\int\phi^*d^{-1}{H}$ where
${H}$ is a closed but (possibly) not exact real 3-form ${H}$ on
the target manifold $M$ in which the two-dimensional field $\phi$
takes values. This may be done in analogy to the one-dimensional
prescription (\ref{holo}). \vskip 0.2cm

Suppose that, for a sufficiently fine covering $(O_i)$, one may
choose the local data $\,(g_{ijk},{A}_{ij},{B}_i)$
with the following properties:
\vskip 0.2cm

1. \ $g_{ijk}=g_{\sigma(i)\sigma(j)\sigma
(k)}^{{\rm sign}(\sigma)}:O_{ijk}\rightarrow U(1)\,$
and on $\,O_{ijkl}$
\qq
g_{jkl}\,g_{ikl}^{-1}\,g_{ijl}
\,g_{ijk}^{-1}=1\,,
\label{toassoc}
\qqq
\vskip 0.2cm

2. \ ${A}_{ij}=-{A}_{ji}\,$ are real 1-forms on
$O_{ij}$ and on $O_{ijk}$
\qq
{A}_{jk}-{A}_{ik}+{A}_{ij}=
i\,g_{ijk}^{-1}dg_{ijk}\,,
\label{toiso}
\qqq
\vskip 0.2cm

3. \ ${B}_i\,$ are real 2-forms on $O_i$ such that
\qq
d{B}_i={H}\,,
\label{to1}
\qqq
\vskip -0.2cm

4. \ on $O_{ij}$, \vskip -0.6cm \qq {B}_{j}-{B}_i=d{A}_{ij}\,.
\label{to2} \qqq \vskip 0.2cm

\noindent Such local data naturally restrict to finer coverings.
Following \cite{Gaw0}, we shall consider two collections
of local data equivalent if, upon restriction to a common sufficiently
fine covering,
\begin{eqnarray}
&&g'_{ijk}=g_{ijk}\,\chi_{jk}^{-1}\,
\chi_{ik}\,\chi_{ij}^{-1}\,,\alabel{eqv2}{a}\cr
&&{A}'_{ij}={A}_{ij}+\Pi_j-\Pi_i
-i\,\chi_{ij}^{-1}d\chi_{ij}\,,\aeqno{b}\cr
&&{B}'_{i}={B}_i+d\Pi_i\aeqno{c}\cr
\end{eqnarray}
\vskip -4.5mm \noindent for
$\,\chi_{ij}=\chi_{ji}^{-1}:O_{ij}\rightarrow U(1)\,$ and real
1-forms $\,\Pi_i\,$ on $O_i$. The equivalence classes
$\,w=[g_{ijk}, {A}_{ij},{B}_i]$ may be viewed as Deligne
(hyper-)cohomology classes in the degree three \cite{Gaw0}. The
set $W(M,{H})$ of the classes corresponding to a given closed
3-form ${H}$ is non empty if and only if ${1\over 2\pi}{H}$ is
integral in the sense that all its periods over closed 3-cycles in
$M$ are integers. If this is the case then $W(M,H)$ is a
$H^2(M,U(1))$-torsor, with the cohomology group $H^2(M,U(1))$
acting on $W(M,{H})$ by \qq (g_{ijk},{A}_{ij},{B}_i)\ \mapsto\
(u_{ijk}g_{ijk},{A}_{ij},{B}_i)\,. \label{ac1} \qqq If
$\,H^3(M,\NZ)\,$ (or $H_2(M,\NZ)$) is without torsion, the above
action is equivalent to \qq (g_{ijk},{A}_{ij},{B}_i)\ \mapsto\
(g_{ijk},{A}_{ij},{B}_i+F)\,, \label{ac2} \qqq where $F$ is a
closed 2-form on $M$. In the latter case, the class in $W(M,{H})$
does not change if and only if $\,{1\over2\pi}F\,$ is an integral
2-form. The equivalence of the two actions follows from the
isomorphism $\,H^2(M,U(1))\cong H^2(M,\NR)/H^2(M,2\pi\NZ)$. \vskip
0.2cm

Let $\,\phi\,$ be a map from a compact oriented surface
$\,\Sigma\,$ to $\,M$. One may triangulate $\Sigma$ in such a way
that for each triangle $\,c\,$ there is an index $\,i_c$ such that
$c\subset O_{i_c}$. We shall also choose indices $\,i_b$ for edges
$b$ and $\,i_v$ for vertices $\,v\,$ so that $\phi(b)\subset O_{i_b}$
and $\phi(v)\in O_{i_v}$. The formal amplitudes
$\,\ee^{i\int\phi^*d^{-1}{H}}\,$ may now be defined, as was first
proposed in \cite{Alva}, by \qq \CA(\phi)\ =\
\exp\Big[i\sum\limits_c\int\limits_c\phi^*{B}_{i_c}
+i\sum\limits_{b\subset c}\int\limits_b\phi^*{A}_{i_ci_b}\Big]
\prod\limits_{v\in b\subset c}g_{i_ci_bi_v}(\phi(v))\,,
\label{ampl} \qqq with the similar orientation conventions as in
(\ref{holo}). It is straightforward to check that if
$\partial\Sigma=\emptyset$ then  $\CA(\phi)$ is independent of the
choices of the triangulation and of the assignment of the covering
indices and does not change under restrictions of the local data
to finer coverings and under the equivalences (\ref{eqv2}).
\vskip 0.2cm

Assume now that $\Sigma$ has a boundary $\partial\Sigma=
\mathop{\sqcup}\limits_s\ell_s$ with the boundary components
$\ell_s$ that may be parametrized by the standard circle $S^1$. In
this case the expression (\ref{ampl}) still does not change if one
modifies the triangulation and the index assignment in the
interior of $\,\Sigma$, but it does change if the changes concern
the boundary data. One may abstract from those changes a
definition of a hermitian line bundle $\CL$ over the space $LM$ of
loops in $M$ (or over the quotient of the latter by
orientation-preserving reparametrizations) in such a way that \qq
\CA(\phi)\ \in\ \mathop{\otimes}\limits_s\CL_{\phi|_{\ell_s}}\,.
\qqq The transition functions of the line bundle $\CL$ have been
constructed in \cite{Gaw0} where it was also shown that $\CL$
carries a natural connection whose curvature 2-form $\Omega$ is
given by \qq
\langle\delta_1\varphi,\delta_2\varphi\,|\,\Omega(\varphi)\rangle\,=\,
\int\limits_{\ell}\varphi^*\iota(\delta_2\varphi)\iota(\delta_1\varphi)
{H}\,. \label{cur} \qqq For completeness, we include the explicit
expressions from \cite{Gaw0} in Sect.\,10.1. For an equivalent
choice of the local data $(g_{ijk},{A}_{ij},{B}_i)$, the bundle
$\CL$ changes to an isomorphic one so that one obtains a natural
map from $W(M,{H})$ to $W(LM,\Omega)$.

\subsection{Bundle gerbes with connections}

As we already mentioned, there are simple geometric objects,
the (hermitian line) bundle gerbes with connection, whose appropriate
isomorphism classes are described by elements of $\,W(M,{H})$.
Let us briefly recall this concept \cite{Murray}\cite{MurrS}.
\vskip 0.2cm

Suppose that we are given a manifold map $\,\pi:Y\rightarrow M\,$
which admits local sections $\sigma_i:O_i\rightarrow Y\,$ over the
sets of a sufficiently fine covering of $M$. Let
$\,Y^{{[n]}}=Y\times_{_M}\hspace{-0.1cm}Y\dots\times_{_M}\hspace{-0.1cm}Y\,$
denote the $n$-fold fiber product of $\,Y$.
$\,Y^{{[n]}}=\{(y_1,\dots,y_n)\in Y^{n}\,|\,
\pi(y_1)=\dots\pi(y_n)\}$. We shall denote by $\pi^{{[n]}}$ the
obvious map from $Y^{{[n]}}$ to $M$ and by $\,p_{n_1\dots n_k}$
the projection of $(y_1,\dots,y_n)$ to $(y_{n_1}, \dots,y_{n_k})$.
A hermitian line bundle gerbe $\CG$ over $M$ with connection of
curvature $H$ (shortly, a gerbe) is a quadruple $(Y,{B},{L},\mu)$
where \vskip 0.2cm

1. \ \ ${B}\,$ is a 2-form on $Y$ such that
\qq
d{B}=\pi^*{H}\,,
\label{1}
\qqq

2. \ \hbox to 11.4cm{ ${L}\,$ is a hermitian line bundle with a
connection $\nabla$ over $Y^{{[2]}}$ with}

\,\ \quad\ \ \hbox to 3cm{curvature\hfil}
\vskip -0.7cm
\qq
curv(\nabla)=p_2^*{B}-p_1^*{B}\,,
\label{2}
\qqq

3. \ \hbox to 11.4cm{\ $\mu$ is an isomorphism of hermitian
line bundles with connection}

\,\,\ \quad\ \ \hbox to 3cm{over $Y^{{[3]}}$\hfil} \vskip -0.8cm
\qq \mu:p_{12}^*{L}\otimes p_{23}^*{L}\longrightarrow
p_{13}^*{L}\,, \label{3} \qqq

4. \ \hbox to 11.4cm{\,\,as isomorphisms of line bundles $p_{12}^*{L}
\otimes p_{23}^*{L}\otimes p_{34}^*{L}$ and $p_{14}^*{L}$}

\,\,\ \quad\ \ \hbox to 3cm{over $Y^{{[4]}}$\hfil} \vskip -0.8cm
\qq \mu\circ(\mu\otimes id)\,=\,\mu\circ(id\otimes\mu)\,.
\label{assoc} \qqq \vskip 0.2cm

\noindent The 2-form ${B}$ is called the curving of the gerbe. The
isomorphism $\mu$ defines a structure of a groupoid on ${L}$ with
the bilinear product $\,\mu:{L}_{(y_1,y_2)}
\otimes{L}_{(y_2,y_3)}\rightarrow{L}_{(y_1,y_3)}$. The
associativity of the product is guaranteed by (\ref{assoc}). The
bundle ${L}$ restricted to the diagonal composed of the elements
$\,(y,y)\,$ may be naturally trivialized by the choice of the
units of the groupoid multiplication and $\mu$ determines a natural
isomorphism between $\kappa^*{L}$ and ${L}^{-1}$, where
$\kappa(y_1,y_2)=(y_2,y_1)$. In order to elucidate the abstract
definition copied from \cite{Murray} (except for fixing the
curving ${B}$ of the gerbe), let us immediately provide examples.
\vskip 0.2cm

First, for an exact 3-form ${H}=d{B}$, the quadruple
$\,(M,{B},\,M^{{[2]}}\times\NC,\,\cdot\,\,)\,$ with $Y=M$, with
the trivial bundle ${L}$ over $M^{{[2]}}\cong M$, and with $\mu$
determined by the product of complex numbers, is a gerbe with
curvature ${H}$. \vskip 0.2cm

Given a map $\pi:Y\rightarrow M$ admitting local sections and a
hermitian line bundle $\,{N}$ over $Y$ with connection of
curvature $F$, a simple example of a gerbe is provided by
$\,\CG_{_N}=(Y,\,F,\,p_1^*{N}^{-1}\hspace{-0.1cm} \otimes
p_2^*{N},\,\mu)\,$ with $\mu$ given by the obvious identification
between
$\,({N}_{y_1}^{-1}\hspace{-0.1cm}\otimes{N}_{y_2})\otimes({N}_{y_2}
^{-1}\hspace{-0.1cm}\otimes{N}_{y_3})\,$ and $\,{N}_{y_1}
^{-1}\hspace{-0.1cm}\otimes{N}_{y_3}$. This is a gerbe with the
vanishing curvature. Following \cite{Murray}\cite{MurrS}, we shall
call gerbes $\,\CG_{_N}$ trivial. Trivial gerbes are useful
to recognize when a bundle $N$ is isomorphic to a pullback $\pi^*P$
of a hermitian bundle with connection on M. This is the case
if and only if there exists a unit length flat section $D$ (called a
descent data) of the trivial gerbe $\,\CG_{_N}$ bundle $\,p_1^*{N}^{-1}
\hspace{-0.1cm}\otimes p_2^*{N}\,$ over $Y^{{[2]}}$ such that
\qq
\mu(D\circ p_{_{12}}\otimes D\circ p_{_{23}})\ =\ D\circ p_{_{13}}\,.
\label{decp}
\qqq
$D(y_1,y_2):N_{_{y_1}}\rightarrow N_{_{y_2}}$
defines then an equivalence relation on $\,\mathop{\sqcup}\limits_{y\in
\pi^{-1}(m)}\hspace{-0.2cm}N_{_y}$. Taking $\,P_{_m}$ as the set of the
equivalence classes, one obtains canonically a bundle $P$ and an isomorphism
of $N$ with $\pi^*P$. We shall say that $P$ is obtained from $N$ and $D$
by the descent principle.
\vskip 0.2cm

The next example will be central to our application of gerbes. Let
$(g_{ijk},{A}_{ij},{B}_i)$ be local data on $M$ as described in
the previous subsection. Take for $\,Y\,$ the disjoint union
$\,\mathop{\sqcup} \limits_{i}O_i\,$ with $\,\pi(x,i)=x$. \,Then
$\,Y^{{[n]}}=\hspace{-0.2cm}
\mathop{\sqcup}\limits_{(i_1,\dots,i_n)}\hspace{-0.2cm}
O_{i_1\dots i_n}$ and the projections $p_{n_1\dots n_k}$ are the
inclusions of $\,O_{i_1\dots i_n}$ into $\,O_{i_{n_1}\dots
i_{n_k}}$. We take as ${L}$ the trivial hermitian line bundle
$\,Y^{{[2]}}\times\NC$. The connection on ${L}$ will be given by
$\,\nabla=d+{1\over i}{A}_{ij}\,$ on $O_{ij}$ and the isomorphism
$\mu$ by the multiplication by $g_{ijk}$ on $O_{ijk}$. The relation
(\ref{1}) is then assured by (\ref{to1}) and the equality
(\ref{2}) by (\ref{to2}). That $\mu$ preserves the connections
follows from (\ref{toiso}) and its associativity (\ref{assoc}) is
a consequence of (\ref{toassoc}). \vskip 0.2cm

Conversely, given a gerbe, one may define local data $(g_{ijk},
{A}_{ij},{B}_i)$ the following way. One first chooses local sections
$\sigma_i:O_i\rightarrow Y$ that induce local sections
$\,\sigma_{i_1\dots i_n}\equiv(\sigma_{i_1},\dots,
\sigma_{i_n})\,$ of $\,Y^{{[n]}}$ defined on intersections
$\,O_{i_1\dots i_n}$. If the covering of $M$ is sufficiently fine,
one may also choose unit length sections $\,s_{ij}:\sigma_{ij}(O_{ij})
\rightarrow{L}\,$ so that $\,s_{ji}=s_{ij}^{-1} \circ\kappa$. One
defines the local data
$\,(g_{ijk}, {A}_{ij},{B}_i)\,$ by the relations \qq
&&{B}_i\,=\,\sigma_i^*{B}\,,\alabel{gtold}{a}\cr
&&\sigma_{ij}^*(\nabla s_{ij})\,=\,{_1\over^i}
{A}_{ij}\,s_{ij}\circ\sigma_{ij}\,,\aeqno{b}\cr
&&\mu\circ(s_{ij}\circ\sigma_{ij} \otimes
s_{jk}\circ\sigma_{jk})\,=\, g_{ijk}\,\,s_{ik}\circ\sigma_{ik}
\,.\aeqno{c}\cr \qqq \vskip -0.4cm \noindent Properties
(\ref{to1}) and (\ref{to2}) follow from (\ref{1}) and (\ref{2}).
Equation (\ref{toiso}) arises by covariantly differentiating
(\ref{gtold},{c}) along directions tangent to $\sigma_{ijk}
(O_{ijk})$ with the use of relations $p_{12}\circ
\sigma_{ijk}=\sigma_{ij}$ etc. and of the fact that $\mu$ preserves
the connections. Finally, the cocycle condition (\ref{toassoc})
follows from the associativity (\ref{assoc}). For a trivial gerbe
$\,\CG_{_N}$, \,we may take $\,s_{ij}=
(\chi_{ij}^{-1}\circ\pi^{{[2]}})(s_i^{-1}\hspace{-0.1cm}\otimes s_j)$,
where $s_i$ are unit length sections
$\,s_i:\sigma_i(O_i)\rightarrow{N}$ and
$\,\chi_{ij}=\chi_{ji}^{-1}:O_{ij}\rightarrow U(1)$. \,One obtains
then the local data \qq
(\chi_{jk}^{-1}\chi_{ik}\,\chi_{ij}^{-1},\,\Pi_j-\Pi_i-i\,
\chi_{ij}^{-1}d\chi_{ij},\,d\Pi_i)\,, \label{lotr} \qqq where
$\Pi_i$ is defined by the relation $\sigma_i^*(\nabla s_i)
={1\over i}\Pi_i\,s_i\circ\sigma_i$. \vskip 0.2cm

Let us show that the class $w\in W(M,{H})$ of the local data
$\,(g_{ijk},{A}_{ij},{B}_i)\,$ depends only on the gerbe $\CG$ and
not of the choices of local sections used in the construction of
the data. First, restricting the sections $\sigma_i$ and,
accordingly, $s_{ij}$ to a finer covering produces the restriction
of the local data to that covering which, by definition, does not
change the class in $W(M,{H})$. For two choices of local
sections, one may assume that they have been already
restricted to a common covering with the sets $O_i$ sufficiently small.
One has then to compare the local data induced by the two families of
sections $\sigma_i,\,s_{ij}$ and $\sigma'_i,\,s'_{ij}$. \,Let $\,\tilde
\sigma_i=(\sigma_i,\sigma_i'):O_i\rightarrow Y^{{[2]}}\,$ and let $\,s_i:
\tilde\sigma_i(O_i)\rightarrow L\,$ be unit length sections of $\,L$.
The relations
\qq
(s_i\circ\tilde\sigma_i)^{-1}
(s_{ij}\circ\sigma_{ij})(s_j\circ\tilde\sigma_j)\ =\ \chi_{ij}\
s'_{ij}\circ\sigma'_{ij}\,,
\label{sss}
\qqq
where on the left hand side the sections of $\,L$ are
multiplied using $\mu$, define $U(1)$-valued functions
$\,\chi_{ij}=\chi_{ji}^{-1}$ on  $\,O_{ij}$. \,Let $\,\Pi_i$ be 1-forms on
$\,O_i$ given by
\qq
\tilde\sigma_i^*\nabla s_i={_1\over^i}\,\Pi_i\,s_i\circ
\tilde\sigma_i\,.
\label{tsp}
\qqq
Relation (\ref{eqv2},{c}) follows then from (\ref{2})
and (\ref{gtold},{a}). Similarly, identity (\ref{eqv2},{b}) is
a consequence of (\ref{gtold},{b}), (\ref{sss}) and the fact that
$\mu$ commutes with the covariant derivation. Finally,
the associativity of the product defined by $\mu$ together
with (\ref{gtold},{c}) and (\ref{sss}) implies (\ref{eqv2},{a}).
This shows that the local data $\,(g_{ijk}, {A}_{ij},{B}_i)\,$ and
$\,(g'_{ijk},{A}'_{ij},{B}'_i)\,$ define the same class $w\in
W(M,{H})$ which, consequently, depends only on the gerbe $\CG$. We
shall denote this class by $w(\CG)$. Clearly, the class of the
gerbe constructed from the local data
$\,(g_{ijk},{A}_{ij},{B}_i)\,$ is the class of those data. \vskip
0.2cm

It is natural to inquire when two gerbes $\CG=(Y,{B},
{L},\mu)$ and $\CG'=(Y',B',L',\mu')$ on $M$ with curvature ${H}$
define the same class $w\in W(M,{H})$. A sufficient condition is
that $Y=Y'$, ${B}=B'$ and that there exists a bundle isomorphism
$\iota:L\rightarrow L'$ preserving the remaining structures. We
shall call such gerbes isomorphic. It is clear, however, that this
is not a necessary condition. For example, two gerbes constructed
from equivalent local data on different open coverings of $M$
define the same class in $W(M,{H})$ but may have spaces $Y$ and
$Y'$ with different numbers of components. The appropriate
geometric notion  of a stable isomorphism of gerbes was introduced
in \cite{MurrS}. It provides a necessary and sufficient condition
for the equality $w(\CG_1)=w(\CG_2)$. We shall describe it now.
\vskip 0.2cm

Let $\CG=(Y,{B},{L},\mu)$ be a gerbe and let $\omega:Z\rightarrow
M$ be another map with local sections. Given also a map
$\sigma:Z\rightarrow Y$ commuting with the projections on $M$, the
pullback gerbe $\,\sigma^*\CG\,$ will be defined as
$\,(Z,\sigma^*{B}, {\sigma^{[2]}}^*{L},{\sigma^{[3]}}^*\mu)$. It
has the same curvature as $\CG$. For two gerbes with the same $Y$,
one may define their tensor product by taking the tensor product
of the hermitian line bundles with connections over $Y^{[2]}$ and
the tensor product $\mu\otimes\mu'$ as the groupoid
multiplication. The curvings and the curvatures add under such
operation. Let $\CG=(Y,{B},{L},\mu)$ and
$\CG'=(Y',{B'},{L'},\mu')$ be two arbitrary gerbes over $M$.
\,Take $\,Z=Y\times_MY'$ and let $\,\sigma:Z\rightarrow Y\,$ and
$\,\sigma':Z\rightarrow Y'\,$ be the projections on the components
in $\,Y\times_MY'$. \,By definition, gerbes $\CG$ and $\CG'$ are
stably isomorphic if there exists a line bundle $N$ over $Z$ and a
line bundle isomorphism \qq {\sigma^{[2]}}^*L\otimes\,
p_1^*N^{-1}\otimes p_2^*N\ \mathop{\longrightarrow}^\iota\
{{\sigma'}^{[2]}}^*L' \label{sism} \qqq defining an isomorphism
between the gerbes $\,\sigma^*\CG\otimes\CG_{_N}$ and
${\sigma'}^*\CG'$. \,In particular, this requires that the
curvature $F$ of $N$ be equal to $\,{\sigma'}^*B'-\sigma^*B$.
\,The stable isomorphism of gerbes is an equivalence relation.
\vskip 0.2cm

The line bundle isomorphism $\,\iota\,$ will be called the stable
isomorphism between $\,\CG\,$ and $\,\CG'$. \,In general, it is
not unique. If $\,\iota'$ is another stable isomorphisms between
$\CG$ and $\CG'$ corresponding to line bundles $\,N'$ over $\,Z\,$
then it necessarily differs from $\,\iota\,$ by an isomorphism
between the trivial gerbes $\,\CG_{_N}$ and $\,\CG_{_{N'}}$.
\,Such an isomorphism defines descent data for the bundle
$\,N^{-1}\otimes N'\,$ so that, canonically, $\,N'\cong
N\otimes\omega^{*}P\,$ for a bundle $P$ on $M$. \,Since $N$ and of
$N'$ have the same curvatures, $P$ has to be a flat bundle.
Conversely, the gerbes $\,\CG_{_N}$ and $\,\CG_{_{N'}}$ for
$\,N'=N\otimes\omega^{*}P\,$ are canonically isomorphic if $P$ is
a flat bundle on $M$. \vskip 0.3cm

\noindent{\bf Remark}. \ It is easy to see that any pullback gerbe
$\sigma^*\CG$ is stably isomorphic to $\CG$. Indeed, taking as the
bundle $N$ over $\,Z\otimes_MY\,$ the pullback of $\,L\,$ by the
map $\,\sigma\times Id\,$ from $Z\otimes_MY$ to $\,Y^{[2]}$, \,we
observe that the groupoid multiplication $\mu$ defines a stable
isomorphism \qq
L_{(\sigma(z_1),\sigma(z_2))}\otimes\,N_{(\sigma(z_1),y_1)}^{-1}\otimes\,
N_{(\sigma(z_2),y_2)}\ \longrightarrow\ L_{(y_1,y_2)} \qqq between
$\sigma^*\CG$ and $\CG$. \,In fact, two gerbes $\CG$ and $\CG'$
are stable isomorphic if and only if they become isomorphic after
the pullback to a common $Z$ (not necessarily equal to
$\,Y\times_MY'$) and the tensor multiplication by a trivial gerbe.
\vskip 0.3cm

Clearly, the stably isomorphic gerbes have the same curvature
${H}$. Moreover, as it is easy to see, they give rise to the same
class $w\in W(M,{H})$. Indeed, under pullbacks of gerbes the local
data do not change if we use the local sections
$\,\sigma\circ\sigma_i: O_i\rightarrow Y$ and $\,s_{ij}$ for the
gerbe before the pullback and $\,\sigma_i:O_i\rightarrow Z\,$ and
$\,s_{ij}\circ \sigma^{[2]}$ for the pullback gerbe. Similarly,
under tensor multiplication by a trivial gerbe, the local data
change by (\ref{lotr}), hence again stay in the same class.
Converse is also true: if $w(\CG)=w(\CG')$ then $\CG$ and $\CG'$
are stably isomorphic. To prove this, it is enough to show that
any gerbe $\CG= (Y,{B},{L},\mu)$ is stably isomorphic to the one
constructed from its local data associated to the sections
$\sigma_i$ and $s_{ij}$. This follows from the fact that the
pullback of $\CG$ by the map $\,\sigma:\mathop{\sqcup}\limits_i
O_i \rightarrow Y\,$ equal to $\sigma_i$ on each $O_i$ is
isomorphic to the local data gerbe, with the corresponding line
bundle isomorphism given by the sections $s_{ij}$ (just recall how
the local data and the corresponding gerbe are defined). \vskip
0.3cm

{\bf Summarizing}: there is a one-to-one correspondence between
the cohomology classes in $W(M,{H})$ and the stable isomorphism
classes of gerbes with curvature ${H}$. \vskip 0.3cm

\nsection{Gerbes on groups $SU(N)$}

Let $G$ be a connected, simply connected, simple compact group and
let $\Ng$ be its Lie algebra. We shall denote by $\,\tr\,$ the
non-degenerate bilinear invariant form on $\Ng$ which allows to
identify $\Ng$ with its dual, by $\Nt$ the Cartan subalgebra of
$\Ng$, by $r$ the rank of $\Ng$, by $\Delta$ the set of the roots
$\alpha$, by $\phi$ the highest root, by $\alpha^\vee$ and
$\phi^\vee$ the coroots, by $e_\alpha$ the step generators
corresponding to roots $\alpha$, by
$\alpha_i,\alpha_i^\vee,\lambda_i,$ and $\lambda_i^\vee$ for
$i=1,\dots,r$, the simple roots, coroots, weights and coweights
and by $Q$, $Q^\vee$, $P$ and $P^\vee$ the corresponding lattices.
In particular, \qq
\Ng^\NC\,=\,\Nt^\NC\oplus\Big(\mathop{\oplus}\limits_{\alpha\in
\Delta}\NC e_{\alpha}\Big) \qqq is the root decomposition of the
complexification of $\Ng$. The standard normalization of $\,\tr\,$
requires that the long roots have length square 2 so that
$\,\alpha^\vee=2\alpha/\tr\,\alpha^2$. \,The highest root
$\phi=\phi^\vee=\sum k_i^\vee\alpha_i^\vee$, where $k_i^\vee$ are
the dual Kac labels. The dual Coxeter number $h^\vee=1+\sum
k_i^\vee$. The positive Weyl chamber $\,C_W\subset\Nt\,$ is
composed $\,\tau\in\Nt\,$ such that $\tr\,\tau\alpha_i\geq0$ for
each $i$ and the positive Weyl alcove $\,A_W\,$ is its subset
restricted by the additional equality $\tr\,\tau\phi\leq1$. \,It
is the $r$-dimensional simplex in $\Nt$ with vertices $0$ and
${1\over k_i^\vee}\lambda_i$. \vskip 0.2cm

On $G$ we shall consider the unique up to normalization left- and
right-invariant real closed 3-form \qq
{H}\,=\,{_1\over^{12\pi}}\,\tr\,(g^{-1}dg)^3\,. \label{chi} \qqq It
will be convenient to parametrize the Lie algebra and the Lie
group elements using the adjoint action of $G$. Elements in $\Ng$
and in $G$ may be written, respectively, as \qq
\gamma\hspace{0.07cm} \tau\gamma^{-1}\qquad{\rm and}\qquad
\gamma\, \ee^{2\pi i\tau}\gamma^{-1} \label{para} \qqq for some
$\gamma\in G$ and $\tau\in\Nt$. The group elements $\gamma$ are
determined up to the right multiplication by $\gamma_0$ in the
isotropy subgroups $G^0_\tau$ and $G_\tau$ composed of elements of
$\,G$ commuting with $\tau$ and with $\ee^{2\pi i\tau}$,
respectively. Clearly, $\,G^0_\tau\subset G_\tau$. The subgroups
$\,G^0_\tau$ and $\,G_\tau$ are connected. They correspond to the
Lie subalgebras $\,\Ng^0_\tau$ and $\,\Ng_\tau$ of $\,\Ng$ with
complexifications \qq
{\Ng^0_\tau}^\NC\,=\,\Nt^\NC\oplus\Big(\mathop{\oplus}\limits_{\alpha
\in\Delta^0_\tau}\NC e_\alpha\Big),\qquad
\Ng_\tau^\NC\,=\,\Nt^\NC\oplus\Big(\mathop{\oplus}\limits_{\alpha\in
\Delta_\tau}\NC e_\alpha\Big), \qqq where \qq
\Delta^0_\tau\,=\,\{\,\alpha\in\Delta\,|\,\tr\,\tau\alpha=0\,\}\,,
\qquad\Delta_\tau\,=\,\{\,\alpha\in\Delta\,|\,\tr\,\tau\alpha\in\NZ\,\}\,.
\label{DgG} \qqq The sets of Lie algebra and group elements
(\ref{para}) with fixed $\tau$ form, respectively, the (co)adjoint
orbit $\CO_\tau\subset\Ng$ and the conjugacy class
$\,\CC_\tau\subset G$. We have \qq \CO_\tau\cong
G/G^0_\tau\qquad{\rm and}\qquad\CC_\tau\cong G/G_\tau \qqq so that
$\CO_\tau$ and $\CC_\tau$ are connected and simply connected.
\vskip 0.2cm

The choice of $\tau$ in the parametrizations (\ref{para}) may be
fixed if we demand that $\tau\in C_W$ or $\tau\in A_W$,
respectively. Consider the open subsets $U_0\subset\Ng$ and
$O_0\subset G$ composed of elements of the form (\ref{para}) for
$\tau\in A_W$ such that $\tr\,\tau\phi<1$. They are related by the
exponential map $\,\Ng\ni X\mapsto\ee^{2\pi iX}\in G$. Using the
parametrization (\ref{para}) it is easy to see that the
exponential map is injective on $U_0$ because $G^0_\tau=G_\tau$ if
$\tr\,\tau\phi<1$. Indeed, the last inequality implies that
$\tr\,\alpha\phi<1$ for all positive roots. Similarly one shows
that the derivative of the exponential map is invertible on $U_0$.
It follows that the exponential map is a diffeomorphism between
$U_0$ and $O_0$. Composing the latter with the homotopy
$(t,X)\mapsto tX$ of $U_0$ and using the Poincare Lemma, one may
obtain a 2-form ${B}_0$ on $O_0$ such that $d{B}_0={H}$.
Explicitly, in the parametrization (\ref{para}), \qq
{B}_0(\gamma\,\ee^{2\pi i\tau}\gamma^{-1})\ =\ Q(\gamma\,\ee^{2\pi
i\tau}\gamma^{-1})\ +\ i\,\tr\, \tau(\gamma^{-1}d\gamma)^2\,,
\label{om0} \qqq where \qq Q(\gamma\,\ee^{2\pi i\tau}\gamma^{-1})\
=\ {_1\over^{4\pi}} \,\tr\,(\gamma^{-1}d\gamma) \,\ee^{2\pi
i\tau}(\gamma^{-1}d\gamma)\,\ee^{-2\pi i\tau}\,. \label{om} \qqq
These 2-forms will be the building blocks for the local data of a
gerbe on $G$ with curvature ${H}$ for $G=SU(N)$. \vskip 0.2cm

The group $SU(N)$ has rank $r=N-1$. It is simply laced so that
$\alpha_i=\alpha_i^\vee$ and $\lambda_i=\lambda_i^\vee$. \,For the
Cartan subalgebra composed of the diagonal $su(N)$ matrices, we
may take \qq &&\alpha_i\,=\,{\rm diag}(0,\dots,1,
-1,\dots,0)\,,\alabel{rowe}{a}\cr &&\lambda_{i}\,=\,{\rm
diag}({_{N-i}\over^{N}},\dots
{_{N-i}\over^{N}},{_{-i}\over^{N}},\dots,
{_{-i}\over^{N}})\aeqno{b}\cr \qqq \vskip -0.4cm \noindent with
$\,1\,$ and the last $\,N-i\over N\,$ at the $\,(i-1)^{\rm th}$
place counting from zero. The highest root is $\phi={\rm
diag}(1,0,\dots,0,-1)$, the Kac labels are $\,k_i^\vee=1$ and the
dual Coxeter number is equal to $\,N$. \,The center of $SU(N)$ is
composed of the elements $z_i=\ee^{2\pi i\lambda_i}$ for $i=0,1,
\dots,r$, where we set $\lambda_0=0$. Let us consider the sets
$O_i=z_i O_0\subset SU(N)$. We may define 2-forms ${B}_i$ on $O_i$
by the pullback of ${B}_0$ from $\,O_0\,$: \qq
{B}_i(g)\,=\,{B}_0(z_i^{-1}g)\,. \label{omi} \qqq Clearly,
$\,d{B}_i={H}$. \,If $\,g=\gamma\,\ee^{2\pi i\tau}\gamma^{-1}$
then $z_i^{-1}g=\gamma\,\ee^{2\pi i(\tau-\lambda_i)}\gamma^{-1}$.
For each $i$ there is an element $w_i$ in the normalizer
$N(T)\subset G$ of the Cartan subgroup $T\subset G$ such that if
$\tau\in A_W$ then also $w_i(\tau-\lambda_i)w_i^{-1}\equiv
\sigma_i(\tau)$ is in $A_W$. Explicitly, the element $w_i$ induces
the Weyl group transformations \qq {\rm diag}(a_0,\dots,a_{r})\
\mapsto&&w_i\,{\rm diag}(a_0,\dots,a_{r}) w_i^{-1}\cr &&={\rm
diag}(a_{i},\dots,a_{r},a_0,\dots,a_{i-1})\quad \qqq and
$\,\sigma_i(\lambda_j)=\lambda_{[j-i]}$, where $\,[j-i]=(j-i)\
{\rm mod} \,N$. It follows that \qq z_i^{-1}\gamma\,\ee^{2\pi
i\tau}\gamma^{-1}\,=\,\gamma\,\ee^{2\pi
i(\tau-\lambda_i)}\gamma^{-1}\,=\,\gamma\,w_i^{-1}\ee^{2\pi i\,
\sigma_i(\tau)}w_i\,\gamma^{-1}\,. \label{zig} \qqq Substituting
into (\ref{om0}) and (\ref{om}), we obtain \qq
{B}_i(\gamma\,\ee^{2\pi i\tau}\gamma^{-1})\ =\ Q(\gamma\,\ee^{2\pi
i\tau}\gamma^{-1})\ +\
i\,\tr\,(\tau-\lambda_i)(\gamma^{-1}d\gamma)^2\,. \label{omi1}
\qqq The sets $O_i$ are composed of group elements $g$ such that
$\tau\in A_W$ and $\,\tr\,\tau\alpha_i>0$ in the parametrization
(\ref{para}). For such $\tau$, $\,G_\tau=G^0_{\tau-\lambda_i}$. In
terms of the eigenvalues of the unitary matrices $\,g\,$ given by
the entries of $\,\ee^{2\pi i\,\tau}={\rm diag}(\ee^{2\pi i\,a_0},
\dots,\ee^{2\pi i\, a_{r}})\,$ such that $\,a_0\geq \cdots\geq
a_{r}\,$ and $\,a_0-a_{r}\leq 1$, \,the sets $O_i$ are defined by
the inequality  $a_i>a_{i+1}$ and $O_0$ by $a_0-a_{r}<1$. Clearly,
$\,\,G\,=\hspace{-0.3cm}\mathop{\cup}\limits_{i=0,1,\dots r}
\hspace{-0.3cm}O_i\,$. \vskip 0.2cm

In the first step in the construction of the gerbe on $SU(N)$
with curvature ${H}$ we set \qq Y\
\,=\mathop{\sqcup}\limits_{i=0,1,\dots,r}\hspace{-0.2cm}O_i
\qquad{\rm and}\qquad{B}|_{_{O_i}}\,=\,{B}_i\,. \label{fsg} \qqq
As discussed before, the fiber products
$\,Y^{[n]}=\sqcup\,O_{i_1\dots i_n}$. \,To continue the
construction of the gerbe, note that on the intersections $O_{ij}$
that form $Y^{[2]}$ (with $i,j=0,1,\dots$), \qq
({B}_j-{B}_i)(\gamma\,\ee^{2\pi i\tau}\gamma^{-1}) \ =\
-i\,\tr\,\lambda_{ij}(\gamma^{-1}d\gamma)^2\,, \label{tobc} \qqq
where $\,\lambda_{ij}\equiv\lambda_j-\lambda_i$. \,The expression
on the right hand side coincides with the one for the
Kirillov-Kostant symplectic form $\,{F}_{_{\lambda_{ij}}}$ on the
(co)adjoint orbit $\,\CO_{_{\lambda_{ij}}}$ passing through
$\,\lambda_{ij}$. \,More exactly, there is a map $\,O_{ij}\ni
g\rightarrow \rho_{ij}(g)\in\CO_{_{\lambda_{ij}}}$ such that \qq
{B}_{j}-{B}_{i}\,=\,\rho_{ij}^*\,{F}_{\lambda_{ij}}\,.
\label{moex} \qqq This map is defined as follows. For $g\in
O_{ij}$, there exist two Lie algebra elements $X_i,\,X_j\in U_0$
such that $\,z_i^{-1}g=\ee^{2\pi i X_i}$ and
$\,z_j^{-1}g=\ee^{2\pi i X_j}$. Then $\,\rho_{ij}(g)=X_{i}-X_{j}$.
\,Explicitly, if $\,g=\gamma\, \ee^{2\pi i\tau}\gamma^{-1}$ then,
as may be seen from (\ref{zig}), $\,X_i=\gamma
w_i^{-1}\sigma_i(\tau)
\,w_i\gamma^{-1}=\gamma(\tau-\lambda_i)\gamma^{-1}$ and similarly
for $X_j$. Hence \qq \rho_{ij}(\gamma\,\ee^{2\pi
i\tau}\gamma^{-1})\,=\,\gamma\, \lambda_{ij}\gamma^{-1}\,.
\label{map} \qqq Another way to see that the map (\ref{map}) is
well defined is to check that if $\,\tr\,\tau\alpha_i>0\,$ and
$\,\tr\,\tau\alpha_j>0\,$ for $\tau\in A_W$ then the isotropy
subgroup $G_\tau$ necessarily is contained in the isotropy
subgroup $\,G^0_{\lambda_{ij}}$. Since for $i<j$ \qq
\lambda_{ij}\,=\,{\rm
diag}(\mathop{{_{i-j}\over^{N}},\dots,{_{i-j}\over
^{N}}}\limits_{i\ {\rm
times}},\mathop{{_{N+i-j}\over^{N}},\dots,
{_{N+i-j}\over^{N}}}\limits_{(j-i)\ {\rm times}},
\mathop{{_{i-j}\over^{N}},\dots,{_{i-j}\over^{N}}}\limits_{(N-j)\
{\rm times}})\,, \label{wedi} \qqq the isotropy subgroup
$\,G^0_{\lambda_{ij}}$ is composed of block matrices $\gamma_0$
that preserve the subspace $V_{ij}\subset\NC^{N}$ of vectors
with vanishing first $i$ and last $N-j$ coordinates and its
orthogonal complement. The coadjoint orbit
$\,\CO_{_{\lambda_{ij}}} \cong G/G^0_{\lambda_{ij}}$ may be
identified with the Grassmannian $\,Gr_{ij}$ of
$(j-i)$-dimensional subspaces $\,\gamma(V_{ij})\,$ in $\NC^{N}$
with $\gamma\in SU(N)$. \vskip 0.2cm

The Kirillov-Kostant theory \cite{Kiril}\cite{Kost} provides an
explicit construction of a hermitian line bundle
$\,{L}_{_{\lambda}}$ over the coadjoint orbit $\,\CO_{_{\lambda}}$
with connection of curvature $\,{F}_{_{\lambda}}$, \,provided
that $\,\lambda\,$ is a weight which holds for $\,\lambda=\lambda_{ij}$.
\,The bundle is
obtained by dividing the trivial line bundle over $G$ by the
equivalence relation \qq {L}_{_{\lambda}}\ =\
(G\times\NC)/\mathop{\sim}\limits_{^\lambda}\,, \qqq where \qq
(\gamma,\,\zeta)\ \mathop{\sim}\limits_{^\lambda}\
(\gamma\gamma_0, \,\chi_{_{\lambda}}(\gamma_0)^{-1}\zeta)\,,
\label{iga} \qqq for $\,\gamma_0\in G_{_{\lambda}}^0$. We shall
denote the corresponding equivalence classes by
$\,[\gamma,\zeta]_{_\lambda}$. \,Above,
$\,\chi_{_\lambda}:G_{_{\lambda}}^0\rightarrow U(1)\,$ stands for
the group homomorphism (character) such that \qq
\partial_t|_{_{t=0}}\,\chi_{_{\lambda}}(\ee^{itX_0})\,=\,
i\,\tr\,\lambda X_0\label{chil} \qqq for
$\,X_0\in\Ng_{_\lambda}^0$. Existence of $\,\chi_{_{\lambda}}$ is
guaranteed if $\,\lambda$ is a weight. The formula \qq
\nabla\,=\,d\,+\,\tr\,\lambda(\gamma^{-1}d\gamma) \label{lcon}
\qqq defines a connection in the trivial bundle over $G$ that
descends to the bundle $\,{L}_{_{\lambda}}$. The curvature of that
connection is equal to $\,F_\lambda= -i\,{\rm
\tr}\,\lambda(\gamma^{-1}d\gamma)^2$. \,For $i<j$, \qq
\chi_{_{\lambda_{ij}}}(\gamma_0)\,=\,{\rm
det}(\gamma_0|_{_{V_{ij}}})\,. \label{chii}\qqq
\vskip 0.2cm

We shall also use an alternative description of the bundle
$\,{L}_{_{\lambda_{ij}}}$. Let, for $i<j$, $\,E_{ij}$ be the
tautological vector bundle over the Grassmannian $\,Gr_{ij}$ whose
fiber at $\gamma(V_{ij})$ is this very subspace in $\,\NC^{N}$.
Clearly, $\,E_{ij}$ is a $(j-i)$-dimensional subbundle of the
trivial bundle $\,Gr_{ij}\times\NC^{N}$ from which it inherits the
hermitian structure. It may be equipped with the connection \qq
\nabla\,=\,P_{_{\gamma(V_{ij})}}d \label{econ} \qqq where
$\,P_{_{\gamma(V_{ij})}}=\gamma\,P_{_{V_{ij}}}\gamma^{-1}$ denotes
the orthogonal projection in $\,\NC^{N}$ on $\gamma(V_{ij})$. The
line bundle $\,{L}_{_{\lambda_{ij}}}$ may be identified with the
top exterior power $\,\wedge^{j-i}E_{ij}$ of the bundle $\,E_{ij}$
by the mapping \qq [\gamma,\,\zeta]_{_\lambda}\ \longmapsto\ \zeta
\,\gamma e_{i}\wedge\dots\wedge\gamma e_{j-1} \,, \label{mapp}
\qqq where $\,e_l$, $l=0,1,\dots,r$, \,are vectors of the
canonical basis of $\,\NC^{N}$. It is easy to see that this
mapping is compatible with the equivalence relation (\ref{iga})
and that it preserves the connection if we equip
$\,\wedge^{j-i}E_{ij}$ with the one inherited from $\,E_{ij}$.
\vskip 0.2cm

We may perform now the next step in the construction of the gerbe
$\,\CG=(Y,{B},{L},\mu)\,$ on $SU(N)$ with $Y$ and ${B}$ given by
(\ref{fsg}). We shall define the line bundle ${L}$ with connection
over $Y^{{[2]}}=\sqcup\,O_{ij}$ by \qq {L}|_{_{O_{ij}}}\ :=\
\rho_{ij}^*{L}_{_{\lambda_{ij}}}\,.\label{lbun} \qqq Equation
(\ref{moex}) guarantees that the curvature of $\,{L}$ satisfies
requirement (\ref{2}). We still have to construct the isomorphism
$\mu$ providing ${L}$ with the groupoid structure. It will be
given by the isomorphisms between the bundles on the triple
intersections $O_{ijk}$ \qq
\mu_{ijk}\,:\,\rho_{ij}^*{L}_{_{\lambda_{ij}}}\otimes\,
\rho^*_{jk}{L}_{_{\lambda_{jk}}}\longrightarrow\,
\rho^*_{ik}{L}_{_{\lambda_{ik}}}\,. \qqq We may assume that
$i<j<k$. Then the isomorphism $\mu_{ijk}$ is determined by the
natural map \qq &&(\gamma e_{i}\wedge\dots\wedge\gamma
e_{j-1})\otimes (\gamma e_{j}\wedge\dots\wedge\gamma
e_{k-1})\cr\cr &&\hspace{3cm}\longmapsto\ \gamma
e_{i}\wedge\dots\wedge\gamma e_{j-1}\wedge \gamma
e_{j}\wedge\dots\wedge\gamma e_{k-1} \qqq and the associativity
(\ref{assoc}) becomes obvious. \vskip 0.2cm

This ends the construction of the gerbe $\,\CG=(Y,{B},{L},\mu)\,$
on the special unitary group $SU(N)$. Since
$\,H^2(G,U(1))=\{1\}\,$ for simply connected groups, $\,\CG\,$ is,
up to stable isomorphism, a unique gerbe on $SU(N)$ with curvature
${H}$ given by (\ref{chi}). The tensor powers
$\,\CG^k=(Y,k{B},{L}^k,\mu^k)\,$ of $\,\CG$ for $k\in\NZ$ give the
gerbes on $SU(N)$ with curvature $k{H}$, again unique up to stable
isomorphism. In particular, $\,\CG^{-1}$ is the gerbe dual to
$\,\CG\,$ ($\mu^{-1}$ is the inverse of the transpose of $\,\mu$).

\nsection{Gerbes on groups covered by $SU(N)$}

Let us consider now the case of non-simply connected groups $G'$,
quotients of simply connected groups $\,G\,$ by a subgroup
$\,\CZ\,$ of their center. The closed 3-form $\,H\,$ of
(\ref{chi}) descends from $G$ to $G'$ to a 3-form $\,H'$ and we
shall be interested in the gerbes on $G'$ with curvature
proportional to $H'$. We shall restrict ourselves to the case when
$\,G=SU(N)\,$ and $\,G'=SU(N)/\CZ$, \,where $\,\CZ\,$ is a cyclic
group of order $\,N'$ such that $\,N=N'N''$. \,More explicitly,
$\, \CZ=\{z_a\,|\,{N''}\ {\rm divides}\ a\}$. \,As was shown in
\cite{FGK}, the forms $\,{1\over2\pi}kH'\,$ on $\,SU(N)/\CZ\,$ are
integral for even $k$ if $N'$ is even and $N''$ is odd and for
integer $k$ in the other cases. In the present Section, we shall
construct the corresponding gerbes $\CG'_k$ on $\,G'$. Not
surprisingly in view of the discussion in Appendix 1 of
\cite{FGK}, the construction reduces to solving a simple
cohomological problem in the group cohomology of
$\,\CZ\cong\NZ_{N'}$, \,see Appendix A for a brief summary on
discrete group cohomology. The resulting gerbe will still be
unique up to stable isomorphism since $\,H^2(G',U(1))=\{1\}\,$ in
the case at hand. \vskip 0.2cm

We shall take $\,Y'=\mathop{\sqcup}O_i$ with $\,O_i\,$ the open subsets
of $\,SU(N)\,$ constructed before.  $\,\pi'$ will be the natural projection
from $\,Y'$ on the quotient group $\,SU(N)/\CZ$. \,For the curving
of the gerbe $\,\CG'_k$, we shall take the 2-form $\,B'$ equal to
$kB_{i}$ on $\,O_{i}$, see (\ref{omi}). We have
\qq
{Y'}^{{[n]}}\,=\,\{\,((g,i),\,(z_{a_1}^{-1}g,i'_1),\,\dots,\,
(z_{a_{n-1}}^{-1}g,i'_{n-1}))\ |
\ z_{a_m}\in \CZ\,\}\,, \qqq
for $\, g\in O_{i\,i_1\dots i_{n-1}}$, where $\,i_m=[i'_m+a_m]$.
\,We may then identify
\qq
{Y'}^{{[n]}}\ \cong\ \mathop{\sqcup}\limits_{a_1,..a_{n-1}}
\,\mathop{\sqcup}\limits_{i,i_1,..,i_{n-1}}O_{i\,i_1..i_{n-1}}\,.
\qqq
The hermitian line bundle $\,L'$ with connection over $O_{ij}\,\subset
\,{Y'}^{{[2]}}$ should have the curvature
\qq p_2^*(kB_{j'})\,-\,p_1^*(kB_{i})\ =\ kB_j\,-\,kB_{i}\ =\,-ik
\,\tr\,\lambda_{ij}(\gamma^{-1} d\gamma)^2\, \qqq
in the parametrization $\,g=\gamma\,\ee^{2\pi i\tau}\gamma^{-1}$.
We set \qq L'|_{O_{ij}}\ =\ \rho_{ij}^*\,L_{_{\lambda_{ij}}}^k\label{L0i}
\qqq
where $\,L_{_{\lambda_{ij}}}$ is the line bundle over the coadjoint
orbit $\,\CO_{_{\lambda_{ij}}}$ described in the previous section and
$\,\rho_{ij}(\gamma \,\ee^{2\pi i\tau}\gamma^{-1})=
\gamma\,\lambda_{ij}\gamma^{-1}$. Recall that the elements in
$\,L_{_{\lambda_{ij}}}^k$ may be viewed as equivalence classes
$\,[\gamma,\zeta]_{_{k\lambda_{ij}}}$, \,see (\ref{iga})
\vskip 0.2cm

In the next step we should construct the isomorphism $\,\mu'$ of
line bundles over $\,{Y'}^{{[3]}}$ defining the groupoid
multiplication in $\,L'$, see (\ref{3}). The elements in
$\,p_{12}^*L'$ over $\,g\in O_{ijl}\subset {Y'}^{{[3]}}$ with
$\,j=[j'+a]\,$ and $\,l=[l'+b]\,$ are given by the classes
$\,[\gamma,\zeta]_{_{k\lambda_{ij}}}$. \,Those in $\,p_{13}^*L'$
by the classes $\,[\gamma,\zeta]_{_{k\lambda_{il}}}$. \,As for the
elements of $\,p_{23}^*L'$, they correspond to the classes
$\,[\gamma w_a^{-1},\zeta]_{_{k\lambda_{j'[l-a]}}}$ since
$\,z_a^{-1}g=\gamma w_a^{-1}\ee^{2\pi
i\sigma_a(\tau)}w_a\gamma^{-1}$, see (\ref{zig}), and
$z_bz_a^{-1}=z_{[b-a]}$. The isomorphism $\mu'$ has then to be
given by \qq
\mu'\Big([\gamma,\zeta]_{_{k\lambda_{ij}}}\otimes[\gamma
w_a^{-1},\zeta']_{_{k\lambda_{j'[l-a]}}} \Big)\,=\,[\gamma,
\,u_{ijl}\,\zeta\zeta']_{_{k\lambda_{il}}}\label{muij}\qqq for
$U(1)$-valued functions $\,u_{ijl}$  on $\,O_{ijl}$ whose
dependence on $\,a\,$ and $\,b\,$ has been suppressed in the
notation. These functions must be constant for $\,\mu'$ to
preserve the connections. Note that $\mu'$ depends on the choice
of matrices $w_{a}$ defined up to the multiplication by elements
of the Cartan subgroup $T$. The latter dependence may, however, be
absorbed in the choice of $\,u_{ijl}$. \vskip 0.2cm

As we show in Appendix B by a direct verification, the
associativity of the product defined by $\,\mu'$ imposes the
condition \qq
u_{j'[l-a][n-a]}\,u_{iln}^{-1}\,u_{ijn}\,u_{ijl}^{-1}\ =\
\chi_{_{k\lambda_{l'[n-b]}}}(w_b\,w_a^{-1}
w_{[b-a]}^{-1})\,.\label{assop} \qqq Upon taking
$\,i=j'=l'=n'=0\,$ and setting $\,u_{0ab}\equiv u_{a[b-a]}$,
\,relation (\ref{assop}) reduces (upon the shift
$\,b\mapsto[a+b],\ c\mapsto[a+b+c]\,$) to the condition \qq
u_{bc}\,\,u_{[a+b]c}^{-1}\,u_{a[b+c]}\,u_{ab}^{-1} \,=\,
\chi_{_{k\lambda_{c}}}(w_{[a+b]}\,w_a^{-1}
w_b^{-1})\,\equiv\,U_{abc}\label{chc0}\qqq which may be
interpreted in terms of the discrete group cohomology
$\,H^*(\CZ,U(1))\,$ with coefficients in $U(1)$, see Appendix A.
The $\,U(1)$-valued 3-cochain $\,(U_{abc})\,$ on the cyclic group
$\,\CZ\,$ satisfies the cocycle condition \qq
\,U_{bcd}\,\,U_{[a+b]cd}^{-1}
\,U_{a[b+c]d}\,\,U_{ab[c+d]}^{-1}\,U_{abc}\ =\ 1\qqq easy to
verify with the use of the relation \qq
\chi_{_{k\lambda_d}}(w_c\,t\,w_c^{-1})\ =\
\chi_{_{k\lambda_{[c+d]}
}}(t)\,\,\chi^{-1}_{_{k\lambda_c}}(t)\label{rtn} \qqq holding for
$\,t\in T$. The condition (\ref{chc0}) requires that
$\,(U_{abc})\,$ be a coboundary. This does not have to be always
the case since $\,H^3(\CZ,U(1))\cong\NZ_{N'}$, \,see Appendix A.
Given a solution $\,(u_{ab})\,$ of \,(\ref{chc0}), \qq u_{ijl}\ =\
u_{a[b-a]}\,\,\chi_{_{k\lambda_{l'}}}(w_b\,
w_a^{-1}w_{[b-a]}^{-1})\label{asa} \qqq solves \,(\ref{assop}), as
a straightforward check with the use of \,(\ref{rtn}) shows.
\vskip 0.3cm

We still have to study when (\ref{chc0}) may be satisfied. In the
action on the vectors of the canonical bases of $\,\NC^N$, \,the
matrices $w_{a}$ take the form $\,w_{a}e_l=u_{a}\,e_{[l-a]}$,
\,where $\,u_{a}$ are diagonal matrices such that $\,{\rm
det}(u_{a})= (-1)^{a(N-a)}\,$ assuring that $\,{\rm det}(w_{a})=1$.
\,In particular, we may take $\,u_a$ proportional to the unit matrix:
\qq u_{a}\ =\ \cases{\hbox to 3,5cm{$\,1$\hfill}{\rm for}\quad N'\
\ \,{\rm odd\ \ \,or}\ \ N''\ \,{\rm even}\,,\cr\hbox to
3,5cm{$\,(-1)^{{N'' \over N'}\,a'(N'-a')}$\hfill}{\rm for}\quad
N'\ \ {\rm even\ \,and}\ \ N''\ \,{\rm odd}\,,}\label{ui} \qqq
where $\,a=a'N''\,$ (here and below, $\,(-1)^x\equiv\ee^{\pi ix}$).
\,For that choice, \qq
w_{[a+b]}w_a^{-1}w_b^{-1}\ =\ \cases{1\,,\cr (-1)^{{N''\over
N'}\,m_{a'b'}}\,,} \qqq respectively, for
$\,m_{a'b'}=(a'+b')(N'-a'-b')-a'(N'-a') -b'(N'-b')\,$ and the
associativity condition (\ref{chc0}) becomes \qq
u_{bc}\,\,u_{[a+b]c}^{-1} u_{a[b+c]}\,u_{ab}^{-1}\ =\
\cases{\,1\,,\cr\,(-1)^{{k(N'')^2\over N'}\,m_{a'b'}\,c'}\,.} \qqq
and may be solved by taking \qq u_{ab}\ =\ \cases{\hbox to
4.2cm{$\,1$\hfill}{\rm for}\ \,N'\ {\rm odd\ \,or}\,\ N''\ \,{\rm
even}\,,\cr\hbox to 4.2cm{$\, (-1)^{-{k({N''})^2\over
N'}\,a'(N'-a')b'}$\hfill}{\rm for}\ \,N'\ {\rm even},\ \,N''\
\,{\rm odd\ and}\,\ k\,\ {\rm even}\,.}\label{uab} \qqq There is
no solution for $N'$ even and  $N''$ and $k$ odd. \vskip 0.3cm

This ends the construction of the gerbes $\,\CG'_k$ on $\,SU(N)/\CZ\,$
with curvature $\,kH'\,$ for all the values of $\,k\,$ where
the latter is an integral 3-form. Clearly, $\,\CG'_k\cong{\CG'_1}^k$
for $N'$ odd or $N''$ even and $\,\CG'_k\cong{\CG'_2}^{k/2}$
for $N'$ even and $N''$ odd.

\nsection{Gerbes on discrete quotients}

The above construction provides an illustration of a more general
one of gerbes on spaces of orbits of a discrete group. The general
case, that we shall briefly discuss in the present section which
is somewhat parenthetical with respect to the main course of the
exposition, sheds more light on the appearance of discrete group
cohomology (the so called "discrete torsion" \cite{Vafa}), as was
noticed first in \cite{Sharpe}. \vskip 0.3cm

Suppose that $\CG$ is a gerbe on $M$ with curvature $H$ and that a
finite group $\Gamma$ acts on $M$ preserving $H$. We may ask the
question if the action preserves $\CG$ in the sense that for each
$\gamma\in\Gamma$ the gerbe $\gamma\CG=(Y_\gamma,\,B,\,L,\,\mu)$,
where $Y_\gamma=Y$ as the space but has the projection on $M$
replaced by $\gamma\circ\pi$, is stably isomorphic to $\CG$.
\,Recall that this means that there exists a hermitian line bundle
$\,N^{^\gamma}$ over $\,Z_\gamma=Y\times_MY_\gamma\,$ with
connection of curvature $\,F^{^\gamma}$ such that \qq
\sigma_\gamma^*B\ =\ \sigma^*B\ +\ F^{^\gamma}\, \label{curr}\qqq
with $\,\sigma,\ \sigma_\gamma$ denoting the projections from
$\,Z_\gamma$ to $Y$ and $Y_\gamma$, and that there exists an
isomorphism \qq \iota_\gamma:\,\,{\sigma^{[2]}}^{*}L\otimes
p_1^*(N^{^\gamma})^{-1} \otimes p_2^*{N^{^\gamma}}\
\longrightarrow\ {\sigma_\gamma^{[2]}}^{*} {L} \qqq of hermitian
line bundles with connection that preserves the groupoid
multiplication. In particular,
 \qq L_{_{(y_1,y_2)}}\otimes
(N^{^{\gamma}})^{-1}_{_{(y_1, y'_1)}}\otimes
N^{^\gamma}_{_{(y_2,y_2')}}\
\mathop{\longrightarrow}^{\iota_\gamma}\ \, L_{_{(y'_1,y'_2)}}\,.
\label{zgg} \qqq if
$\,\pi(y_1)=\pi(y_2)=\gamma\,\pi(y'_1)=\gamma\,\pi(y'_2)$. \,For
$\gamma=1$ one may take $\,N^1=L\,$ and $\,\iota_{_1}$ defined by
$\,\mu$.  \vskip 0.2cm

Suppose now that $\Gamma$ acts without fixed points so that
$M/\Gamma$ is non-singular. We would like to construct a gerbe
$\,\CG_{_\Gamma}=(Y_{_\Gamma},\,B_{_\Gamma},\,L_{_\Gamma},\,\mu_{_\Gamma})$
on $M/\Gamma$ with the curvature equal to the projection $H'$ of $\,H$.
\,We shall set $\,Y_{_\Gamma}=Y\,$ with the projection
$\,\pi_{_\Gamma}$ on $\,M/\Gamma$ given by the composition of
$\,\pi:Y\rightarrow M\,$ with the canonical projection on the
quotient space. The curving $\,B_{_\Gamma}$ will be taken equal to
$B$. Note that \qq Y_{_\Gamma}^{{[2]}}\ =\
\mathop{\sqcup}\limits_{\gamma\in\Gamma}Z_\gamma\,.\label{yyg}\qqq
Let us take \qq L_{_\Gamma}|_{{Z_\gamma}}\ =\ N^{^\gamma}\otimes
\pi_{_1}^*P^{^\gamma}\,,\label{lg} \qqq where $\,P^{^\gamma}$ is a
flat line bundle on $M$ and $\,\pi_{_1}(y,y')=\pi(y)$. \,Relation
(\ref{curr}) assures then that the curvature of $\,L_{_\Gamma}$ is
related to the curving by (\ref{2}). \vskip 0.2 cm

We still have to define the groupoid multiplication
$\,\mu_{_\Gamma}$ in $\,L_{_\Gamma}$ that is an isomorphism of
line bundles over \qq Y_{_\Gamma}^{{[3]}}\ =
\mathop{\sqcup}\limits_{\gamma_1,\gamma_2\in\Gamma}\{\,(y,y',y''\in
Y^{3}\,\,|\,\,\pi(y)=\gamma_1\pi(y'),\
\pi(y')=\gamma_2\pi(y'')\,\}\,.\qqq On the $(\gamma_1,\gamma_2)$
component of $\,Y_{_\Gamma}^{{[3]}}$, \qq
\mu_{_{\Gamma}}:\,p_{_{12}}^*N^{^{\gamma_1}}\otimes\pi_{_1}^*P^{^{\gamma_1}}
\otimes p_{_{23}}^*N^{^{\gamma_2}}\otimes\pi_{_2}^*P^{^{\gamma_2}}
\ \longrightarrow\
\,p_{_{13}}^*N^{^{\gamma_1\gamma_2}}\otimes\pi_{_1}^*
P^{^{\gamma_1\gamma_2}}\,,\label{mug}\qqq where $\pi_{_n}=\pi\circ
p_{_n}$. A necessary condition for existence of $\,\mu_{_\Gamma}$
is that the bundle \qq p_{_{12}}^*(N^{^{\gamma_1}})^{-1}\otimes
p_{_{23}}^*(N^{^{\gamma_2}})^{-1}\otimes
p_{_{13}}^*N^{^{\gamma_1\gamma_2}}\ \equiv\ \tilde R^{^{\gamma_1,
\gamma_2}} \label{tofl} \label{nnr}\qqq be isomorphic to a
pullback $\,\pi_{_1}^*R^{^{\gamma_1,\gamma_2}}$ of a flat bundle
over $\,M$. That this condition is fulfilled may be seen the
following way. First note that the map $\,\iota_\gamma$ of
(\ref{zgg}) defines isomorphisms \qq
(N^{^\gamma})_{_{(y_1,y'_1)}}\ \longrightarrow\
N^{^\gamma}_{_{(y_2,y'_2)}}\otimes L_{_{(y_1,y_2)}}\otimes
L^{\,-1}_{_{(y'_1,y'_2)}}\,. \label{iso3} \qqq Combining the
latter with the groupoid multiplication in $L$ one obtains
canonical isomorphisms between the fibers of the bundle $\,\tilde
R^{^{\gamma_1,\gamma_2}}$ over the triples $(y_1,y'_1,y''_1)$ and
$(y_2,y'_2,y''_2)$ in $\,Y_{_\Gamma}^{{[3]}}$ with the same
projections on $M$. Such isomorphisms define the descent data for
the bundle $\,\tilde R^{^{\gamma_1,\gamma_2}}$, \,see the
discussion around (\ref{decp}). The existence of canonical bundle
$\,R^{^{\gamma_1,\gamma_2}}$ and of the canonical isomorphism
$\,\tilde
R^{^{\gamma_1,\gamma_2}}\cong\pi_{_1}^*R^{^{\gamma_1,\gamma_2}}$
follows then by the descent principle. Besides, there exists a
canonical isomorphism \qq R^{^{\gamma_1,\gamma_2}}\otimes
R^{^{\gamma_1\gamma_2, \gamma_3}}\ \cong\
R^{^{\gamma_1,\gamma_2\gamma_3}}\otimes
(\gamma_1^{-1})^*R^{^{\gamma_2,\gamma_3}}\,, \label{123} \qqq as
may be easily seen on the level of $\,\tilde R\,$ bundles. \vskip
0.2cm

To construct isomorphisms $\mu_{_\Gamma}$ of (\ref{mug}) becomes then
equivalent to specifying a family of isomorphisms
$\,\iota_{_{\gamma_1,\gamma_2}}$ of flat bundles over $M$ \qq
\iota_{_{\gamma_1,\gamma_2}}:\,P^{^{\gamma_1}}\otimes
(\gamma_1^{-1})^*P^{^{\gamma_2}}\ \rightarrow\
P^{^{\gamma_1\gamma_2}}\otimes
R^{^{\gamma_1,\gamma_2}}\,.\label{iogg} \qqq The associativity of
$\,\mu_{_\Gamma}$ becomes the condition \qq
\iota_{_{\gamma_1\gamma_2,\gamma_3}}\,\iota_{_{\gamma_1,\gamma_2}}
\ =\ \iota_{_{\gamma_1,\gamma_2\gamma_3}}
\,(\gamma_1^{-1})^*\iota_{_{\gamma_2,\gamma_3}}\label{assoi} \qqq
for isomorphisms between the bundles $\,
P^{^{\gamma_1}}\otimes(\gamma_1^{-1})^*
P^{^{\gamma_2}}\otimes(\gamma_1^{-1})^*(\gamma_2^{-1})^*P^{^{\gamma_3}}\,$
and the target bundles
$$\,\,P^{^{\gamma_1\gamma_2\gamma_3}}\otimes
R^{^{\gamma_1,\gamma_2}}\otimes
R^{^{\gamma_1\gamma_2,\gamma_3}}\,\quad\ {\rm and}\
\quad\,\,P^{^{\gamma_1\gamma_2\gamma_3}}\otimes
R^{^{\gamma_1,\gamma_2\gamma_3}}\otimes(\gamma_1^{-1})^*
R^{^{\gamma_2,\gamma_3}}\,$$ naturally identified due to
(\ref{123}). \vskip 0.2cm

Given a family $\,\iota_{_{\gamma_1,\gamma_2}}$ of isomorphisms
(\ref{iogg}) such that (\ref{assoi}) holds, we may obtain another
family by multiplying $\,\iota_{_{\gamma_1,\gamma_2}}$ by an
$U(1)$-valued 2-cocycle $\,u_{_{\gamma_1,\gamma_2}}$ satisfying
\qq u_{_{\gamma_1,\gamma_2}}\,u_{_{\gamma_1\gamma_2,\gamma_3}}\,
u_{_{\gamma_1,\gamma_2\gamma_3}}^{-1}\,u_{_{\gamma_2,\gamma_3}}^{-1}
\ =\ 1\,. \qqq The new family gives another solution for
associative $\,\mu_{_{\Gamma}}$. The coboundary choice
$\,u_{_{\gamma_1,\gamma_2}}=v_{_{\gamma_1\gamma_2}}
\,v_{\gamma_1}^{-1}v_{_{\gamma_2}}^{-1}\,$ with $U(1)$-valued
$\,v_{_{\gamma}}$ leads to an isomorphic gerbe on $\,M/\Gamma\,$
whereas the choices leading to non-trivial elements of
$\,H^2(\Gamma, U(1))\,$ may give stably non-isomorphic gerbes.
\vskip 0.2cm

The construction of gerbes $\,\CG'_k$ on $\,SU(N)/\CZ\,$ in the
preceding section is an illustration of the general procedure
described here, as we explain in detail in Appendix C. \vskip
0.2cm

Although above we have assumed that $\Gamma$ acts without fixed
points on $M$, the above construction still goes through for
general orbifolds provided that we redefine the spaces
$\,Y_{\Gamma}^{{[n]}}$ as \qq
Y_{\Gamma}^{{[n]}}\,:=\,\{\,(y,y_1,\gamma_1,\dots,y_{n-1},\gamma_{n-1})
\ |\ \pi(y)=\gamma_m\pi(y_m)\,\}\,, \qqq i.e.\,\,keeping track of
$\,\gamma_m\in\Gamma$ (which could be recovered from $y_m$'s for
free action of $\Gamma$). The resulting ``orbifold gerbes''
provide a natural tool for the treatment of strings on orbifolds
in the background of closed 3-forms, see also \cite{Sharpe1}.

\nsection{Bundle gerbes and loop-space line bundles}

The construction of the amplitudes $\CA(\phi)$ described in
Sect.\,\,2.2 with the use of the local data may be easily
translated to the language of gerbes. In particular, the
construction of an isomorphism class of hermitian line bundles
with connection over the loop space $LM$ from a class $w\in
W(M,{H})$ may be lifted to a canonical assignment of a hermitian
line bundle with connection to a gerbe on $M$. In the present
subsection, we shall describe those constructions that gain in
simplicity when formulated with use of gerbes. \vskip 0.2cm

Let $\CG=(Y,{B},{L},\mu)$ be a gerbe on $M$ with curvature ${H}$
and, as before, $\,\sigma_i:O_i\rightarrow Y\,$ be local sections.
Let $\phi$ be a map from a compact surface $\Sigma$ to $M$. For a
sufficiently fine triangulation of $\,\Sigma\,$ and a label assignment
$(c,b)\mapsto(i_c,i_b)$ such that $\phi(c)\subset O_{i_c}$ and
$\phi(b)\subset O_{i_b}$, let us set \qq
\,\phi_c=\sigma_{i_c}\circ\phi|_c\,,\qquad\phi_b=\sigma_{i_b}\circ\phi|_b\,,
\qquad\phi_{cb}=\sigma_{i_ci_b}\circ\phi|_b\,, \qqq the latter for
$\,b\subset c$. These are lifts to $Y$ or to $Y^{{[2]}}$ of
restrictions of $\phi$ to the elementary cells. Denoting by
$\CH(\,\cdot\,)$ the parallel transport in ${L}$, we define: \qq
\CA(\phi)\,=\,\exp\Big[i\sum\limits_c\int\limits_c\phi_{c}^{\,*}
{B}\Big]\,\mathop{\otimes}\limits_{b\subset c}\CH(\phi_{cb})\,.
\label{amplg} \qqq Since
$\,\CH(\phi_{cb})\in\hspace{-0.15cm}\mathop{\otimes}
\limits_{v\in\partial b}{L}_{(y_c,y_b)}\hspace{0.05cm}$,
\,where $y_c=\phi_{c}(v)$ and $y_b=\phi_{b}(v)$ (with the convention
that the dual fiber is taken if $\,v$ is the beginning of $\,b$),
\qq \CA(\phi)\
\,\in\ \mathop{\otimes}\limits_{v\in b\subset c}
{L}_{(y_c,y_b)}\,. \label{prli} \qqq The point is that
if $\,\partial\Sigma=\emptyset\,$ then there is a canonical
isomorphism, defined by the gerbe multiplication $\mu$, between
the line in (\ref{prli}) and the complex line $\NC$  so that the
amplitude $\CA(\phi)$ may be naturally interpreted as a number.
Indeed, fixing a vertex $v$ and going around as in Fig.\,\,1,

\leavevmode\epsffile[-75 -20 245 215]{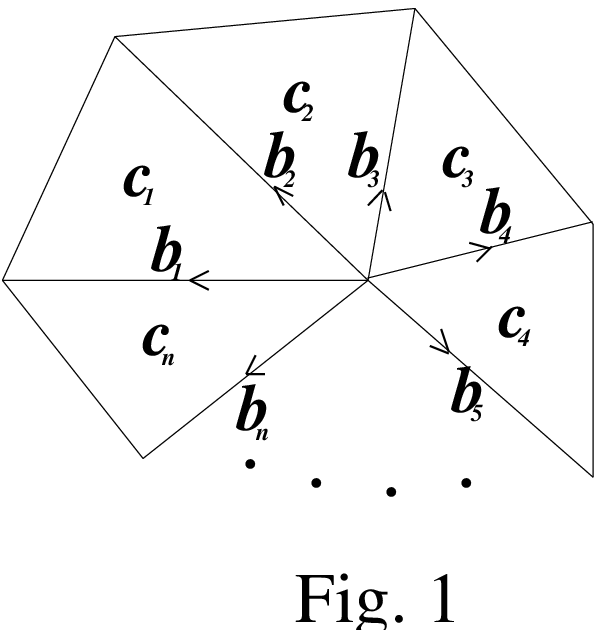}

\noindent we gather the contribution \qq
{L}_{(y_{b_1},y_{c_1})}\otimes\,{L}_{(y_{c_1},y_{b_2})}\otimes\
\cdots\ \otimes\,{L}_{(y_{c_n},y_{b_1})} \label{chain} \qqq to the
line in (\ref{prli}) which is trivialized by subsequent
application of $\mu$. That the result does not depend on where we
start numbering the cells may be seen by choosing $\,y_v\in Y$
with $\,\pi(y_v)=\phi(v)\,$ and inserting
$\,{L}_{(y_{b_r},y_v)}\otimes\,{L}_{(y_v,y_{b_r})} \cong\NC\,$ at
every second place in the chain (\ref{chain}). We may now use
$\mu$ to trivialize the blocks \qq
{L}_{(y_v,y_{b_r})}\otimes\,{L}_{(y_{b_r},y_{c_r})}\otimes\,
{L}_{(y_{c_r},y_{b_{r+1}})}\otimes\,{L}_{(y_{b_{r+1}},y_v)}\,.
\label{chain1} \qqq It is easy to check that the number obtained
for $\CA(\phi)$ coincides with the one defined by the expression
(\ref{ampl}) with the use of the local data obtained from the
sections $\sigma_i$ and $s_{ij}$. We give the proof in Appendix D.
From the results of \cite{Gaw0}, it follows now that $\CA(\phi)$
does not depend on the choices of the local sections
$\sigma_i$, of the lifts $\phi_c$, $\phi_b$
and $\phi_{cb}$, nor of the triangulation of $\Sigma$.
Additionally, $\CA(\phi)$ is invariant under the composition of $\phi$
with orientation-preserving diffeomorphisms of $\Sigma$ and it goes to
its inverse for diffeomorphisms reversing the orientation.
\vskip 0.2cm

Suppose now that $\partial\Sigma=\sqcup\ell_s$. Consider the same
expression (\ref{amplg}). Proceeding as before, we may canonically
reduce the line in (\ref{prli}) to \qq
\mathop{\otimes}\limits_s\mathop{\otimes}\limits_{v\in
b\subset\ell_s} {L}_{(y_v,y_b)}\,, \label{prlb} \qqq see
Fig.\,\,2 which replaces Fig.\,\,1 in the boundary situation.

\leavevmode\epsffile[-55 -20 265 170]{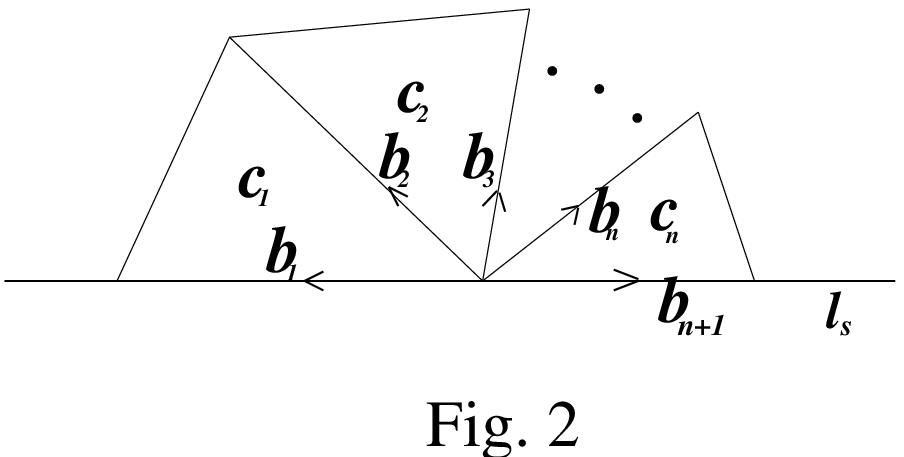}

\noindent Let, for a closed loop $\,\varphi:\ell\rightarrow M\,$
and a sufficiently fine split of $\,\ell$, \qq
\CL_\varphi\,=\,\mathop{\otimes}\limits_{v\in b\subset\ell}
{L}_{(y_v,y_b)} \label{lp} \qqq with $y_v\in Y$ such that
$\pi(y_v)=v$, and  $\,y_b=\varphi_b(v)$, where
$\,\varphi_b\,$ lift $\,\varphi|_b\,$ to $\,Y$. \,Let us show that
the lines (\ref{lp}) are canonically
isomorphic for different choices of $\,y_v$ and $\varphi_b$. Let
$\varphi'_b$ and $y'_v$, $y'_b=\varphi'_b(v)$ be another choice.
Note that the parallel transport in ${L}$ along
$\,(\varphi_b,\varphi'_b): b\rightarrow Y^{{[2]}}$ defines
a canonical trivialization of the line
$\mathop{\otimes}\limits_{v\in b\subset\ell} {L}_{(y_b,y'_b)}$.
\,Similarly, the line $\mathop{\otimes}\limits_{v\in
b\subset\ell}{L}_{(y'_v,y_v)}$ is canonically trivial since each
factor is accompanied by its dual. Using also the product $\mu$,
we obtain a chain of canonical isomorphisms \qq
\mathop{\otimes}\limits_{v\in b\subset\ell}{L}_{(y_v,y_b)}\,
\cong\hspace{-0.1cm}\mathop{\otimes}\limits_{v\in b\subset\ell}
\Big({L}_{(y'_v,y_v)}\otimes{L}_{(y_v,y_b)}\otimes
{L}_{(y_b,y'_b)}\Big)
\,\cong\hspace{-0.1cm}\mathop{\otimes}\limits_{v\in b\subset\ell}
{L}_{(y'_v,y'_b)}\,. \label{iso0} \qqq Associativity of $\mu$
assures that the resulting isomorphisms are transitive so that we
may free ourselves from the choice of local lifts in the
definition of $\,\CL_\varphi$ by passing to equivalence classes of
elements related by the isomorphisms (\ref{iso0}). Similarly, if
we pass to a finer split of $\,\ell\,$ and use the restrictions of
maps $\varphi_b$ to the new intervals, setting also
$y_v=\varphi_b(v)$ for new vertices in the interior of the old
intervals, then the net result on $\CL_\varphi$ is to add trivial
factors on the right hand side of (\ref{lp}). Dropping them is
compatible with isomorphisms (\ref{iso0}). In order to loose
memory of the split used in (\ref{lp}), one may then define the
projective limit $\CL(\CG)_\varphi$ over trivializations of the
lines obtained for fixed trivializations. All in all, we obtain
this way a canonical hermitian line bundle $\CL(\CG)$ over the
loop space $LM$. Note that, by construction, $\CL(\CG)_\varphi$ is
invariant under orientation-preserving reparametrizations of
$\,\ell$ and that the change of orientation gives rise to the dual
line. \vskip 0.2cm

Comparing the lines (\ref{prlb}) and (\ref{lp}) we infer that if
$\partial\Sigma=\sqcup\ell_s$ then the amplitude (\ref{ampl}) may
be canonically defined as an element of the product of lines of
$\CL(\CG)\,$: \qq \CA(\phi)\ \in\ \mathop{\otimes}\limits_s
\CL(\CG)_{\phi|_{\ell_s}}\,. \label{amplgb} \qqq The hermitian
line bundle $\CL(\CG)$ may be equipped with a (hermitian)
connection such that the parallel transport along the curve in the
loop space $LM$ defined by $\phi:[0,1]\times\ell\rightarrow M$ is
given by $\CA(\phi)$. The curvature of this connection is equal to
the 2-form $\Omega$ on $LM$ defined in (\ref{cur}), The amplitudes
$\CA(\phi)$ for arbitrary surfaces $\Sigma$ provide a
generalization of the parallel transport in the loop space. \vskip
0.2cm

How does the line bundle $\CL(\CG)$ and the amplitude $\CA(\phi)$
depend on the gerbe?  First, the line bundles $\CL(\CG)$ and
$\CL(\sigma^*\CG)$, where $\sigma^*\CG$ is a pullback gerbe, are
canonically isomorphic. Second, an isomorphism between gerbes
$\CG_1$ and $\CG_2$ induces an isomorphism of the bundles
$\CL(\CG_1)$ and $\CL(\CG_2)$. Third, the line bundle
$\,{\CL}(\CG_1\otimes\CG_2)\,$ is canonically isomorphic to
$\,{\CL}(\CG_1)\otimes \CL(\CG_2)$. \,Finally, for a trivial gerbe
$\,\CG_{_N}$, \qq \CL_\varphi\,=\,\mathop{\otimes}\limits_{v\in
b\subset\ell}\Big( {N}_{y_v}^{-1}\otimes{N}_{y_b}\Big)\ \cong\
\mathop{\otimes}\limits_{v\in b\subset\ell} {N}_{y_b}\ \cong\
\NC\, \label{tris} \qqq where the last isomorphism is given by the
parallel transport in ${N}$ along $\,\varphi_b$. It follows that a
stable isomorphism between gerbes induces an isomorphism of the
corresponding line bundles over $LM$. If the gerbe $\CG$ is
constructed from the local data then $\CL(\CG)$ is canonically
isomorphic to the line bundle $\CL$ over $LM$ constructed from the
local data described in Section.\,\,10.1. \vskip 0.2cm

In the language of the trivialization (\ref{tris}), the
isomorphism between the lines $\,\CL_\varphi$ corresponding to the
isomorphic trivial gerbes $\,\CG_{_{N'}}$ and $\,\CG_{_N}$ for $\,N'\cong
N\otimes\pi^*P\,$ with $P$ a flat bundle on $M$ is given by the
multiplication by the holonomy of $P$ along $\varphi$. It follows
that the change of stable isomorphism between two gerbes obtained
by composition with the isomorphism between the trivial gerbes
$\,\CG_{_N}$ and $\,\CG_{_{N'}}$ multiplies the isomorphism of the line
bundles over $LM$ by the holonomy of $P$.

\nsection{Gerbes and branes}

We have shown in the previous section that, given a gerbe
$\,\CG=(Y,{B},{L}, \mu)\,$ on $M$ of curvature ${H}$, the formal
amplitudes $\,\ee^{i\int\phi^*d^{-1}{H}}\,$ of a classical fields
$\,\phi:\Sigma\rightarrow M\,$ defined on the worldsheet $\Sigma$
with boundary may be given sense as elements in the tensor product
of lines of the line bundle $\CL(\CG)$ canonically associated to
$\CG$, see (\ref{amplgb}). In general, $\CL(\CG)$ is a non-trivial
bundle so that the amplitude cannot be naturally defined as
numbers. Suppose, however, that the field $\phi$ is restricted by
the boundary conditions \qq \phi(\ell_s)\ \in\ D_s\subset M\,,
\label{bc} \qqq forcing its values on the boundary loops $\ell_s$
of $\Sigma$ to belong to submanifolds $D_s$ of $M$. Suppose
moreover that $D_s$ are chosen so that the line bundle $\CL(\CG)$
restricted to the space $LD_s$ of loops in $D_s$ becomes trivial.
Upon a choice of trivializations of $\,\CL(\CG)|_{_{LD_s}}$, the
amplitude $\CA(\phi)$ may then be assigned a numerical value. Note
that it is not necessary to require that the trivializations of
$\,\CL(\CG)|_{_{LD_s}}$ flatten the connection. \vskip 0.2cm

Let us assume that a submanifold $D\subset M$ is such that the
restriction of the 3-form ${H}$ to $D$ is exact: $\,{H}|_{_D}=dQ$.
To the 2-form $\,Q$, we may associate a gerbe
$\,\CK=(D,Q,D\times\NC,\,\cdot\,\,)\,$ over $D$ with curvature
$\,{H}|_{_D}$. \,Note that the corresponding hermitian line bundle
$\,\CL(\CK)$ over $\,LD$ is trivial but that its connection has a
non-trivial curvature if ${H}|_{_D}\not=0$. A natural way to
assure the triviality of the restricted bundle
$\,\CL(\CG)|_{_{LD}}$ and to provide for its trivializations is to
assume that the restriction $\,\CG_{_D}=(Y_{_D},{B}_{_D},
{L}_{_D},\mu_{_D})\,$ of the gerbe $\CG$ to $D$, where $\,Y_{_D}
=\pi^{-1}(D)$, $\,{B}_{_D}={B}|_{_{Y_{_D}}}$ etc., is stably
isomorphic to the gerbe $\,\CK$.  \,Explicitly, this means that
there exist: a line bundle $\,{N}$ over $\,Y_{_D}$ with connection
of curvature $F$ such that \qq {B}_{_D}\,+\,F\,=\,\pi_{_D}^*Q
\label{B+F} \qqq and an isomorphism \qq
\iota:\,{L}|_{Y_D^{[2]}}\otimes
p_1^*{N}^{^{-1}}\hspace{-0.1cm}\otimes p_2^*{N}
\,\rightarrow\,Y_{_D}^{{[2]}}\times\NC \label{sti} \qqq of line
bundles with connection over $Y_{_D}^{{[2]}}$, compatible with the
groupoid multiplication. By definition, a brane $\,\CD\,$ of $\,\CG\,$
with support $\,D\,$ and curving $\,Q\,$ is the quadruple
$\,(D,Q,{N},\iota)$. We shall consider two branes represented by
collections $\,(D,Q,{N},\iota)\,$ and $\,(D,Q,{N}',\iota')\,$ 
equivalent if the line bundles ${N}$ and ${N}'$ are isomorphic and 
$\iota$ and $\iota'$ are intertwined by the induced isomorphism of 
the trivial gerbes (note that such isomorphism is not unique). 
Non-equivalent branes with fixed support and curving correspond 
to $\,N'\cong N\otimes\pi_{_D}^*P$, \,where $P$ 
is a non-trivial flat bundle over $D$. \,A choice of $\,{N}\,$ and
$\,\iota\,$ induces canonically an isomorphism between
$\,\CL(\CG)|_{_{LD}}$ and the trivial hermitian line bundle
$\,\CL(\CK)$, see (\ref{tris}), with equivalent choices leading to
the same isomorphism. Non-equivalent choices give rise to isomorphisms
differing by multiplication by holonomy in a flat line bundle $P$
over $D$.  \vskip 0.2cm

Given a gerbe $\CG$ with curvature $H$, we may ask which submanifolds
$D$ with $\,{H}|_{_D}=dQ$ support branes with curving $Q$. The
obstructions to stable isomorphism of the gerbes $\CG_{_D}$ and $\CK$
lie in the cohomology group $\,H^2(D,U(1))\,$ that acts freely and
transitively on the set $\,W(D,{H}|_{_D})\,$ of stable isomorphism
classes of gerbes on $D$ with curvature ${H}|_{_D}$. \,If
the obstruction vanishes, then the cohomology group
$\,H^1(D,U(1))\,$ (the group of isomorphism classes of flat
hermitian bundles on $D$)  acts freely and transitively on the set
of equivalence classes (moduli) of branes $\CD$ with curving $Q$ 
supported by $D$. We shall see this in work in the next two sections. 
\vskip 0.3cm

Let $IM$ be the space of open curves (strings)
$\,\varphi:[0,\pi]\rightarrow M\,$ and $\,\CG=(Y,{B},{L},\mu)\,$ a
gerbe on $M$ with curvature ${H}$. The same construction that
associated to $\CG$ a line bundle with connection over the loop
space $LM$, when applied to open curves, induces a hermitian line
bundle with connection $\,\CN\,$ over the space \qq
\CY\,=\,\{\,(\varphi,y_0,y_1)\in IM\times Y^2\,|\,\pi(y_0)=
\varphi(0),\ \pi(y_1)=\varphi(\pi)\,\}\,. \qqq The line bundle
$\,\CN\,$ is composed from the fibers $\,\CL_\varphi\,$ of
(\ref{lp}), with all the identifications as before except that one
has to keep the memory of $y_v=y_0$ and $y_v=y_1$ for the end
point vertices. The parallel transport in $\,\CN\,$ along a curve
in $\,IM\,$ is still determined by the amplitude $\,\CA(\phi)\,$
defined by (\ref{amplg}) for $\,\phi:[0,1]\times[0,\pi]\rightarrow
M$. \,The curvature of $\,\CN\,$ is given by the closed 2-form \qq
\Omega_{_{IM}}(\varphi,y_0,y_1)\ =\ \Omega(\varphi)\,
+\,{B}(y_0)\,-\,{B}(y_1)\,, \label{curIM} \qqq on $\,\CY$, where
$\,\Omega\,$ defined by (\ref{cur}) with $\ell=[0,\pi]$. \vskip
0.2cm

Given two branes $\CD_{0}$ and $\,\CD_{1}$ with supports $\,D_0\,$
and $\,D_1\,$ of gerbe $\CG$, we may
consider in the space $\,IM\,$ of open strings the subspace
 \qq I_{_{D_0D_1}}M=\{\,\varphi:[0,\pi]\rightarrow
M\,\,|\, \,\varphi(0)\in D_0,\ \varphi(\pi)\in D_1\}\,. \label{ID}
\qqq  A slight modification of the construction described above
permits now to define over $\,I_{_{D_0D_1}}M\,$ a hermitian line
bundle $\,\CL_{_{\CD_0\CD_1}}(\CG)\equiv \CL_{_{\CD_0\CD_1}}\,$
with connection by setting \qq
(\CL_{_{\CD_0\CD_1}})_{_{\varphi}}\,=\ (N_0)_{_{y_0}}\otimes\,
\CL_{_{\varphi}}\otimes\,(N_1)^{-1}_{_{y_1}}\,,\label{ldd} \qqq
where $\,\CL_\varphi\,$ is given by (\ref{lp}). Due to the
isomorphism (\ref{sti}), the lines obtained this way are
canonically isomorphic also for different choices of $y_0$ and
$y_1$, giving rise upon their identification to the fibers of
$\,\CL_{_{\CD_0\CD_1}}$. A choice of equivalent branes leads to
to (non-canonically) isomorphic bundles. \,The
parallel transport in $\,\CL_{_{\CD_0\CD_1}}$ is determined
by \qq \CA_{_{\CD_0\CD_1}}(\phi)\ =\
\CA(\phi)\,\otimes\Big(\mathop{\otimes}
\limits_{b\subset\ell_s}\CH_s(\phi_b)\Big) \label{AD} \qqq for
$\,\phi:[0,1]\times[0,\pi]\rightarrow M$, \,where $\,\ell_s$
denotes the piece of the boundary of $\,[0,1]\times[0,\pi]\,$
mapped into $\,D_s$ for $s=0,1$, $\,\CH_s(\phi_b)\,$ stands for
the parallel transport in $\,N_s$ along a lift $\,\phi_b$ of
$\,\phi|_b$ to $\,Y\,$ and $\,\CA(\phi)\,$ is given by
(\ref{amplg}). The curvature of $\,\CL_{_{\CD_0\CD_1}}$ is
given by the 2-form on $\,I_{_{D_0D_1}}M\,$ \qq
\Omega_{_{D_0D_1}}\ =\ \Omega\,+\,e_1^*Q_1\,-\,e_0^*Q_0\,,
\label{curD} \qqq where $\,\Omega\,$ is as in (\ref{curIM}) and
$e_s$ are the evaluation maps, \qq
\varphi\,\mathop{\longrightarrow}^{e_0}\,\varphi(0)\,,\qquad
\varphi\,\mathop{\longrightarrow}^{e_1}\,\varphi(\pi)\,.
\label{e01} \qqq \vskip 0.2cm

\leavevmode\epsffile[-62 -20 258 220]{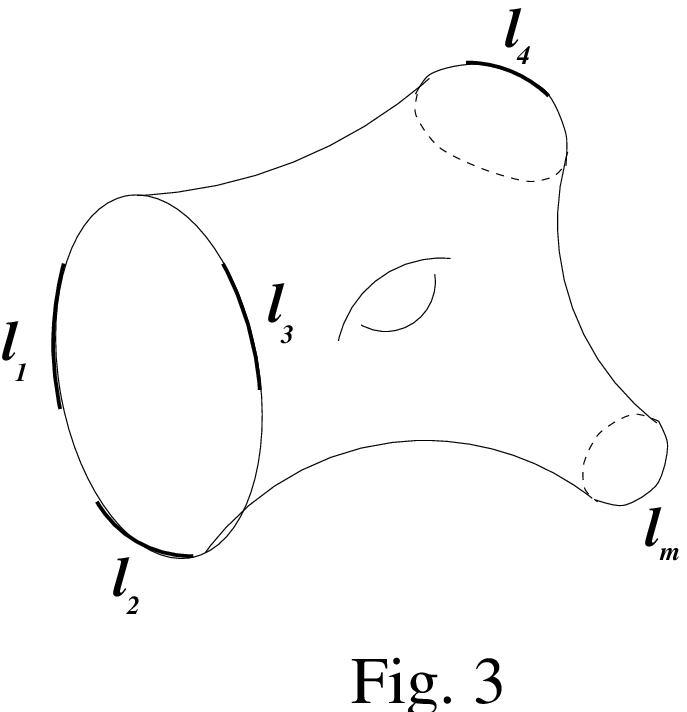}

More generally, suppose that $\,\phi:\Sigma\rightarrow M\,$
satisfies the boundary conditions (\ref{bc}) for $\,\ell_s$ being
closed disjoint subintervals of the boundary loops of $\,\Sigma$,
\,see Fig.\,\,3. Then the amplitude defined by (\ref{AD})
satisfies \qq \CA_{_{(\CD_s)}}(\phi)\ \in\
\Big(\mathop{\otimes}\limits_{(s,s')}(\CL_{_{\CD_s\CD_{s'}}})
_{_{\phi|_{\ell_{(s,s')}}}}\Big)\otimes\Big(\mathop{\otimes}\limits_m
\CL_{_{\phi|_{\ell_m}}}\Big), \label{genamp} \qqq where
$\,\ell_{(s,s')}$ are the boundary intervals bordering $\ell_s$
and $\ell_{s'}$ and $\,\ell_m$ are the boundary loops that do not
contain intervals $\,\ell_s$. The curves
$\,\phi|_{_{\ell_{(s.s')}}}$ stretch between the submanifolds
$D_s$ and $D_{s'}$ and the line bundles $\,\CL_{_{\CD_s\CD_{s'}}}$
correspond to that geometry. The expression (\ref{AD}) generalizes
the definition of the amplitude of a field to the case when local
boundary conditions are imposed on pieces of the boundary of
$\,\Sigma$. \vskip 0.2cm

In the quantum field theory, the amplitudes
$\,\CA_{_{\CD_0\CD_1}}(\phi)\,$ are ``summed'' (with additional
scalar weights) over all fields satisfying the boundary conditions
(\ref{bc}) resulting, at least formally, in a vector in the tensor
product of Hilbert spaces of states, see Sect.\,\,10 below. To each 
interval $\,\ell_{(s,s')}$ there corresponds a factor
$\,\NH_{_{\CD_s\CD_{s'}}}$, the Hilbert space of states of the
string stretching between the branes $\,\CD_s$ and
$\,\CD_{s'}$ and  to each $\,\ell_m$ a factor $\,\NH$, the closed
string space of states. Geometrically, spaces
$\,\NH_{_{\CD_0\CD_1}}$ are formed of sections of the
corresponding line bundles $\,\CL_{_{\CD_0\CD_1}}$ and the space
$\,\NH\,$ of sections of $\,\CL\,$ (more precisely, they are
Hilbert space completions of spaces of sections). Even without
going into the detailed construction of such spaces of states, the
geometric classification of branes discussed above allows to
obtain the spectrum of branes. We shall illustrate that in the
next section on the example of the $SU(N)$ WZW theory and of its
versions with groups covered by $SU(N)$.

\nsection{Branes in the WZW model}

In the WZW model, the candidate for the simplest form of the
boundary condition that guarantees the conservation of half of the
current algebra symmetries is to require that the values of the
field $\,g:\Sigma\rightarrow G\,$ on the boundary loops
$\ell_s\subset\partial\Sigma$ belong to the conjugacy classes
$\,\CC_{\tau_s}\subset G$. The closed 3-form $\,H\,$ of
(\ref{chi}) becomes exact when restricted to a conjugacy class:
$\,H|_{{\CC_\tau}}=dQ_\tau$, \,where $\,Q_\tau$ is given by the
expression (\ref{om}) with constant $\tau$. The preservation of
the current algebra symmetries requires that one sticks to that
choice (or to its multiplicities) for the curving of branes supported
by $\,\CC_\tau$, see \cite{ASchom} or \cite{GTT}. Now it is
easy to check for which conjugacy classes the restriction of the
gerbe on $\,G\,$ with curvature $\,kH\,$ is stably isomorphic to the
gerbe $\,\CK=(\CC_\tau,\,kQ_\tau, \,\CC_\tau\times\NC,\,\cdot\,)$.

\subsection{$SU(N)$ \ groups}
\vskip 0.2cm

Recall that for $\,G=SU(N)$, \,the gerbe
$\,\CG^k=(Y,kB,\CL^k,\mu^k)\,$ with curvature $kH$ (unique, up to
stable isomorphism) has $\,Y=\mathop{\sqcup}\limits_{i=0}^r O_i$.
\,We shall choose a coadjoint orbit $\,\CC_\tau\,$ and denote
$\,Z=\pi^{-1}(\CC_\tau)=Y_{\CC_\tau}$, where $\,\pi\,$ is the
projection from $\,Y\,$ to $SU(N)$. Thus $\,Z=\sqcup Z_i\,$ where
$\,Z_i=\CC_\tau\cap O_i$.  \,Since the sets $\,O_i\subset SU(N)\,$
are invariant under conjugations, $\,Z_i$ are either empty or
equal to $\,\CC_\tau$. \,Let $\,\omega=\pi|_Z\,$ and
$\,\sigma:Z\rightarrow Y\,$ be the natural inclusion. We have to
compare the restriction of the gerbe $\,\CG^k$ to $\,\CC_\tau\,$
with the pullback gerbe $\,\sigma^*\CK$. \,On $\,Z_i\,$ the
difference of the two curvings
$\,kQ_\tau-\,kB_i|_{{\CC_\tau}}\,\equiv\,F_{\tau i}\,$ has the
form \qq F_{\tau i}(\gamma\,\ee^{2\pi
i\tau}\gamma^{-1})\,=\,-ki\,\tr\,(\tau-\lambda_{i})
(\gamma^{-1}d\gamma)^2\,, \label{fmu} \qqq see (\ref{omi1}). If
the two gerbes $\,\CG^k|_{{\CC_\tau}}$ and $\,\CK\,$ are stably
isomorphic then the closed forms $\,F_{\tau i}$ must be curvature
forms of hermitian line bundles $\,N|_{Z_i}$ and hence
$\,{1\over2\pi}F_{\tau i}\,$ must be integral. This condition is
equivalent to the requirement that $k\tau$ be a weight. Indeed, in
our case, $\,Z_i=\CC_\tau$ may be identified with the coadjoint
orbits $\,\CO_{k(\tau-\lambda_{i})}$ by the maps
$\,\gamma\,\ee^{2\pi i\tau}\gamma^{-1}\mapsto
\gamma\,k(\tau-\lambda_{i})\, \gamma^{-1}$ since the isotropy
groups satisfy $\,G_\tau=G^0_{\tau-\lambda_{i}}$. Upon this
identification, $\,F_{\tau i}\,$ becomes the Kirillov-Kostant
symplectic form and integrality of $\,{1\over2\pi}F_{\tau i}\,$
requires that $\,k(\tau-\lambda_{i})\,$ be a weight. \vskip 0.2cm

In the latter case, the line bundles $\,N|_{Z_i}\,$ may be taken
to be the pullbacks of the Kirillov-Kostant bundles
$\,L_{k(\tau-\lambda_{i})}\,$ by the identification of $\,Z_i$
with $\,\CO_{k(\tau-\lambda_{i})}$. Since the conjugacy classes
are simply connected, the resulting line bundle $\,N\,$ over
$\,Z\,$ is unique up to isomorphism. The mapping \qq
[\gamma,\zeta]_{k\lambda_{ij}}\otimes\,[\gamma,\zeta']^{-1}_{k(\tau
-\lambda_i)}\otimes\,[\gamma,\zeta'']_{k(\tau-\lambda_j)}\ \
\mathop{\longmapsto}^\iota\ \ (y_1,y_2,\,\zeta\zeta'\zeta'') \qqq
for $\,y_1=(g,i)$, $\,y_2=(g,j)$ and $\,g=\gamma\,\ee^{2\pi i\tau}
\gamma^{-1}\,\in\,\CC_\tau$, \,which is well defined because
$\,\chi_{k\lambda_{ij}}
(\gamma_0)\,\chi_{k(\tau-\lambda_i)}^{-1}(\gamma_0)\,
\chi_{k(\tau-\lambda_j)}(\gamma_0)=1\,$ for $\,\gamma_0\in
G_\tau$, \,determines then the unique isomorphism (\ref{sti}) that
commutes with the groupoid multiplication. It provides a stable
isomorphism between the gerbes $\,\CG^k|_{\CC_\tau}\,$ and
$\,\CK$. Other choices of $N$ and $\iota$ lead to equivalent branes
in the present case. We thus obtain for the $SU(N)$ WZW model a
family of branes labeled by the weights $\,\lambda\in kA_W$,
supported by the conjugacy classes $\,\CC_{\tau}\,$ with
$\,\lambda=k\tau$. The weights in the dilated Weyl alcove $kA_W$
are called ``integrable at level k'' \cite{Kac} and they also
label the irreducible highest-weight representations of the level
$k$ current algebra and the bulk primary fields of the model with
conformal weights $\,h(\lambda)={\tr\,\lambda(\lambda+2\rho)
\over2(k+h^\vee)}$.

\subsection{Groups covered by $SU(N)$}
\vskip 0.2cm

We shall consider branes supported by the conjugacy classes
in $\,G'=SU(N)/\CZ\,$ for $\,\CZ\cong\NZ_{N'}$. \,Each conjugacy
class in $\,G'\,$ is an
image under the canonical projection from $G=SU(N)$ to $G'$ of a
conjugacy class $\,\CC_\tau\subset G\,$ with $\,\tau\in A_W$.
\,Recall that the elements $\,z_i$ of the center of $\,SU(N)\,$
act on $\,A_W$ by $\,\tau\mapsto\sigma_i(\tau)$. \,The conjugacy
classes $\,\CC_{\sigma_a(\tau)}$ in $G$ for different
$\,z_a\in\CZ\,$ project to the same class in $G'$. This way the
conjugacy classes in $\,G'\,$ may be labeled by the
$\,\CZ$-orbits $\,[\tau]\,$ of elements in $\,A_W$. \,We
shall denote the class in $\,G'$ corresponding to $\,[\tau]\,$ by
$\,\CC'_{[\tau]}$. \,For any $\,\tau\in[\tau]$, $\,\CC'_{[\tau]}$
may be canonically identified with $\,\CC_\tau/\CZ_{\tau}$,
where $\,\CZ_{\tau}$ is the subgroup of $\,\CZ\,$ leaving
$\,\tau\,$ unchanged (it depends only on the orbit $\,[\tau]$).
\,The 3-form $\,H'\,$ restricted to $\,\CC'_{[\tau]}\,$ still
satisfies $\,H'|_{\CC'_{[\tau]}}=dQ'_{\tau}$, \,with
$\,Q'_{\tau}$ denoting the the projection of $\,Q_\tau$ to
the quotient space $\,\CC_\tau/\CZ_{\tau}\,$ and defining a 2-form
on $\,\CC'_{[\tau]}$ that does not depend on $\,\tau\in[\tau]$. 
\,We obtain then the
gerbe $\,\CK'=(\CC'_{[\tau]},\,kQ'_{\tau},\,\CC'_{[\tau]}\times
\NC,\,\cdot\,\,)\,$ on $\,\CC'_{[\tau]}$. \vskip 0.2cm

Let us consider the space
$\,Z'={\pi'}^{-1}(C'_{[\tau]}))=\mathop{\sqcup}\limits_{\tau\in
[\tau]}\mathop{\sqcup}\limits_iZ_{\tau i}\,$, \,with $\,Z_{\tau
i}=\CC_\tau\cap O_i$. \,Let $\,\sigma':Z'\rightarrow Y'\,$ be the
inclusion map and $\,\omega'=\pi'|_{Z'}$. \,We have to compare the
restriction to $\,\CC'_{[\tau]}$ of the gerbe $\,\CG'_k=
(Y',B',L',\mu')\,$ on $\,G'$ constructed in Section 4 to the
pullback gerbe $\,{\sigma'}^*\CK'$. \,Over $\,Z_{\tau i}$ the
difference of the curvings is $\,F_{\tau i}$, see (\ref{fmu}), and
the existence of the stable isomorphism between the two gerbes
requires again that $\,k\tau\,$ be a weight, similarly as in the
simply connected case. Let $\,N'$ denote the line bundle over
$\,Z'$ that over $\,Z'_{\tau i}$ coincides with the pullback of
the Kirillov-Kostant bundle $\,L_{k(\tau-\lambda_i)}\,$ by the
identification of $\,Z_{\tau i}$ with the coadjoint orbit $\,\CO_{
k(\tau-\lambda_i)}$. \,In order to construct the primed version of
the bundle isomorphism (\ref{sti}), let us consider the pairs
$\,(y_1,y_2)\in {Z'}^{[2]}\,$ with $\,y_1=(g,i)$, $\,y_2
=(z_a^{-1}g,j'))$, $\,g=\gamma\,\ee^{2\pi i\tau}\gamma^{-1}$,
$\,z_a^{-1}g=\gamma w_a^{-1}\,\ee^{2\pi i\sigma_a(\tau)} w_a
\gamma^{-1}\,$ and the mapping \qq
[\gamma,\zeta]_{k\lambda_{ij}}\otimes\,[\gamma,\zeta']^{-1}_{k(\tau
-\lambda_i)}\otimes\,[\gamma
w_a^{-1},\zeta'']_{k(\sigma_a(\tau)-\lambda_{j'})}\
\hspace{2cm}\cr \mathop{\longmapsto}^{\iota'}\ \ (y_1,y_2,
\,v_{\tau,a}\,\zeta\zeta'\zeta'')\,,\quad\label{triv} \qqq where
$\,j=[j'+a]\,$ and $\,v_{\tau,a}\in U(1)$. \,Note that
$\,\iota'\,$ is well defined since the isotropy subgroups satisfy
$$G_\tau=\,G^0_{\tau-\lambda_i}=\,G^0_{\tau-\lambda_j}=\,w_a^{-1}\,
G^0_{\sigma_a(\tau)-\lambda_{j'}}w_a\ \subset\
G^0_{\lambda_{ij}}$$ and for $\,\gamma_0\in G_\tau$,
the product $\,\,\chi_{k\lambda_{ij}}(\gamma_0)\,\chi_{k(\tau-\lambda_i)}^{-1}
(\gamma_0)\,\chi_{k(\sigma_a(\tau)-\lambda_{j'})}(w_a\,\gamma_0
w_a^{-1})\,$ is equal to $\,1\,$ since $\,\chi_{k(\sigma_a(\tau)
-\lambda_{j'})}(w_a\,\gamma_0w_a^{-1})=\chi_{k(\tau-\lambda_j)}(\gamma_0)$.
\vskip 0.3cm

We have to choose $\,v_{\tau,a}\,$ so that the
isomorphism $\,\iota'\,$ of hermitian line bundles with
connections preserves also the groupoid multiplication. \,Let \qq
V_{\tau,ab}\ =\
\chi_{k\sigma_{[a+b]}(\tau)}(w_{[a+b]}w_a^{-1}w_b^{-1})
\,u_{ab}\,,\label{vtab}\qqq where $\,(u_{ab})$, a solution of
(\ref{chc0}), enters via (\ref{asa}) the definition of the
groupoid multiplication $\mu'$, see (\ref{muij}). \,As we prove in
Appendix E, the requirement to preserve the groupoid
multiplication imposes the cohomological relation \qq V_{\tau,ab}\
=\
v_{\sigma_a(\tau),b}\,v_{\tau,[a+b]}^{-1}\,v_{\tau,a}\,.\label{frel}
\qqq \,In Appendix F, we show that the 2-cochain
$\,(V_{\tau,ab})\,$ on $\CZ\cong\NZ_{N'}$ with values in the group
$\,U(1)^{^{[\tau]}}\,$ of $\,U(1)$-valued functions on the
$\CZ$-orbit $\,[\tau]\,$ is a 2-cocycle, i.e.\,\,that \qq
V_{\sigma_a(\tau),bc}\,\,V_{\tau,[a+b]c}^{-1}\,V_{\tau,a[b+c]}
\,\,V_{\tau,ab}^{-1}\ =\ 1\,.\label{cocy}\qqq Equation
(\ref{frel}) requires that it be a coboundary, i.e.\,\,that it
defines a trivial element in the cohomology group
$\,H^2(\CZ,U(1)^{^{[\tau]}})$. This always holds since
$\,H^2(\CZ,U(1)^{^{[\tau]}})=\{1\}$, see Appendix A.
\vskip 0.3cm

The multiplication of a solution of (\ref{frel}) by 1-cocycles
\,$(v'_{\tau,a})\,$ satisfying the relation
$\,v'_{\sigma_a(\tau),b}\,{v'_{\tau,[a+b]}}
^{\hspace{-0.3cm}-1}\,v'_{\tau,a}=1\,$ gives all other solutions.
Solutions differing by 1-boundaries $\,v''_{\sigma_a(\tau)}
{v''_\tau}^{-1}\,$ lead to equivalent branes and the set of
equivalence classes of branes supported by $\,\CC'_{[\tau]}\,$ forms 
a $\,H^1(\CZ,U(1)^{^{[\tau]}})$-torsor. Since \,$H^1(\CZ,
U(1)^{^{[\tau]}})\cong\CZ_{\tau}$, see Appendix A, and
$\,\CZ_{\tau}\cong H^1(\CC'_{[\tau]},U(1))\,$ and describes the
moduli of flat line bundles on $\,\CC'_{[\tau]}$, \,this agrees
with the general result about the classification of branes, see
Sect.\,\,7. Let $\,\CZ_{\tau}\,$ be a cyclic subgroup of
$\,\CZ\,$ of order $\,n'\,$ and let $\,n''$ and $\,m''$ be
such that $\,n'n''=N$, $n'm''=N'$ and $\,n''=m''N''$.
\,Explicitly, \qq \CZ_{\tau}\ =\ \{\,z_a\,\,|\,\,a=a''n''\quad{\rm
for}\quad a''=0,1,\dots,n'-1\,\}\,. \qqq In order to generate all
classes in $\,H^1(\CZ,U(1)^{^{[\tau]}})\,$ it is enough to
take \qq v'_{\tau,a}\ =\ (-1)^{2ar\over N} \label{fgs} \qqq
i.e.\,\,$\,\tau$-independent and equal
to the characters of $\,\CZ$. \,Besides $\,r\,$ above may be
restricted to integers between $\,0\,$ and $\,n'-1\,$ since
there exists $\,(v''_\tau)\,$ such that
$\,(-1)^{2an'\over N}=(-1)^{2a'\over m''}=v''_{\sigma_a(\tau)}
\,{v''_\tau}^{-1}$. \vskip 0.4cm

It remains to describe explicitly a single solution of
(\ref{frel}). Let $\,\tau_0\in[\tau]$. Note that \qq k\tau_0\ =\
\sum\limits_{i'=0}^{n''-1}n_{i'}\sum\limits_{a''=0}^{n'-1}
\lambda_{i'+a''n''} \qqq with
$\,\sum\limits_{i'=0}^{n''-1}n_{i'}={k\over n'}\,$ so that $n'$
has to divide $\,k$. The complicated case is when $\,N'$ is even and $\,N''$
is odd and we shall deal with it first. Here, for $\,k\over n'\,$ odd,
$\,n'\,$ must be even since $\,k\,$ is necessarily even.
Let us choose the elements $\,w_a\in SU(N)\,$ inducing the Weyl group
transformations as at the end of Sect.\,\,4. \,Let for $\,u\in
U(1)$, \qq &&\hbox to 1.3cm{$\chi_{\lambda}(u)$\hfill} \ =\
u^{\sum\limits_{i=0}^{N-1}in_i}\qquad{\rm
if}\qquad\lambda=\sum\limits_in_i\lambda_i\,,\alabel{81}{a}\cr\cr
&&\hbox to 1.3cm{$\psi(a,b)$\hfill}\ =\
\chi_{k\lambda_a}^{\,-1}(u_b)\, \cdot\,\cases{\,\hbox to
2cm{$1$\hfill}{\rm for}\ \ {k\over n'} \ \ {\rm even}\,,\cr
\,\hbox to 2cm{$(-1)^{-{ab\over n''}}$\hfill}{\rm for}\ \ {k\over
n'} \ \ {\rm odd}\,,}\aeqno{b}\cr \qqq \vskip -0.1cm \noindent
where $\,u_a\in U(1)\,$ are given by (\ref{ui}). The first formula
may be viewed as extending characters $\,\chi_\lambda\,$ to
constant diagonal
$\,U(N)$-matrices. The following properties of $\,\psi(a,b)\,$ are
straightforward to verify: \qq &&\psi(a+n'', \,b)\ =\ \psi(a,b+N)\
=\ \psi(a,b)\,,\alabel{etv}{a}\cr &&\psi(0,b) \ =\ 1\,,\
\quad\psi([a+b],c)\ =\ \psi(a,c)\,\psi(b,c)\,,\aeqno{b}\cr
&&\psi(a,[b+c])^{\,-1}\,\psi(a,b)\,\psi(a,c)\ =\
\chi_{k\lambda_a}(u_{[b+c]}u_b^{-1}u_c^{-1})\,.\quad\aeqno{c}\cr
\qqq \vskip -0.4cm \noindent To each fixed weight
$\,\lambda_0=k\tau_0\,$ with $\,\tau_0\in[\tau]$, we may assign a
solution $\,(v_{\tau,a}^{\lambda_0})\,$ of (\ref{frel}) given by
\qq &&v_{\tau_0,a}^{\lambda_0}\ \ =\ \chi_{k\tau_0}^{-1}(u_a)\,\,
\chi_{k\lambda_a}(u_a)\,\cdot\,\cases{\,\hbox to 1.6cm{$1$\hfill}\
\ {\rm for}\ \ {k\over n'}\ \ {\rm even}\,,\cr \,\hbox to 1.6 cm{$
(-1)^{a^2\over 2n''}$\hfill}\ \ {\rm for}\ \ {k\over n'}\ \ {\rm
odd}\,,}\alabel{solu}{a}\cr &&v_{\sigma_c(\tau_0),a}^{\lambda_0}\
=\
\,\,\psi(c,a)^{-1}\,\,v_{\tau_0,a}^{\lambda_0}\,.\aeqno{b}\cr\qqq
\vskip -0.2cm \noindent We shown in Appendix G that
$\,(v_{\tau,a}^{\lambda_0})\,$ solves indeed (\ref{frel}) for
$\,N'$ even and $\,N''$ odd. \,For $\,N'$ odd or $\,N''$ even,
since $\,\,V_{\tau,ab}=1$,  \,we may take the trivial solution of
(\ref{frel}) $\,v_{\tau,a}^{\lambda_0}=1\,$ (with a superfluous
dependence on $\,\lambda_0$). \vskip 0.3cm

As we mentioned above, a general solution of (\ref{frel}) is
obtained by multiplying the particular solution $\,(v_{\tau,a}^{\lambda_0})\,$
by $\,(v'_{\tau,a})\,$ of (\ref{fgs}). \,The bundle isomorphisms
$\,\iota'\,$ obtained this way determine stable isomorphisms
between the restriction of gerbe $\,\CG'_k$ to
$\,\CC'_{[\tau]}$ and gerbe $\,\CK'$. \,Consequently, they
determine the branes $\,(\CC'_{[\tau]},\,kQ'_{\tau},
\,N',\,\iota')\,$ supported by the conjugacy class
$\,\CC'_{[\tau]}\,$ in $\,SU(N)/\CZ$. \,As discussed in
Sect\,\,7, using the above structure, one may define the
Wess-Zumino amplitudes (\ref{genamp}) for the fields
$\,\phi:\Sigma\rightarrow SU(N)/\CZ\,$ satisfying boundary
conditions (\ref{bc}) on subintervals of the boundary. \vskip
0.3cm

Note that the solutions $\,(v_{\tau,a})\,$ of (\ref{frel})
differing by 1-coboundaries
$\,(v''_{\sigma_a(\tau)}{v''_\tau}^{-1})\,$ and giving rise to
equivalent branes coincide for $\,a\in\CZ_{\tau}$.
\,Conversely, two solutions coinciding on $\,\CZ_{\tau}\,$
necessarily differ by a 1-coboundary
$\,(v''_{\sigma_a(\tau)}{v''_\tau}^{-1})$. \,Indeed, they differ
by a 1-cocycle $\,(v'_{\tau,a})\,$ such that $\,v'_{\tau,a}=1\,$
for $\,a\in\CZ_{\tau}$. \,Setting $\,v''_{\tau_0}=1\,$ for
fixed $\,\tau_0\in[\tau]\,$ and
$\,v''_{\sigma_a(\tau_0)}=v'_{\tau_0,a}\,$ assures then that
$\,v'_{\tau,a}=v''_{\sigma_a(\tau)}{v''_\tau}^{-1}$. \,Hence there
is a one-to-one correspondence between the moduli of branes 
supported by $\,\CC'_{[\tau]}\,$ and the restrictions of
the solutions $\,(v_{\tau,a})\,$ of (\ref{frel}) to
$\,a\in\CZ_{\tau}$. \,The latter may be taken as products of the
restrictions to $\,\CZ_{\tau}\,$ of the special solutions
$\,(v_{\tau,a}^{\lambda_0})\,$ assigned to $\,\lambda_0=k\tau_0\,$
with $\,\tau_0\in[\tau]\,$ by characters $\,\psi_{\lambda_0}\,$
of $\,\CZ_{\tau}\,$ given by the right hand side of (\ref{fgs})
with $\,a\in\CZ_{\tau}$. \,As follows from (\ref{solu}), two pairs
$\,(\lambda_0,\psi_{\lambda_0})\,$ and $\,(\lambda'_0,\psi'_{\lambda'_0})
\,$ give rise to the same restricted solution if \qq
\lambda'_0\,=\,{}^b\hspace{-0.07cm}\lambda_0\qquad{\rm
and}\qquad\psi_{\lambda'_0}(a)\,=\,\phi_{\lambda_0}(b,a)\,\,
\psi_{\lambda_0}(a)\,,\label{eqr}\qqq
with  $\,{}^b\hspace{-0.06cm}\lambda_0=k\sigma_b^{-1}(\tau_0)\,$ and
\qq
\phi_{\lambda_0}(b,a)\ =\ \psi(b,a)\,\,v_{\tau_0,a}^{\lambda_0}/
v_{\tau'_0,a}^{\lambda'_0}\,.\label{eqr1}
\qqq
\,Note that for any $\,b\in\CZ$, $\,\,\phi_{\lambda_0}(b,a)\,$ must
be a character
of $\,\CZ_\tau\,$ in its dependence on $\,a\,$ and that it satisfies
a cocycle condition $\,\phi_{\lambda_0}(b,a)\,\phi_{\hspace{0.03cm}
{}^b\hspace{-0.07cm}
\lambda_0}(c,a)=\phi_{\lambda_0}([b+c],a)$.
\vskip 0.3cm

The upshot of the above discussion is that the set of moduli
of symmetric branes in the $\,SU(N)/\CZ\,$ WZW theory may be identified 
with the set of equivalence classes $\,[\lambda_0,\psi_{\lambda_0}]\,$ 
where $\,\lambda_0\,$ runs through the integrable weights and
$\,\psi_{\lambda_0}\,$ through the characters of
$\,\CZ_{\tau_0}\,$ for $\,\lambda_0=k\tau_0$, \,with the
equivalence relation given by (\ref{eqr}). This description of
branes, obtained here from the Lagrangian considerations, agrees
with the description of symmetric branes in simple current
extension conformal field theories conjectured in
\cite{FHSSW}\cite{SchwFW}. The general classification of the
branes proposed there, basing on consistency considerations,
involves equivalence classes of primary fields and characters of
their ``central stabilizers'' that, for the $\,SU(N)\,$ WZW
theory, reduce to the ordinary stabilizers $\,\CZ_{\tau}$ in the
simple current group $\,\CZ$. \,The cocycle
$\,\phi_{\lambda_0}(b,a)\,$ is not unique. If we multiply the
special solution $\,(v_{\tau,a}^{\lambda_0})\,$ by a
$\,\lambda_0$-dependent character $\,\rho_{\lambda_0}(a)\,$ of
$\,\CZ$, \,then \qq \phi_{\lambda_0}(b,a)\ \longmapsto\
\phi_{\lambda_0}(b,a)\,\,
\rho_{\lambda_0}(a)/\rho_{{}^b\hspace{-0.06cm}\lambda_0}(a)\,.\label{mcc}
\qqq As we show in Appendix H, upon an appropriate choice of
$\,\rho_{\lambda_0}(a)$, $\,\,\phi_{\lambda_0}(b,a)\,$ may be
reduced to $\,1$. \,In other words, it is possible to choose the
solution $\,(v_{\tau,a}^{\lambda_0})\,$ so that, when restricted
to $\,\CZ_\tau$, it does not depend on $\,\lambda_0$.

\nsection{Partition functions}

Among the elementary quantum amplitudes of the WZW model
are the partition functions. We shall describe them here in the simplest
geometries: those of a torus for the bulk theory and of annulus for
the boundary one, relating in the latter case the Lagrangian description
with the use of gerbes to what was known from previous work.
Although we shall concentrate on the example of the WZW model based on
groups covered by $\,SU(N)$, the general picture should be similar
for other WZW models.

\subsection{Bulk case}

The toroidal level $\,k\,$ partition functions are formally given
by the functional integral over the toroidal amplitudes \qq
Z(\tau)\ =\ \int \ee^{-\,S_\sigma(\phi)}\,\,\CA(\phi)\ D\phi\,,
\label{bupf} \qqq where fields $\,\phi\,$ map the torus
$\,\CT_\tau=\NC/(2\pi\NZ +\tau\NZ)\,$ with the modular parameter
$\,\tau\,$ from the upper half plane\footnote{The modular
parameter, for which we use the traditional notation, should not
be confused with the Weyl alcove elements also denoted by $\tau$
in the present paper} to the group $\,G$, \,the sigma model action
functional \qq S_\sigma(\phi)\ =\ {_k\over^{4\pi
i}}\int\limits\tr\,(\phi^{-1}\partial\phi)(\phi^{-1}\bar\partial\phi)\,,
\label{ssi}\qqq and the amplitude $\,\CA(\phi)\,$ is obtained with
the use of gerbe on $\,G\,$ with curvature $\,kH$. \,In the
Hamiltonian language, \qq Z(\tau)\ =\ \tr_{_{\,\NH}}\,\,\ee^{2\pi
i\tau\,(L_0-{c\over 24})-2\pi i\bar\tau(\bar L_0-{c\over 24})}\,,
\label{bupf1} \qqq where $\,\NH\,$ is the closed-string Hilbert
space composed of sections of the bundle $\,\CL\,$ over the loop
group $\,LG$, \,operators $\,L_0,\ \bar L_0\,$ are the Virasoro
generators and $\,c={k\,{\rm dim}(G)\over k+h^\vee}\,$ is the
Virasoro central charge of the theory. For the connected,
simply-connected simple compact groups, \qq \NH\ \cong\
\mathop{\oplus}\limits_{\lambda}\,\hat V_{\lambda}\otimes
\overline{\hat V_{\lambda}}\,,\label{hdc} \qqq where the sum is
over the integrable weights, i.e.\,\,such that $\,\lambda\in
kA_W$, and $\,\hat V_{\lambda}$ carries the unitary level $k$
irreducible representation of the current algebra $\hat\Ng$
associated with group $G$ and the related action of the Virasoro
algebra given by the Sugawara construction. Integrable weights
$\,\lambda\,$ label the primary fields of the model with the
fusion rule \qq \lambda_0*\lambda_1\ =\ \sum\limits_\lambda
N_{\lambda_0\lambda_1}^{\,\,\lambda}\,\lambda\,.\qqq The
decomposition (\ref{hdc}) implies that \qq Z(\tau)\ =\
\sum\limits_\lambda|\hat\chi_{_\lambda}(\tau)|^2\,,\qqq where
$\,\hat\chi_{_\lambda}(\tau)={\rm tr}_{_{\hat
V_{\lambda}}}\ee^{2\pi i(L_0-{c\over 24})}\,$ are the (restricted)
level $\,k\,$ affine characters of $\,\hat\Ng\,$ satisfying \qq
\hat\chi_{_\lambda}(\tau)\ =\ \ee^{-2\pi
i\,h(\lambda)}\,\,\hat\chi_{_\lambda}(\tau+1)\ =\
\sum\limits_{\lambda'}S_\lambda^{\lambda'}\,\,\hat\chi_{_{\lambda'}}
(-{_1\over^\tau})\,,\label{mtr}\qqq where $\,h(\lambda)\,$ is the
conformal weight of the primary field corresponding to
$\,\lambda\,$ and
$\,S_\lambda^{\lambda'}=S_{\lambda'}^\lambda=\overline{S_{\bar\lambda}
^{\lambda'}}\,$ are the elements of a unitary modular matrix
$\,S\,$ which enter the Verlinde formula for the fusion
coefficients: \qq N_{\lambda_0\lambda_1}^{\,\,\lambda}\ =\
\sum\limits_{\lambda'}
{_{S_{\lambda_0}^{\lambda'}\,\,S_{\lambda_1}^{\lambda'}\,\,
\overline{S_{\lambda}^{\lambda'}}}\over^{S_{0}^{\lambda'}}}\,.\label{Vf}
\qqq \vskip 0.3cm

The toroidal partition functions for all connected non-simply
connected simple compact groups were first obtained in \cite{FGK}.
For $\,G'=SU(N)/\CZ\,$ with $\,\CZ\cong\NZ_{N'}$, \,the integer
level $\,k\geq0\,$ is restricted by the condition of integrality
of the 3-form $\,{k\over2\pi}H'\,$ which reads \qq
{_{kN'}\over^2}\,\tr\,\lambda_{N''}^2=
{_1\over^2}k{N''}(N'-1)\in\NZ\,.\label{rsk}\qqq The closed string
Hilbert space \qq \NH'\ \cong\ \mathop{\oplus}\limits_{z_a\in\CZ}
\mathop{\oplus}\limits_{\lambda\in C_a} \hat
V_{\lambda}\otimes\overline{\hat
V_{\hspace{0,02cm}{}^{^a}\hspace{-0.07cm}\lambda}}\,,
\label{hpr}\qqq where, as before, for integrable weights
$\,\lambda= k\tau=\sum\limits_{i=0}^{N-1}n_i\lambda_i\,$ with
$\,\sum n_i=k$, the transformed weight \qq
{}^{^a}\hspace{-0.06cm}\lambda\ =\ k\,\sigma_a^{-1}(\tau) =\sum
n_i\lambda_{[i+a]}\,,\qqq and where \qq C_a&=&\{\,\lambda\ |\
\tr\,\lambda\lambda_{N''}
+{_{ka}\over^{2N''}}\,\tr\,\lambda_{N''}^2\,\in\NZ\,\}\cr
&=&\{\,\lambda\ |\ -{_{\sum
in_i}\over^{N'}}+{_{ka(N'-1)}\over^{2N'}}\in\NZ\,\}\,.\qqq
Consequently, the partition function \qq Z'(\tau)\ =
\sum\limits_{z_a\in\CZ} \sum\limits_{\lambda\in
C_a}\,\hat\chi_{_\lambda}(\tau)\,
\,\overline{\hat\chi_{_{\hspace{0,02cm}{}^{^a}\hspace{-0.07cm}
\lambda}}(\tau)}\,\,\hspace{1.4cm}\cr
=\ Z'(\tau+1)\ =\ Z'(-{_1\over^\tau})\,.\label{Z'}\qqq One obtains
this a way a family of modular invariant sesquilinear combinations
of characters $\,\hat\chi_{_\lambda}(\tau)$, \,for example for the
case of the $\,SU(2)\,$ group, the $\,A\,$ and $\,D\,$ series in
the $ADE$ classification \cite{CIZ} of modular invariants. \vskip
0.3cm

The above expressions for the toroidal partition functions
coincide with the ones for the ``simple current extensions''
\cite{Schellek} of the $\,SU(N)\,$ WZW theory by the simple
current group $\,\CZ\,$ generated by the simple current $\,J\,$
corresponding to the integrable weight $\,k\lambda_{N''}\,$ with
the fusion rule \qq J^{^{\,a'}}\hspace{-0.13cm}*\lambda\ =\
{}^{^a}\hspace{-0.06cm}\lambda\,. \qqq for $\,a=a'N''$. \,The
restriction on the level $k$ is expressed by demanding that the
conformal weight $\,h(J)={k\over2}\,\tr\, \lambda_{N''}^2\,$
multiplied by the order $N'$ of $\,J\,$ be an integer, which
coincides with condition (\ref{rsk}). \,The requirement
$\,\lambda\in C_a\,$ is expressed by the monodromy charge with
respect to the simple current $\,J\,$ and its modulo 2 refinement.
The monodromy charge \qq
Q_{_J}(\lambda)&=&h(\lambda)+h(J)-h(J*\lambda)\ \,{\rm
mod}\,\,1\,\cr &=&-\tr\,\lambda\lambda_{N''}\ {\rm mod}\,\,1\,=\,
{_{\sum in_i} \over^{N'}}\ \,{\rm mod}\,\,1\label{Q}\qqq is an
important quantity of the conformal field theory. It is conserved
in fusion and it relates the matrix elements of the modular matrix
$\,S_{\lambda'}^{\lambda}\,$ along the orbits of $\,\CZ\,$: \qq
S_{\hspace{0,01cm}{}^{^a}\hspace{-0.07cm}\lambda'}^{\,\lambda}\
=\ \ee^{\,2\pi i\,a'Q_J(\lambda)} \,S_{\lambda'}^{\lambda}\,.
\label{alor} \qqq
The condition $\,\lambda\in C_a\,$ is equivalent to demanding that
\qq Q_{_J}(\lambda)+a'X\in\NZ\,,\label{QX} \qqq where \qq
X\,=\,-{_k\over^2}\,\tr\,\lambda_{N''}^2\ {\rm
mod}\,\,1\,=\,-{_{kN''(N'-1)} \over^{2N'}}\ {\rm
mod}\,\,1\label{X}\qqq so that $\,2X=Q_J(J)\ {\rm mod}\,\,1$.
\vskip 0.3cm

When $\,h(J)={k\over2}\,\tr\,\lambda_{N''}^2\,$ is an integer, the
sets $\,C_a\,$ coincide for different $\,a\,$ and are preserved by
the action of $\,\CZ\,$ on the set of integrable weights. In
this case, the ``pure simple current extension'' in the
terminology of \cite{Schellek}, the partition function (\ref{Z'})
may be rewritten as \qq Z'(\tau)\
=\,{1\over{N'}}\sum\limits_{\lambda\in C_0}\Big|
\sum\limits_{z_a\in\CZ}\hat\chi_{_{\hspace{0,01cm}{}^{^a}
\hspace{-0.07cm}\lambda}}(\tau)\Big|^2
\ =\sum\limits_{[\lambda]\atop\lambda\in C_0}|\CZ_{\lambda}|\,\Big|
\sum\limits_{\lambda\in[\lambda]}\hat\chi_{_\lambda}(\tau)\Big|^2\,,
\quad\label{Z's} \qqq where $\,[\lambda]\,$ runs through the set of 
the $\,\CZ$-orbits in the set of integrable weights and
$\,\CZ_{\lambda}\,$ denote the corresponding isotropy subgroups of
$\,\CZ\,$ (we shall use this notation also below).

\subsection{Boundary case}

For the WZW model based on the simply-connected group $\,G=SU(N)$,
the open string annular partition function corresponding to branes
$\,\CD_s\,$ supported by the conjugacy classes $\,\CC_{\tau_s}$
with $\,s=0,1\,$ is formally given by the functional integral
expression \qq Z_{_{\CD_0\CD_1}}(T)\ =\ \int
\ee^{-\,S_\sigma(\phi)}\,\,\CA_{_{\CD_0\CD_1}}(\phi)\ D\phi\,,
\label{bnpf} \qqq where $\,\phi:[0,T]\times[0,\pi]\rightarrow G\,$
with \qq \phi(t,0)\in
\CC_{\tau_0}\,,\qquad\phi(t,\pi)\in\CC_{\tau_1}\,,\qquad \phi(T,x)
=\phi(0,x)\,, \qqq i.e. $\phi\,$ is periodic in the time
direction. In the Hamiltonian language, \qq Z_{_{\CD_0\CD_1}} (T)\
=\ \tr_{_{\,\NH_{_{\CD_0\CD_1}}}}\ee^{-T(L_0-{c\over24})}\,,
\label{Hl}\qqq where the open string Hilbert space
$\,\NH_{_{\CD_0\CD_1}} \,$ is composed of sections of the bundle
$\,\CL_{_{\CD_0\CD_1}}$ over the space of open paths in $\,G$.
\,Space $\,\NH_{_{\CD_0\CD_1}}$ carries a unitary representation
of the current algebra $\,\hat g\,$ that decomposes into the
irreducible representations according to \qq \NH_{_{\CD_0\CD_1}}\
\cong\
\mathop{\oplus}\limits_{\lambda}\,W_{\,\lambda_0\lambda}^{\,
\lambda_1}\otimes \hat V_{\lambda}\,,\label{decc}\qqq where
$\,\lambda_s=k\tau_s\,$ and the multiplicity spaces
$\,W_{\lambda_0\lambda}^{\,\lambda_1}\,$ may be naturally
identified \cite{GTT}\cite{GawGH} with the spaces of 3-point
conformal blocks of the bulk group $SU(N)$ level $k$ WZW theory on
the sphere with insertions of the primary fields corresponding to
the integrable weights $\,\lambda_0$, $\,\bar\lambda_1$ and
$\,\lambda$. \,In particular, the dimension of the multiplicity
spaces is given by the fusion coefficients: \qq {\rm
dim}(W_{\lambda_0\lambda}^{\,\lambda_1})\ = \
N_{\lambda_0\,\lambda}^{\,\,\lambda_1}\,. \qqq As the result, the
annular partition functions of the $SU(N)$ WZW theory take the
form  \qq Z_{_{\CD_0\CD_1}}(T)\ =\
\sum\limits_{\lambda}N_{\lambda_0\,\lambda}^{\,\,\lambda_1}\,\,\hat\chi_{
_\lambda}({_{Ti}\over^{2\pi}})\,.\label{sbpf}\qqq \vskip 0.3cm

In the dual Hamiltonian description, the annular partition
functions may be described as closed string matrix elements \qq
Z_{_{\CD_0\CD_1}}(T)\ =\ <\hspace{-0.08cm}\CD_1|
\,\ee^{-{2\pi^2\over T}(L_0+\bar
L_0-{c\over12})}\,|\CD_0\hspace{-0.08cm}>\,.\label{dubpf}\qqq
The states $\,|\CD_s\hspace{-0.08cm}>\,$ corresponding
to branes $\,\CD_s\,$ belong to (a
completion of) the closed string Hilbert $\,\NH$. They are
combinations of the so called Ishibashi states
$\,|\lambda\hspace{-0.08cm}>\,$ representing the identity operators
of the representation spaces $\,\hat V_{\lambda}\,$: \qq
|\lambda\hspace{-0.08cm}>\ =\ \sum\limits_{i}
e^\lambda_i\otimes\overline{e^\lambda_i}\qqq for any orthonormal
basis $\,(e^\lambda_i)\,$ of $\,\hat V_{\lambda}$. \,Explicitly
\cite{Cardy}, for the branes $\,\CD_s\,$ supported by the
conjugacy classes $\,\CC_{\tau_s}\,$ with $\,\lambda_s=k\tau_s$,
\qq |\CD_s\hspace{-0.08cm}>\ =\ \sum\limits_{\lambda}
{_{S_{\lambda_s}^\lambda}\over
^{\sqrt{S_{0}^\lambda}}}\,\,|\lambda\hspace{-0.08cm}>\,. \qqq Since \qq
<\hspace{-0.08cm}\lambda|\,\ee^{-{2\pi^2\over T}(L_0+\bar L_0-{c\over12})}\,
|\lambda'\hspace{-0.08cm}>\
=\ \delta_{\lambda\lambda'}\,\,\hat\chi_{_{\lambda'}}
({_{2\pi i}\over^T})\,, \label{spis} \qqq the right hand side of
(\ref{dubpf}) is then equal to \qq
\sum\limits_{\lambda'}{_{\overline{S_{\lambda_1}^{\lambda'}}\,
\,S_{\lambda_0}^{\lambda'}}
\over^{S_{0}^{\lambda'}}}\,\,\hat\chi_{_{\lambda'}}({_{2\pi
i}\over^T})\,, \qqq which indeed coincides with the right hand
side of (\ref{sbpf}) in virtue of the modular property (\ref{mtr})
of the affine characters and the Verlinde formula (\ref{Vf}).
\vskip 0.4cm

For the non-simply-connected group $\,G'=SU(N)/\CZ$, the annular
partition function corresponding to branes $\,\CD'_s$ supported by
the conjugacy classes $\,\CC'_{[\tau_s]}\subset G'\,$ is given by
the primed versions of (\ref{bnpf}) and (\ref{Hl}): \qq
Z'_{_{\CD'_0\CD'_1}}(T)\ =\ \int
\ee^{-k\,S_\sigma(\phi')}\,\,\CA'_{_{\CD'_0\CD'_1}}(\phi')\ D\phi'\
=\ \tr_{_{\,\NH'_{_{\CD'_0\CD'_1}}}}
\ee^{-T(L_0-{c\over24})}\,.\quad\label{bnpf'} \qqq The functional
integral is now over fields $\,\phi':[0,T]\times[0,\pi]\rightarrow
G'\,$ such that \qq \phi'(t,0)\in\CC'_{[\tau_0]}\,,\qquad
\phi'(t,\pi)\in\CC'_{[\tau_1]}\,,\qquad \phi'(T,x)=\phi'(0,x)\,.
\qqq  Each $\,\phi'\,$ may be lifted in $\,N'\,$ different ways to
a twisted periodic map
$\,\phi:[0,T]\times[0,\pi]\rightarrow G\,$ such that \qq
\phi(t,0)\in\CC_{\tau_0}\,,\qquad
\phi(t,\pi)\in\CC_{\tau_1}\,,\qquad \phi(T,x)=z_a\,\phi(0,x) \qqq
for $\,\tau_s\in[\tau_s]\,$ and
$\,z_a\in\CZ_{\tau_0}\cap\CZ_{\tau_1}\subset\CZ$. \,Expressing the
functional integral over fields $\,\phi'\,$ in terms of the one
over their lifts leads to a natural representation for the boundary
partition functions of $\,G'\,$ WZW theory which does not seem to have
appeared in the literature, although it is related to the well studied 
string theory construction of the branes in orbifold 
theories, see e.g.\,\,\cite{DiaDG}\cite{DiaG}. \,Unlike the amplitudes
$\,\CA'_{_{\CD'_0\CD'_1}}(\phi')$, \,which are complex numbers,
the ones of field $\,\phi\,$ are line-bundle valued: \qq
\CA_{_{\CD_0\CD_1}}(\phi)\ \in\ (\CL_{_{\tilde\CD_0\tilde\CD_1}}
)_{_{\tilde\varphi}}\otimes\,(\CL_{_{\CD_0\CD_1}}
)^{^{-1}}_{_{\varphi}} \,,\label{line}\qqq where
$\,\varphi(x)=\phi(T,x)$, $\,\tilde\varphi=\phi(0,x)=z_a^{-1}
\varphi(x)$ and $\,\CD_s\,$ are the branes of the group $\,G\,$
theory supported by the conjugacy classes $\,\CC_{\tau_s}$. 
\vskip 0.3cm

How do the amplitudes
$\,\CA_{_{\CD_0\CD_1}}(\phi)\,$ relate to $\,\CA'_{_{\CD'_0\CD'_1}}
(\phi')\,$? \,The point is that the bundle gerbe $\,\CG'_k\,$ on $\,G'$ 
together with the branes $\,\CD'_s,\,$ determine canonically for any 
$\,\varphi \in I_{_{\CC_{\tau_0}\CC_{\tau_1}}}$, $\,z_a\in\CZ$, \,and
$\,\tilde\varphi=z_a^{-1}\varphi\,$ a non-zero element \qq
\Phi(\tilde\varphi,\varphi)\,\in\,
(\CL_{_{\tilde\CD_0\tilde\CD_1}})^{^{\,-1}}_{_{\tilde\varphi}}
\otimes\,(\CL_{_{\CD_0\CD_1}})_{_{\varphi}}\,, \label{nzem}\qqq
where the branes $\,\CD_s=(\CC_{\tau_s},kQ_{\tau_s},
N_s,\iota_s)\,$ are obtained by the restriction of the branes 
$\,\CD'_s=(\CC'_{[\tau_s]},kQ'_{\tau_s},N'_s,\iota'_s)\,$
to the conjugacy classes $\,\CC_{\tau_s}\,$ for $\,\tau_s\in[\tau_s]$.
We shall call such group $\,G\,$ theory branes $\,D_s\,$ compatible 
with $\,\CD'_s$.
\,Now, for $\,\varphi=\phi(T,\,\cdot\,)\,$ and
$\,\tilde\varphi=\phi(0,\,\cdot\,)$, \,with
$\,z_a\in\CZ_{\tau_0}\cap\CZ_{\tau_s}$, \qq
\CA'_{_{\CD'_0\CD'_1}}(\phi')\ =\
\langle\,\CA_{_{\CD_0\CD_1}}(\phi)\,,\,\Phi(\tilde\varphi,\varphi)
\,\rangle\label{aa'}\qqq in the natural pairing. We describe a
construction of the elements $\,\Phi(\tilde\varphi,\varphi)$, 
\,possessing the multiplicative property \qq
\Phi(\tilde{\hspace{-0.03cm}\tilde\varphi},\tilde\varphi)
\otimes\Phi(\tilde\varphi,\varphi)\ =\ \Phi(\tilde{\hspace{-0.03cm}
\tilde\varphi},\varphi)\label{mp} \qqq 
for $\,\tilde\varphi=z_a^{-1}\varphi\,$ and
$\,\tilde{\hspace{-0.03cm}\tilde\varphi}=z_b^{-1}\tilde\varphi\,$, 
\,\,in Appendix I. \,The construction relies on the concept of branes 
as formulated in the preceding sections. 
\vskip 0.3cm

The multiplication by $\,\Phi\,$ determines an isomorphism between
the line bundles $\,\CL_{_{\tilde\CD_0\tilde\CD_1}}\,$ and
$\,\CL_{_{\CD_0\CD_1}}\,$ covering the map
$\,z_a^{-1}\varphi\mapsto\varphi\,$ between the spaces
$\,I_{_{\CC_{\sigma_a(\tau_0)}\CC_{\sigma_a (\tau_1)}}}$ and
$\,\,I_{_{\CC_{\tau_0} \CC_{\tau_1}}}$ of open paths in $\,G$. 
\,Altogether, one obtains the action of $\,\CZ\,$ on the
bundle 
\qq
{\tilde\CL}'_{_{\CD'_0\CD'_1}}\ =\ \mathop{\cup}\limits_{(\CD_s)}
\CL_{_{\CD_0\CD_1}}
\qqq
where the union is taken over the branes $\,\CD_s\,$ compatible
with $\,\CD'_s$, $\,s=0,1$. \,The multiplicativity of this
action follows from (\ref{mp}). \,The line 
bundle $\,\CL'_{_{\CD'_0\CD'_1}}\,$ over the space $\,I'_{D'_0D'_1}\,$
of open paths in $\,G'\,$ is canonically isomorphic with 
the quotient bundle $\,{\tilde\CL}'_{_{\CD'_0\CD'_1}}/\CZ$. \,\,By the 
formula \qq ({}^{a}\Psi)(\varphi)\ =\
\Psi(z_a^{-1}\varphi)\otimes\Phi(z_a^{-1}\varphi,\varphi)\,,\label{bb'}
\qqq the action of $\,\CZ\,$ may be carried to sections of
the line bundle $\,{\tilde\CL}'_{_{\CD'_0\CD'_1}}\,$ in the way that
maps sections of $\,\CL_{_{\tilde\CD_0\tilde\CD_1}}\,$ to those of
$\,\CL_{_{\CD_0\CD_1}}$. \,Finally, sections of the line bundle
$\,\CL'_{_{\CD'_0\CD'_1}}\,$ may be identified with sections 
of $\,{\tilde\CL}'_{_{\CD'_0\CD'_1}}\,$ invariant under the action of
$\,\CZ$. \vskip 0.3cm

The above gives rise to the following simple picture of the open string 
space of states $\,\NH'_{_{\CD'_0\CD'_1}}\,$ of the group $\,G'\,$
WZW theory. \,The maps $\,\Psi\mapsto{}^{a}\Psi\,$ 
induce the (unitary)
transformations \qq
U(a):\,\NH_{_{\tilde\CD_0\tilde\CD_1}}}\,\longrightarrow\,\NH_{_{\CD_0
\CD_1}\label{unim}\qqq between the open string spaces of states of
the group $\,G\,$ WZW theory. It may be shown that those transformations 
commute with the current algebra and hence also Virasoro algebra actions. 
\,Put together, they define a
representation $\,U\,$ of the group $\,\CZ\,$ in the Hilbert
space
\qq
{\tilde\NH}'_{_{\CD'_0\CD'_1}}\ =\ 
\,\mathop{\oplus}\limits_{\tau_0\in[\tau_0]\atop\tau_1\in[\tau_1]}
\NH_{_{\CD_0\CD_1}}\,.\label{bis} \qqq
When restricted to
$\,\CZ_{\tau_0}\cap\CZ_{\tau_1}\subset\CZ$, \,this representation
acts diagonally, i.e.\,\,within the group $\,G\,$ 
open string spaces $\,\NH_{_{\CD_0\CD_1}}$ with fixed $\,\CD_s$.
\,The open string space states for the group $\,G'\,$ theory may be
naturally identified with the $\,\CZ$-invariant families of states
of the group $\,G\,$ theory:
\qq \NH'_{_{\CD'_0\CD'_1}}\ \simeq\
P\,\,{\tilde\NH}'_{_{\CD'_0\CD'_1}}\,,\label{opo}
\qqq for \qq P\ =\
{_1\over^{N'}}\sum\limits_{z_a\in\CZ}U(a)\label{P}\qqq denoting 
for the projector on the $\,\CZ$-invariant subspace.
The scalar product in the space $\,\NH'_{_{\CD'_0\CD'_1}}\,$ 
should, however, be divided by $\,N'\,$ with respect to the one 
inherited from $\,{\tilde\NH}'_{_{\CD'_0\CD'_1}}\,$ to avoid the 
overcount. \vskip 0.3cm

For the annular partition function of the group $\,G'\,$ theory
one obtains this way the Hamiltonian expression:
\qq
Z'_{_{\CD'_0\CD'_1}}(T)\ =\ {_1\over
^{N'}}\sum\limits_{\tau_0\in[\tau_0]\atop\tau_1
\in[\tau_1]}\,\sum\limits_{z_a\in\CZ_{\tau_0}\cap\CZ_{\tau_1}}
\tr_{_{\,\NH_{_{\CD_0\CD_1}}}}\,\ee^{-T(L_0-{c\over24})}\,\,U(a)\,.
\label{148}\qqq 
This indeed is compatible with the functional integral formula
if rewrite the functional integral over fields $\,\phi'\,$ in
(\ref{bnpf'}) in terms of the one over fields $\,\phi\,$ using 
relation (\ref{aa'}) and the equality of the sigma model actions 
$\,S_\sigma(\phi')=S_\sigma(\phi)$. The factor $\,{1\over N'}\,$
takes care of the $\,N'$-fold overcount due to the fact that there 
are $\,N'\,$ fields $\,\phi\,$ corresponding to each $\phi'$. 
\vskip 0.3cm

The commutation of the maps $\,U(a)\,$ of (\ref{unim}) with the 
the current algebra action implies that they descend to the multiplicity 
spaces in the decomposition (\ref{decc}):
\qq U(a):\,W_{\lambda_0\lambda}^{\,\lambda_1}\,\longrightarrow\,
W_{\hspace{0.02cm}^{^a}\hspace{-0.07cm}\lambda_0\lambda}^{{}^{^a}
\hspace{-0.07cm}\lambda_1}\label{wact}\qqq  
We may then rewrite expression (\ref{148}) for the
open string partition function as \qq
Z'_{_{\CD'_0\CD'_1}}(T)\ =\
{_1\over^{N'}}\sum\limits_{\tau_0\in[\tau_0]\atop\tau_1
\in[\tau_1]}\sum\limits_{z_a\in\CZ_{\tau_0}\cap\CZ_{\tau_1}}
\hspace{-0.1cm}\sum\limits_\lambda\Big(\tr_{_{W_{\lambda_0
\lambda}^{\,\lambda_1}}}U(a)\Big)\,\,\hat\chi_{_\lambda}({_{Ti}
\over{2\pi}})\,,\quad\label{149}\qqq
i.e. in terms of the traces of the action of simple currents
on the spaces of 3-point conformal blocks.   
\,The actions of simple currents on spaces of genus zero conformal blocs 
have been defined, up to phases, in \cite{FS}. In the action 
(\ref{wact}), the phase freedom is fixed by the choice of brane 
structures $\,\CD'_s\,$ on the conjugacy classes $\,\CC'_{[\tau_s]}\,$ 
in $\,G'$. \,Under the change of the brane structures $\,\CD'_s\,$
twisting the isomorphisms $\,\iota'\,$ of
(\ref{triv}) by the multiplication of $\,v_{\tau_s,a}\,$ by
$\,v'_{\tau_s,a}\,$ of (\ref{fgs}) with $\,r=r_s$, \qq
\Phi(\tilde\varphi,\varphi)\ \longmapsto\ (-1)^{{2ar_0\over
N}} \,(-1)^{-{2ar_1\over N}}\,\,\Phi(\tilde\varphi,\varphi)
\qqq inducing the transformation 
 \qq U(a)\ &\longmapsto&\
(-1)^{{2ar_0\over N}} \,(-1)^{-{2ar_1\over N}}\,\,U(a)\,,\quad
\qqq i.e.\,\,multiplying the representation $\,U\,$  by the ratio 
of characters of $\,\CZ$.
\vskip 0.3cm

The annular partition functions for the simple current extension
conformal field theories have been described in
\cite{FHSSW}\cite{SchwFW}, see also \cite{PZ0}. They fit into the 
general scheme identified in \cite{BPPZ}. In the dual Hamiltonian 
description
they are given by the primed version of the matrix elements
(\ref{dubpf}). The states $\,|\CD'_s\hspace{-0.08cm}>\,$
in (a completion of) the
closed string boundary space $\,\NH'\,$ may now be expressed as
combinations of the Ishibashi states $\,|\lambda,z_a\hspace{-0.08cm}>
=\sum e^\lambda_i\otimes \overline{e^\lambda_i}\,$ in the diagonal
components of $\,\NH'$, \,see (\ref{hpr}). Let us denote by
$\,\CE\,$ the corresponding set of labels, i.e. \qq
\CE\,=\,\{\,(\lambda,a)\,\,|\,\,\lambda\in C_a,\ z_a\in
\CZ_\lambda\,\}\,. \qqq Explicitly, for the branes $\,\CD'_s\,$
supported by the conjugacy classes $\,\CC'_{[\tau_s]}\,$
corresponding to the equivalence classes
$\,[\lambda_s,\psi_{\lambda_s}]$, \,where $\,\lambda_s=k\tau_s\,$
and $\,\psi_{\lambda_s}\,$ are characters of $\,\CZ_{\tau_s}$,
\,see the end of Sect.\,\,8.2, \qq |\CD'_s\hspace{-0.08cm}>\ =\
\sum\limits_{(\lambda,a)\in\CE}
{_{\Psi_{\,\CD'_s}^{(\lambda,a)}}\over^{\sqrt{S_0^\lambda}}}
\,\,|\lambda,a\hspace{-0.08cm}>\label{cds0} \qqq with \cite{FHSSW} \qq
\Psi_{\,\CD'_s}^{(\lambda,a)}\ =\
{_{\sqrt{N'}}\over^{|\CZ_{\lambda_s}|}}
\,S_{\lambda_s}^{\lambda}(a)\,\,\psi_{\lambda_s}(a)
\,.\label{cds}\qqq Here $\,S_{\lambda'}^{\lambda}(a)\,$ are the
matrix elements of modified unitary modular matrices non-zero only
if $\,z_a\in\CZ_\lambda\cap\CZ_{\lambda'}$. \,They satisfy
the identity \qq
S_{\hspace{0.02cm}{}^b\hspace{-0.06cm}\lambda'}^{\,\,\lambda}(a)\
=\ \phi_{\lambda'} (b,a)^{-1}\,\,\ee^{2\pi
i\,b'(Q_J(\lambda)+a'X)}\,\,S_{\lambda'}
^{\lambda}(a)\,,\label{941} \qqq see \cite{FSS}\cite{FHSSW}. The
last relation, together with (\ref{QX}), assures that the right
hand side of (\ref{cds}) is independent of the choice of
$\,\lambda_s\in[\lambda_s]\,$ if $\,\phi_\lambda(b,a)\,$ is the
same as the one used in the definition (\ref{eqr}) of the
equivalence classes $\,[\lambda_s,\psi_{\lambda_s}]$. \,Recall
that the latter was fixed up to the transformations (\ref{mcc})
with the help of which, as shown in Appendix H, it could be
reduced to $\,1$. \,As described in \cite{FSS}, up to a phase that
does not depend on $\,\lambda\,$ and $\,\lambda'$, the matrix
elements $\,\,S_{\lambda}^{\lambda'}(a)\,$ are equal to the
entries $\,\check{S}_{\check{\lambda}}^{{\check{\lambda}}'} \,$ of
the modular matrix of the WZW theory based on the so called
"orbit Lie algebra". In our case the latter theory is the 
level $\,{k\over{\check{n}}'}\,$ one with group $\,SU({\check{n}}'')$,
\,where $\,{\check{n}}'\,$ is the order
of the subgroup $\,\CZ_a\,$ of $\,\CZ\,$ generated by $\,z_a\,$
and $\,N={\check{n}}'{\check{n}}''$. \,Writing \qq \lambda\ =\
\sum\limits_{\check{\iota}=0}^{{\check{n}}''-1}{n}_{\check{\iota}}
\sum\limits_{\check{a}=0}^{{\check{n}}'-1}\lambda_{\check{\iota}+\check{a}
{\check{n}}''} \label{45}\qqq with
$\,\sum\limits_{\check{\iota}=0}^{{\check{n}}''-1}{n}_{\check{\iota}}
={k\over{\check{n}}'}$, \,the corresponding weight of the orbit
Lie algebra is \qq \check{\lambda}\ =\
\sum\limits_{\check{\iota}=0}^{\check{n}-1}
{n}_{\check{\iota}}\check{\lambda}_{\check{\iota}}\,,\label{46}
\qqq where $\,\check{\lambda}_{\check{\iota}}\,$ are the
fundamental weights of $\,SU({\check{n}}'')$. \,One has \cite{FSS}
\qq S_{\lambda'}^\lambda(a)\ =\
(-1)^{-{k(({\check{n}}')^2-1){\check{n}}''
\over4{\check{n}}'}}\,\,{\check{S}}_{{\check{\lambda}}'}^{\check{\lambda}}
\,. \qqq In particular, $\,S_{\lambda'}^\lambda(a)\,$ depends on
$\,a\,$ only through $\,\CZ_a$. \,For fixed $\,a$, \,the action 
of $\,\CZ\,$ on the
integral weights $\,\lambda\,$ such that $\,z_a\in\CZ_\lambda\,$
descends to the one of the quotient group
$\,\check{\CZ}\equiv\CZ/\CZ_a\,$ on the weights
$\,\check{\lambda}$. \,The induced action is generated by the fusion 
with the simple current $\,\check{J}\,$ of the $\,SU({\check{n}''})\,$ 
theory with the weight $\,{k\over {\check{n}}'}\check{\lambda_1}$.  
\,The identity (\ref{alor}) for the orbit group implies now that for
$\,z_b\in\CZ$, \qq
\check{S}^{\,{\check{\lambda}}}_{({}^{^b}\hspace{-0.06cm}\lambda'
\check{)\,}}\ =\ \ee^{\,2\pi i
b\,\check{Q}_{\check{J}}({\check{\lambda}})}\,\,
\check{S}_{{\check{\lambda}}'}^{{\check{\lambda}}}\,. \label{ioc}
\qqq In Appendix H we show that \qq \ee^{\,2\pi i
b\,\check{Q}_{\check{J}}({\check{\lambda}})}\ =\ \ee^{2\pi
i\,b'(Q_J(\lambda)+a'X)}\,, \qqq so that (\ref{941}) holds with
$\,\phi_{\lambda'}(b,a)\equiv 1\,$, which is compatible with the
results of Sect.\,\,8.2. \vskip 0.3cm

Relations (\ref{cds0}) and (\ref{cds}), with the use of
(\ref{spis}) and (\ref{mtr}), lead to the following expression for
the annular partition function: \qq Z'_{_{\CD'_0\CD'_1}}(\tau)\ =\
\sum\limits_\lambda\,{\CN}^{\,\CD'_1} _{\CD'_0\lambda}\
\hat\chi_{_\lambda}({_{Ti} \over^{2\pi}})\,,\label{150}\qqq where
\qq
{\CN}^{\,\CD'_1}_{\CD'_0\lambda}&=&\sum\limits_{(\lambda',a)\in\CE}
{_{\overline{\Psi_{\CD'_1}^{(\lambda',a)}}\,\,
\Psi_{\CD'_0}^{(\lambda',a)}\,\,
S^{\lambda'}_\lambda}\over^{S_0^{\lambda'}}}\cr &=&
{_{N'}\over^{|\CZ_{\lambda_0}|\,|\CZ_{\lambda_1}|}}
\sum\limits_{z_a\in\CZ_{\lambda_0}
\cap\CZ_{\lambda_1}}\sum\limits_{\lambda'\in C_a}\,
{_{\overline{S_{\lambda_1}^{\lambda'}(a)}\,\,
S_{\lambda_0}^{\lambda'}(a)\,\,S_\lambda^{\lambda'}}\over
^{S_0^{\lambda'}}} \,\,{\psi_{\lambda_0}(a)
\over\psi_{\lambda_1}(a)}\,,\label{151} \qqq where, again, the
right hand side does not depend on the choice of
$\,\lambda_s\in[\lambda_s]$. \,The orthogonality relations
\cite{FHSSW} \qq \sum\limits_{\CD'}
\overline{\Psi_{\CD'}^{(\lambda,a)}}\,\,
\Psi_{\CD'}^{(\lambda',{a'})} \ =\
\delta_{\lambda\lambda'}\,\delta_{a{a'}} \qqq guarantee that
the matrices $\,\CN_\lambda=({\CN}^{\,\CD'_1}_{\CD'_0\lambda})$,
\,whose entries have to be nonnegative integers, represent the
fusion algebra \cite{BPPZ}: \qq \sum\limits_{\CD'_1}
\CN_{\CD'_0\lambda}^{\,\CD'_1}\,\CN_{\CD'_1\lambda'}^{\,\CD'_2}\
=\ \sum\limits_{\lambda''}N_{\lambda\lambda'}^{\,\lambda''}
\,\,\CN_{\CD'_0\lambda''}^{\,\CD'_2}\,. \qqq Search for the
representations of the fusion algebra by matrices with entries
that are nonnegative integers (the so called ``NIM's'') has been
the basis of the approach to classification of boundary conformal
field theories developed in \cite{BPPZ}, see also \cite{PZ}.
\vskip 0.3cm

Expressions (\ref{150}) with (\ref{151}) are compatible with
relation (\ref{149}) if we assume the following formula for the
traces of the action of $\,\CZ_{\lambda_0}\cap\CZ_{\lambda_1}\,$
on the spaces of 3-point conformal blocks conjectured (up to
multiplication by characters) in \cite{FSS}: \qq {\rm
tr}_{W_{\lambda_0\lambda}^{\,\lambda_1}}U(a)\ =\
\sum\limits_{\lambda'}
{_{\overline{S_{\lambda_1}^{\lambda'}(a)}\,\,
S_{\lambda_0}^{\lambda'}(a)\,\,S_{\lambda}^{\lambda'}}\over
^{S_0^{\lambda'}}}\,\,{\psi_{\lambda_0}(a)
\over\psi_{\lambda_1}(a)} \label{fftr}\qqq  Indeed, the summation
of (\ref{fftr}) over the weights in fixed orbits gives \qq
\sum\limits_{\lambda_0\in[\lambda_0]\atop\lambda_1\in[\lambda_1]}
{\rm tr}_{W_{\lambda_0\lambda}^{\lambda_1}}U(a)\ =\ {_1\over
^{|\CZ_{\lambda_0}|\,|\CZ_{\lambda_1}|}}\sum\limits_{\lambda'}
\sum\limits_{z_b,z_c\in\CZ}{_{\overline{S_{{}^{^b}\hspace{-0.06cm}
\lambda_1}^{\,
\lambda'}(a)}\,\,S_{{}^{^c}\hspace{-0.06cm}\lambda_0}^{\,\lambda'}(a)\,\,
S_{\lambda}^{\lambda'}}\over^{S_0^{\lambda'}}}\,\,{\psi_{\lambda_0}(a)
\over\psi_{\lambda_1}(a)}\,.\quad\label{iwf}\qqq
Note that the sum over $\,\lambda'\,$ is effectively restricted
to integrable weights fixed by $\,z_a$. \,With the use of
transformation property (\ref{941}), the sums over $\,z_b,\,z_c\,$
may be factored out as $\,\Big|\sum\limits_{z_b\in\CZ}\ee^{2\pi
i\,b'(Q_J(\lambda')+a'X)}\Big|^2$. Since the sum inside that factor
divided by $\,N'$ represents the characteristic function of
$\,C_a$, the identity (\ref{iwf}) reduces to the relation \qq
\sum\limits_{\lambda_0\in[\lambda_0]\atop\lambda_1\in[\lambda_1]}
{\rm tr}_{W_{\lambda_0\lambda}^{\lambda_1}}U(a)\ =\
{_{(N')^2}\over^{|\CZ_{\lambda_0}|\,|\CZ_{\lambda_1}|}}\sum
\limits_{\lambda'\in C_a}{_{\overline{S_{\lambda_1}^{\,\lambda'}
(a)}\,\,S_{\lambda_0}^{\,\lambda'}(a)\,\,S_{\lambda}^{\lambda'}}
\over^{S_0^{\lambda'}}}\,\,{\psi_{\lambda_0}(a)
\over\psi_{\lambda_1}(a)}\,,\label{rtr}\qqq  where on the right hand
side the weights $\,\lambda_s\in[\lambda_s]\,$ are fixed and
the sum over $\,\lambda'\,$ is additionally constraint by the 
requirement $\,\lambda'\in C_a$. \,Equation (\ref{rtr}),
when inserted into (\ref{149}), reproduces (\ref{150}). \vskip 0.4cm

One may also consider partition functions that do not resolve
different branes with the same support. Summing
$\,Z'_{_{\CD'_0\CD'_1}}(\tau)\,$ over the different brane
structures $\,\CD'_0\,$ and $\,\CD'_1\,$ supported by the
conjugacy classes $\,\CC'_{[\tau_0]}\,$ and $\,\CC'_{[\tau_1]}$,
\,respectively, \,freezes $\,z_a\,$ in (\ref{149}) to $\,1\,$ and
leads to the unresolved partition functions  \qq
Z'_{_{\CC'_{[\tau_0]}\CC'_{[\tau_1]}}}(T)\,&=&\,
\sum\limits_{\lambda_0\in[\lambda_0]\atop\lambda_1
\in[\lambda_1]}{_{|\CZ_{\lambda_0}|\,\,|\CZ_{\lambda_1}|}\over^{N'}}
\,\,\tr_{_{\,\NH_{_{\CD_0\CD_1}}}}\,\ee^{-T(L_0-{c\over24})}\cr
&=&\,\sum\limits_{\lambda_0\in[\lambda_0]\atop\lambda_1
\in[\lambda_1]}{_{|\CZ_{\lambda_0}|\,\,|\CZ_{\lambda_1}|}\over^{N'}}
\sum\limits_\lambda\,N_{\lambda_0\,\lambda}^{\,\,\lambda_1}\
\hat\chi_{_\lambda}({_{Ti}\over^{2\pi}})\,,\qqq where, as usually,
$\,\lambda_s=k\tau_s$. \,Rewriting the sums over the orbits
$\,[\lambda_s]\,$ as sums over the group $\,\CZ\,$ and using the
symmetry of the fusion coefficients
$\,N_{{}^{^a}\hspace{-0.06cm}\lambda_0\,\lambda}^{{}^{^a}\hspace{-0.06cm}
\lambda_1}= \,N_{\lambda_1\,\lambda}^{\,\,\lambda_1}$, we finally
obtain \qq Z'_{_{\CC'_{[\tau_0]}\CC'_{[\tau_1]}}}(T)\ =\
\sum\limits_{z_a\in\CZ}\sum\limits_\lambda
N_{\,\lambda_0\,\,\lambda}^{{}^{^a}\hspace{-0.06cm}\lambda_1}\,\,
\hat\chi_{_\lambda} ({_{Ti}\over^{2\pi}})\,, \qqq where on the
right hand side, $\,\lambda_s\,$ are arbitrary elements in the
$\,\CZ$-orbits $\,[\lambda_s]$. \,The relation between the annular
partition functions with resolved and unresolved branes was
discussed in \cite{MSchomS} for the case of pure simple current
extensions. The geometric approach based on gerbes should allow
to recover within the Lagrangian framework  similar relations to 
the ones discussed above for general orbifold conformal field
theories.

\nsection{Quantum amplitudes}

The general quantum amplitudes of the WZW theory based on group $\,G\,$
are formally given by the functional integrals
\qq
A_{_{(\CD_s)}}(\Sigma)\ =\ \int \ee^{-\,S_\sigma(\phi)}\,\,\CA_{(\CD_s)}
(\phi)\,\,D\phi
\qqq
over fields $\,\phi:\Sigma\rightarrow G\,$ satisfying boundary conditions
(\ref{bc}) on closed disjoint subintervals $\,\ell_s\,$ of the boundary
loops of $\,\Sigma$, \,with the amplitude $\,\CA_{(\CD_s)}(\phi)\,$ as 
in (\ref{genamp}). Accordingly, we should have
\qq
A_{_{(\CD_s)}}(\Sigma)\ \,\in\ \,
\Big(\mathop{\otimes}\limits_{(s,s')}\NH_{_{\CD_s\CD_{s'}}}
\Big)\otimes\Big(\mathop{\otimes}\limits_m
\,\NH\Big).
\qqq
The (purely) open string amplitudes have no external closed string factors
$\,\NH\,$ (although they may have closed strings states propagating 
in loops). In particular, if $\,\Sigma\,$ is a disc $\,O\,$ and fields 
$\,\phi\,$ are constrained to map three disjoint subintervals of the 
boundary into the supports of three branes $\,\CD_s$, $\,s=0,1,2$, 
\,see Fig.\,\,4, 

\leavevmode\epsffile[-58 -20 262 230]{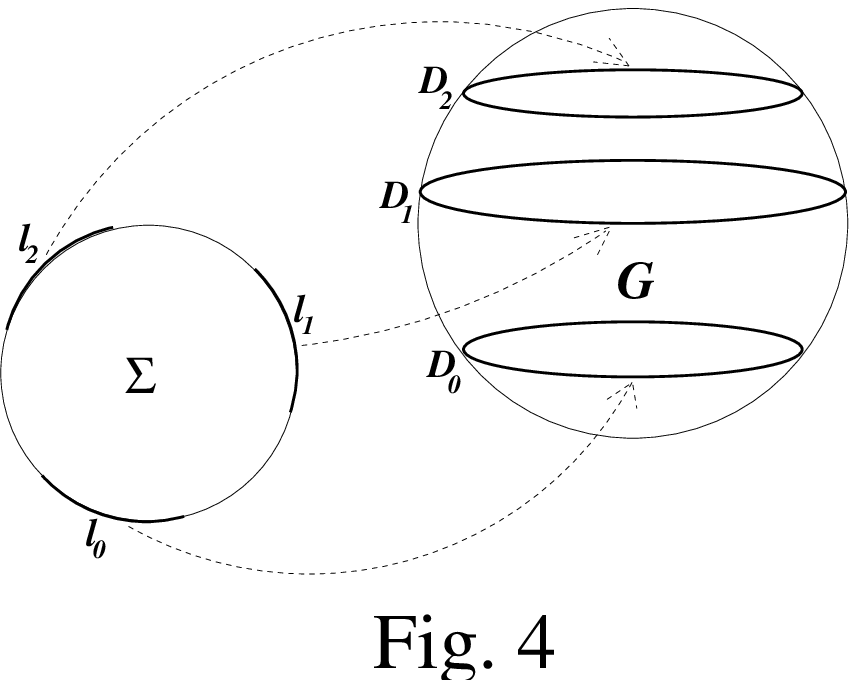}

\noindent one obtains the quantum open string amplitude 
\qq
A_{_{\CD_0\CD_1\CD_2}}(O)\ \,\in\ \,\NH_{_{\CD_0\CD_{1}}}\otimes\,
\NH_{_{\CD_1\CD_{2}}}\otimes\,\NH_{_{\CD_2\CD_{0}}}
\label{3pam}
\qqq
that encodes the operator product expansion of the boundary 
operators. \,The functional integral representation, together with
the geometric interpretation of the relation (\ref{opo}) between 
the spaces of open string states for group $\,G\,$ and group $\,G'\,$
WZW models leads to the following relation between the amplitudes
(\ref{3pam}) for the two cases:
\qq
A'_{_{\CD'_0\CD'_1\CD'_2}}(O)\ =\ (N')^2\,\,P\otimes P\otimes P\ \,
{\tilde A}'_{_{\CD'_0\CD'_1\CD'_2}}(O)\,
\qqq
with  $\,P\,$ is given by (\ref{P}) and
\qq
{\tilde A}'_{_{\CD'_0\CD'_1\CD'_2}}(O)\ =
\mathop{\oplus}\limits_{(\CD_s)}A_{_{\CD_0\CD_1\CD_2}}(O)
\ &\in&\mathop{\oplus}\limits_{(\CD_s)}
\NH_{_{D_0D_1}}\otimes\,\NH_{_{D_1D_2}}\otimes\,\NH_{_{D_2D_0}}\cr
&\subset&\quad\ \,\,{\tilde\NH}'_{_{D'_0D'_1}}\otimes\,
{\tilde\NH}'_{_{D_1D_2}}\otimes\,{\tilde\NH}'_{_{D_2D_0}}\,,
\qqq
see (\ref{bis}). The direct sums above are over branes $\,\CD_s\,$ 
compatible with $\,\CD'_s$, $\,s=0,1,2$. \,An analogous 
relation, with $\,(N')^2\,$ replaced by $\,(N')^{M-1}$, \,holds for 
the disc amplitudes with $\,M\,$ subintervals mapped into brane supports. 
The latter amplitudes permit to reconstruct the general open string 
amplitudes of the WZW theory by gluing. This way one obtains a simple 
relation between the quantum open string amplitudes for the theory based 
on the simply connected group $\,G\,$ and for its simple current orbifolds.
\,We postpone a more detailed discussion of that point to a future 
publication.

\nsection{Local constructions}

Let us describe at the end the local versions of the constructions
inducing from a bundle gerbe on manifold $M$ various geometric
structures on the spaces of closed or open curves (strings) in $M$.

\subsection{Closed strings}
\vskip 0.2cm

We start by recalling from \cite{Gaw0} the construction of local
data for a line bundle on $LM$ from local data
$\,(g_{ijk},{A}_{ij},{B}_i)\,$ of a gerbe on an open
covering $(O_i)$ of $M$. \vskip 0.2cm

Given a split of the circle $S^1$ into closed intervals $b$ with
common vertices $v$ and an assignment $b\mapsto i_b$ and $v\mapsto
i_v$, that we collectively abbreviate as $I$, consider the open
subset \qq O_I\,=\,\{\varphi\in LM\,\,|\,\,\varphi(b)\subset
O_{i_b},\ \varphi(v)\in O_{i_v}\} \label{OI} \qqq of the loop
space $LM$. Sets $O_I$ for different choices of $I$ cover $LM$ (we
may discard $I$ for which $O_I=\emptyset$). Let us define 1-forms
$A_I$ on $O_I\subset LM$ by \qq
\langle\,\delta\varphi,\,A_I(\varphi)\,\rangle\ =\
\sum\limits_b\int
\limits_b\varphi^*\,\iota(\delta\varphi)\,{B}_{i_b}\,+\,\sum
\limits_{v\in b}\langle\,\delta\varphi(v),\,{A}_{i_vi_b}\,\rangle
\label{EI} \qqq with the usual sign convention in
$\sum\limits_{v\in b}$. \vskip 0.2cm

If $\,I\,$ is the collection of $(b,v,i_b,i_v)$ and $J$ the one of
$(b',v',j_{b'},j_{v'})$, consider the split of $S_1$ into
intersections $\bar b$ of the intervals $b$ and $b'$ and denote by
$\bar v$ its vertices. The new split inherits two label
assignments from the original ones. We set \qq &&i_{\bar
b}=i_b\quad{\rm if}\ \ \ \bar b\subset b\,,\hspace{1.8cm} j_{\bar
b}=j_{b'}\quad{\rm if}\ \ \ \bar b\subset b'\,,\cr\cr &&i_{\bar
v}=\cases{i_v\quad{\rm if}\ \ \bar v=v,\cr i_b\quad{\rm if}\ \
\bar v\subset int(b),}\qquad j_{\bar v}=\cases{j_{v'}\quad{\rm
if}\ \ v=v',\cr j_{b'}\quad{\rm if}\ \ \bar v\subset int(b'),}
\qqq where $\,int(b)\,$ denotes the interior of $\,b$. \,Let us
define functions $\,g_{IJ}:O_{IJ}\rightarrow U(1)$, \qq
g_{IJ}(\varphi)\ =\ \exp\Big[i\sum\limits_{\bar
b}\int\limits_{\bar b} \varphi^*{A}_{j_{\bar b}i_{\bar
b}}\Big]\,\prod\limits_{\bar v\in\bar b} \Big(g_{i_{\bar
v}j_{\bar v}j_{\bar b}}(\varphi(\bar v))\,/\, g_{i_{\bar v}i_{\bar
b}j_{\bar b}}(\varphi(\bar v))\Big). \label{GIJ} \qqq
The collection $\,(g_{IJ},A_I)\,$ provides local data for a
hermitian line bundle $\CL$ with connection over $LM$. The curvature
of $\,\CL\,$ is given by the 2-form $\Omega$ on $LM$, see
(\ref{cur}). If $\,(g'_{ijk},{A}'_{ij}, {B}'_i)\,$ are equivalent
local data related to the original ones by (\ref{eqv2}), then \qq
g'_{IJ}\,=\,g_{IJ}f_Jf_I^{-1}\,\qquad A'_I\,=\,A_I+i\,
f_I^{-1}df_I \label{PIJ} \qqq for \qq
f_I^{-1}(\varphi)\,=\,\exp\Big[i\sum\limits_b\int\limits_b\varphi^*
\Pi_{i_b}\Big]\,\prod\limits_{v\in
b}\chi_{i_vi_b}(\varphi(v))\,, \label{FI} \qqq i.e. the
local data $\,(g_{IJ},A_I)\,$ change to equivalent ones, see
(\ref{eq1}). \vskip 0.2cm

\subsection{Open strings}
\vskip 0.2cm

We may apply the previous constructions to the case of open
curves. Using the same formulae (\ref{OI}), (\ref{EI}) and
(\ref{GIJ}) to define an open covering $(O_I)$ of $IM$, 1-forms
$A_I$ on $O_I$ and $U(1)$-valued functions on $O_{IJ}$, we obtain,
however, a different structure. Now on $\,O_{IJK}$, $\,O_I$ and
$\,O_{IJ}$, respectively, \qq &&(g_{IJ}\,g_{IK}^{-1}\,g_{IJ})\ =\
g_{i_1j_1k_1}\circ e_1\,/ g_{i_0j_0k_0}\circ e_0\,,\cr\cr &&dA_I\
=\ \Omega\,+\,e_0^*{B}_{i_0}\,-\, e_1^*{B}_{i_1}\,,\cr\cr
&&A_J-\,A_I-\,{i}\,g_{IJ}^{-1}dg_{IJ}\ =\
e_0^*{A}_{i_0j_0}\,-\,e_1^*{A}_{i_1j_1}\, \qqq with  $\Omega$
given by (\ref{cur}) and  $e_s$ being the evaluation maps of
(\ref{e01}). \vskip 0.2cm

Let $D$ be a submanifold of $M$ and let $Q$ be a 2-form on $D$
such that $\,dQ={H}|_{_D}$. \,Suppose, moreover, that $\Pi_i$ are
one-forms on $\bar O_i=O_i\cap D$ and that $\,\chi_{ij}=\chi_{ji}^{-1}\,$
are $U(1)$-valued functions on $\,\bar O_{ij}$ such that
\vskip 0.2cm

1. \ On $\,\bar O_i$,
\qq
Q\,=\,{B}_i\,+\,d\Pi_i\,,
\label{11}
\qqq

2. \ On $\,\bar O_{ij}$
\qq
0\,=\,{A}_{ij}\,+\,\Pi_j-\Pi_i-i\,\chi_{ij}^{-1}d\chi_{ij}\,,
\label{12}
\qqq

3. \ On $\,\bar O_{ijk}$,
\qq
1\,=\,g_{ijk}\,\chi_{jk}^{-1}\chi_{ik}\,\chi_{ij}^{-1}\,.
\label{13}
\qqq
\vskip 0.1cm

\noindent Provided that the covering $(O_i)$ is sufficiently fine,
the existence of $\,(\chi_{ij},\Pi_i)\,$ with the above properties
is equivalent to the existence of a stable isomorphism between the
restriction to $D$ of the gerbe $\CG$ constructed from the local
data $\,(g_{ijk},{A}_{ij},{B}_i)$ and the gerbe on $D$ obtained
from the local data $\,(Q|_{_{\bar O_i}},\,0,\,1)$. \,The choice
of $\,(\chi_{ij},\Pi_i)\,$ providing the stable isomorphism is
determined up to the local data $\,(h_{ij},R_i)\,$ of a flat
hermitian line bundle over $D$. The $U(1)$-valued functions
$\,f_I$ defined on the open subsets $\bar O_I\subset LD$ by
(\ref{FI}) satisfy now \qq g_{IJ}|_{_{\bar O_{IJ}}}=\ f_J^{-1}f_I
\qqq and define a trivialization of the line bundle $\,L|_{_{LD}}$
defined on $LD$ from the local data $\,(g_{IJ}|_{_{\bar
O_{IJ}}})$. They permit to assign a numerical value \qq
\CA_{_{\CD}}(\phi)\ =\ \CA(\phi)\,f^{-1}_I(\phi|_{_{\ell}})
\label{AB} \qqq to the field $\,\phi:\Sigma\rightarrow M\,$ if
$\,\partial\Sigma\,$ is composed of a single loop $\ell$ mapped by
$\,\phi\,$ into $\,D$. \,Here $\,\CA(\phi)\,$ is given by the
expression (\ref{ampl}) and the collection $I$ is obtained by
restricting the triangulation of $\Sigma$ and corresponding label
assignment to the boundary loop $\ell$. The result does not depend
of the choices of the triangulation of $\Sigma$ neither on the
label assignments. It does not change under the passage to
equivalent local data $\,(g_{ijk},{A}_{ij},{B}_i)\,$ on $M$ if we
absorb the transformations (\ref{eqv2}) in the choice of
$\,(\chi_{ij},\Pi_i)\,$ in (\ref{11}) to (\ref{13}). We may,
however, always modify that choice by the local data
$\,(h_{ij},R_i)\,$ of a flat hermitian bundle $P$ on $D$. Such a
modification multiplies the amplitude $\,A_{_{\CD}}(\phi)\,$ by
the holonomy of $P$ along $\phi|_\ell$. It gives the local
description of the change of the stable isomorphism between the
restricted gerbe and the gerbe constructed from the 2-form $Q$, as
discussed in Sect.\,\,7. \vskip 0.2cm

The generalization of the above discussion to the case when
$\Sigma$ has multiple boundary components with $\phi$ mapping
(some of) them into branes is straightforward.
\vskip 0.2cm

Let $D_0$ and $D_1$ be two submanifolds of $M$ and
$\,{H}|_{_{D_s}} =dQ_s$ with the choices of the data
$\,(\chi^s_{ij},\Pi_i^s)\,$ as above for each $D_s$. Consider the
subspace $\,I_{_{D_0D_1}}\subset IM\,$ of curves ending on the branes,
see (\ref{ID}). We may adapt the definitions (\ref{EI}) and
(\ref{GIJ}) to the present case by defining \qq
&&A_I^D\,=\,A_I\,+\,e_0^*\Pi^0_{i_0}\,-\,e_1^*\Pi^0_{i_0}\,,\cr
&&g_{IJ}^D\,=\,g_{IJ}\,\left(\chi^0_{i_0j_0}\circ e_0\,/\,
\chi^1_{i_1j_1}\circ e_1\right)\,. \qqq One obtains this way local
data $\,(g^D_{IJ},A^D_I)\,$ of a hermitian line bundle $\,\CL_{_{\CD_0
\CD_1}}$ with connection over $\,I_{_{D_0D_1}}M$. The terms added in
(\ref{EI}) and (\ref{GIJ}) are sensitive to the modification of
$\,(\chi^s_{ij},\Pi^s_{i})\,$ by local data of flat bundles $P_s$
on $D_s$. The net result is the multiplication of $\,\CL_{_{\CD_0\CD_1}}$ by
$\,e_0^*P_0\otimes e_1^*P_1^{-1}$. This does not effect the
curvature $\,\Omega_{_D}$ of $\,\CL_{_{\CD_0\CD_1}}$ given by (\ref{curD}).
Upon a change to the equivalent local data $\,(g'_{ijk},{A}'_{ij},
{B}'_i)$, the relations (\ref{PIJ}) still hold for $f_I$ given by
(\ref{FI}), provided we absorb the changes in the choice of
$\,({\chi^s}'_{\hspace{-0.17cm}ij},{\Pi^s}'_{\hspace{-0.1cm}i})$.
The line bundle $\,\CL_{_{\CD_0\CD_1}}$ constructed from the local data
$\,(g^D_{IJ},A^D_I)\,$ is canonically isomorphic to the line
bundle $\CL_{_{\CD_0\CD_1}}(\CG)$ for the gerbe $\,\CG\,$ obtained from
the local data $\,(g_{ijk},{A}_{ij},{B}_i)$, provided that one uses
the data $\,(\chi^s_{ij},\Pi^s_i)\,$ to construct the stable
isomorphism $\iota_s$ of (\ref{sti}) between the restrictions
of $\CG$ and  $\CK_s$.

\nsection{Conclusions}

We have shown how the concept of a bundle gerbes with connection
may be applied to resolve Lagrangian ambiguities in defining sigma 
models in the presence of the antisymmetric tensor field $\,B\,$
determined locally up to closed form contributions. This was done
both in closed string geometry and for open strings stretching between 
branes. Application of that approach to the WZW models based on
groups covered by $\,SU(N)\,$ has permitted to recover within
the Lagrangian approach the classification of symmetric branes.
It has also allowed to make precise a straightforward relation 
between the quantum open string amplitudes for the WZW models 
based on simply connected groups and for their simple current
orbifolds. Those relations are simpler than the ones for closed
string amplitudes where the appearance of twisted sectors complicates
the analysis. They should extend to more general orbifold theories.
\vskip 0.3cm

There are further problems to which one may try to apply the geometric 
methods based on gerbes within the Lagrangian approach to conformal 
field theory. The case of non-orientable worldsheets 
has not been discussed in the present paper. The analysis of gerbes 
entering the WZW models with other groups, with applications to 
the classification of symmetric and symmetry-breaking branes in general 
non-simply connected groups, including the $SO(2N)/\NZ_2$ case with
discrete torsion, is an open problem. Neither have the coset 
models been treated within this framework. Supersymmetric extensions 
require a modification of the approach presented here to take care of 
the fermionic anomalies \cite{FreWit}. Finally, the bundle gerbes should 
be useful in analyzing open string models coupled to non-abelian 
Chan-Patton degrees of freedom, including the fractional branes in 
general orbifolds \cite{DiaDG}, and in the description of the Ramond-Ramond 
brane charges \cite{BCMMS}. We plan to return to some of those 
issues in the future.
\vskip 0.7cm

%

\nappendix{A}
\vskip 0.3cm

\noindent For reader's convenience, we gather here the basic facts
about discrete group cohomology, see \cite{Eilen}\cite{Cartan}\cite{Brown}.
Let $\,\Gamma\,$ be a discrete group with elements
$\,\gamma_0,\gamma_1, \dots\,$ and $\,\CU\,$ an abelian group on
which $\,\Gamma\,$ acts (possibly trivially). We shall use the
multiplicative notation
for the product both in $\,\Gamma\,$ and $\,\CU\,$ and for the
action of $\,\Gamma\,$ on $\,\CU$. \,In our applications,
$\,\Gamma\,$ will be a subgroup $\,\CZ\,$ of the center of
$\,SU(N)\,$ and $\,\CU\,$ will be equal to $\,U(1)\,$ or to the
group of $\,U(1)$-valued functions on the orbit of $\,\CZ\,$ in
the Weyl alcove of $\,su(N)$. \vskip 0.3cm

In general, the abelian group $\,C^n(\Gamma,\CU)\,$ of $n$-cochains
on $\,\Gamma\,$ with values in $\,\CU\,$ is composed of maps \qq
\Gamma^n\ni(\gamma_1,\dots,\gamma_n)\ \longmapsto\
u_{\gamma_1,\dots,\gamma_n}\in\CU\,.\qqq Consider the group homomorphisms
$\,d:C^n(\Gamma,\CU)\rightarrow C^{n+1}(\Gamma,\CU)\,$ defined
by the formula \qq (du)_{\gamma_1,\dots,\gamma_{n+1}}\ =\
(\gamma_1u_{\gamma_2,\dots,\gamma_{n+1}})\,\Big(\prod\limits_{m=1}^n
u_{\gamma_1,..,\gamma_m\gamma_{m+1},..,\gamma_{n+1}}^{\,(-1)^m}\Big)\,
u_{\gamma_1,\dots,\gamma_n}^{(-1)^{n+1}}\,.\qqq For $\,n=0,1,2,3$,
\,the cases relevant for this paper, this gives \qq
(du)_{\gamma}&=&(\gamma u)\,u^{-1}\,\cr
(du)_{\gamma_1,\gamma_2}&=&(\gamma_1u_{\gamma_2})\,u_{\gamma_1\gamma_2}^{-1}\,
u_{\gamma_1}\,,\cr
(du)_{\gamma_1,\gamma_2,\gamma_3}&=&(\gamma_1u_{\gamma_2,\gamma_3})\,u_{\gamma_1\gamma_2,
\gamma_3}^{\,-1}\,u_{\gamma_1,\gamma_2\gamma_3}\,u_{\gamma_1,\gamma_2}^{\,-1}\,,\cr
(du)_{\gamma_1,\gamma_2,\gamma_3,\gamma_4}&=&(\gamma_1u_{\gamma_2,\gamma_3,\gamma_4})
\,u_{\gamma_1\gamma_2,
\gamma_3,\gamma_4}^{\,-1}\,u_{\gamma_1,\gamma_2\gamma_3,\gamma_4}
\,u_{\gamma_1,\gamma_2,\gamma_3\gamma_4}^{\,-1}\,u_{\gamma_1,\gamma_2,\gamma_3}\,.\qqq
The square of $\,d\,$ vanishes and the cohomology groups of $\,\Gamma\,$
with values in $\,\CU\,$ are defined as \qq H^n(\Gamma,\CU)\ =\ {\{\,u\in
C^n(\Gamma,\CU)\,\,|\,\,du=1\,\}\over
d\,C^{n-1}(\Gamma,\CU)}\,.\qqq For the special case of $\,\Gamma\cong\NZ_p\,$
with generator $\,\gamma_0\,$ and $\,n=1,2,\dots$, \qq
H^{2n}(\Gamma,\CU)\ \,\, &\cong& {\{\,u\in\CU\,\,|\,\,\gamma
u=u\,\ {\rm for}\ \,\gamma\in\Gamma\,\}\over
\{\,\prod\limits_{r=0}^{p-1}(\gamma_0^{\,r}u)\,\,|\,\,u\in\CU\,\}}\,,\cr
H^{2n-1}(\Gamma,\CU)&\cong&
{\{\,u\in\CU\,\,|\,\,\prod\limits_{r=0}^{p-1}(\gamma_0^{\,r}u)=1\,\}\over\{\,
(\gamma_0u)\,u^{-1}\,\,|\,\,u\in\CU\,\}}\,,\qqq see \cite{Eilen}\cite{Cartan}.
In particular, for the trivial action of $\,\Gamma\,$ on $\,U(1)$,
\qq H^{2n}(\Gamma,U(1))\ \cong\ \{1\}\,,\qquad
H^{2n-1}(\Gamma,U(1))\ \cong\ \NZ_p\,\qqq and for $\,U(1)^\CT\,$
being the group of $\,U(1)$-valued function on the set $\,\CT\,$
with the action of \ $\,\Gamma\,$ induced from that on
$\,\CT$, \qq H^{2n}(\Gamma,U(1)^\CT)\ \cong\ \{1\}\,,\qquad
H^{2n-1}(\Gamma,U(1)^{\CT})\ \cong\
\mathop{\large{\times}}\limits_{[\tau]\in\CT/\Gamma}\hspace{-0.1cm}
\CZ_\tau\,,\qqq where $\,\CZ_\tau\,$ denotes the stabilizer
subgroup of $\tau\in\CT$ (which depends only on the $\Gamma$-orbit
$\,[\tau]\,$ of $\,\tau$).
\vskip 0.2cm

The discrete group cohomology should be distinguished from the one
for Lie groups appearing also in the paper. The latter is defined 
as for general topological spaces, e.g. by the $\check{\rm C}$ech
construction.
\vskip 0.2cm

\nappendix{B}
\vskip 0.3cm

\noindent To check the associativity of the product $\,\mu'$ of
(\ref{muij}) over $O_{ijln}\subset
{Y'}^{{[4]}}$ with $\,j=[j'+a]$, $\,l=[l'+b]\,$ and $\,n=[n'+c]$,
we first calculate \qq
&&\mu'\Big(\mu'\Big([\gamma,\zeta]_{k\lambda_{ij}}\otimes [\gamma
w_a^{-1},\zeta']_{k\lambda_{j'[l-a]}}\Big)\otimes [\gamma
w_b^{-1},\zeta'']_{k\lambda_{l'[n-b]}}\Big)\,\,\cr
&&\quad=\ \mu'\Big([\gamma,\,u_{ijl}\zeta\zeta']_{k\lambda_{il}}\otimes
[\gamma w_b^{-1},\zeta'']_{k\lambda_{l'[n-b]}}\Big)\ =\
[\gamma,\,u_{ijl}u_{iln}\zeta\zeta'\zeta'']_{_{k
\lambda_{in}}}\,.\label{lhs2} \qqq On the other
hand, using the identity \qq [\gamma
w_b^{-1},\zeta'']_{k\lambda_{l'[n-b]}}\ =\ [\gamma
w_a^{-1}w_{[b-a]}^{-1} w_{[b-a]}w_a
w_b^{-1},\,\zeta'']_{k\lambda_{l'[n-b]}}\,\,\cr =\ [\gamma
w_a^{-1}w_{[b-a]}^{-1},\,
\chi_{_{k\lambda_{l'[n-b]}}}(w_{[b-a]}w_a
w_b^{-1})\,\zeta'']_{_{k\lambda_{l'[n-b]}}} \qqq that follows
from the equivalence relation (\ref{iga}) since $\,w_{[b-a]}w_a
w_b^{-1}\,$ lies in the Cartan subgroup $T$, we infer that \qq
&&\mu'\Big([\gamma w_a^{-1},\zeta']_{k\lambda_{j'[l-a]}}\otimes
[\gamma w_b^{-1},\zeta'']_{k\lambda_{l'[n-b]}}\Big)\cr
&&\hspace{0,4cm}=\ [\gamma w_a^{-1},\,u_{j'[l-a][n-a]}\,
\chi_{_{k\lambda_{l'[n-b]}}}(w_{[b-a]}w_a w_b^{-1})
\,\zeta'\zeta'']_{_{k\lambda_{j'[n-a]}}}\,. \qqq Another
application of $\,\mu'$ gives then: \qq
&&\mu'\Big([\gamma,\zeta]_{k\lambda_{ij}}\otimes\mu'\Big(
[\gamma w_a^{-1},\zeta']_{k\lambda_{j'[l-a]}}\otimes [\gamma
w_b^{-1},\zeta'']_{k\lambda_{l'[n-b]}}\Big)\Big)\cr &&=\
[\gamma,\,u_{ijn}\,u_{j'[l-a][n-a]}\,
\chi_{_{k\lambda_{l'[n-b]}}}(w_{[b-a]}w_a w_b^{-1})
\,\zeta\zeta'\zeta'']_{_{k\lambda_{in}}}\,.\label{rhs2} \qqq
Comparing (\ref{lhs2}) and (\ref{rhs2}), we find
condition (\ref{assop}) for the associativity of $\,\mu'$.
\vskip 0.2cm

\nappendix{C}
\vskip 0.3cm

\noindent We describe here the construction of the ``quotient gerbe''
$\,\CG'_k$
on $\,G'=SU(N)/\CZ\,$ along the lines of Section\,\,5. The resulting
gerbe coincides with the gerbe on $\,G'$ constructed in Section 4.
\vskip 0.2cm

With $\,M=SU(N)\,$ and the gerbe $\,\CG^k=(Y,\,kB,\,L^k,\,\mu^k)\,$
constructed in Section\,\,3 we shall take as $\,\Gamma\,$ the subgroup
$\,\CZ\cong\NZ_{N'}$ of the center of $SU(N)$ acting on $\,M\,$ by the
(left or right) multiplication. Recall that
$\,Y=\mathop{\sqcup}\limits_{i=0}^rO_i$. \,For
$\,\gamma=z_{a}\in\Gamma$, where $\,a\,$ is divisible by $N''=N/N'$,
\qq Z_\gamma\ =\
Y\times_{_M}Y_\gamma&=&\{\,((g,i),(z_{a}^{-1}g,j')) \ |\ g\in
O_i,\ z_{a}^{-1}g\in O_{j'}\,\}\cr\cr
&\cong&\mathop{\sqcup}\limits_{i,j} O_{ij}\,, \qqq where
$\,j=[j'+a]$. We may take the bundle $\,N^{^\gamma}$ over
$\,Z_\gamma$ to be equal to $\,\rho_{ij}^*L^k_{_{\lambda_{ij}}}$ over
the $\,O_{ij}$ component. Since \qq
kB_{j'}(z_a^{-1}g)\,-\,kB_i(g)\ =\ kB_j(g)\,-\,kB_i(g) \ =\
k\,\rho_{ij}^*F_{_{\lambda_{ij}}}(g)\,, \qqq the relation
(\ref{curr}) is satisfied. The isomorphism
$\,\iota_\gamma$ of (\ref{zgg}), upon taking \qq y_1=(g,i_1),\,\
y_2=(g,i_2),\,\ y'_1=(z_a^{-1}g,j'_1),\,\
y'_2=(z_a^{-1}g,j'_2) \qqq with $\,g=\gamma\,\ee^{2\pi
i\tau}\gamma^{-1}\in O_{i_1i_2j_1j_2}$, \,may be defined by \qq
\iota_\gamma\Big([\gamma,\zeta]_{_{k\lambda_{i_1i_2}}}\hspace{-0.1cm}
\otimes\,[\gamma,\zeta']^{\,-1}_{_{k\lambda_{i_ij_1}}}\hspace{-0.1cm}
\otimes\,[\gamma,\zeta'']_{_{k\lambda_{i_2j_2}}}\Big)=\,[\gamma
w_a^{-1},\zeta\zeta'\zeta'']_{_{k \lambda_{j'_1j'_2}}}\,.
\qqq For $\,Y_{_\Gamma}=Y\,$ with the projection on $\,G'$ and
$\,(y,y',y'')\in Y_{_\Gamma}^{{[3]}}$,\qq y=(g,i),\,
\ y'=(z_a^{-1}g,j'),\,\ j=[j'+a],\ \,
y''=(z_b^{-1}g,l'),\,\ l=[l'+b],
\qqq the map \qq
[\gamma,\zeta]_{_{k\lambda_{ij}}}^{-1}\otimes\,[\gamma
w_a^{-1},\zeta']_{_{k\lambda_{j'[l-a]}}}^{-1}\otimes\,
[\gamma,\zeta'']_{_{k\lambda_{il}}}\ \longmapsto\ \,
[g,\zeta^{-1}{\zeta'}^{-1}\zeta'',\,ijl]\qqq defines for
$\,\gamma_1=z_a$ and $\,\gamma_2=z_a^{-1}z_b$ an isomorphism
between the bundle $\,{\tilde R}^{^{\gamma_1,\gamma_2}}$, see
(\ref{nnr}), and the pullback of the flat bundle \qq
R^{^{\gamma_1,\gamma_2}}\ =\
\Big(\mathop{\sqcup}\limits_{i,j',l'}O_{ijl}\times\NC\Big)\Big/\sim\,
\qqq over $\,SU(N)$, \,where the equivalence relation $\,\sim\,$ is
defined by \qq (g,\zeta_1,i_1j_1l_1)\ \sim\ (g,\zeta_2,i_2j_2l_2)
\quad{\rm
if}\quad\zeta_1=\zeta_2\,\chi_{_{k\lambda_{l'_1l'_2}}}(w_b
\,w_a^{-1}w_{[b-a]}^{-1})\,. \qqq For $\,(y,y',y'',y''')\in
Y_{_\Gamma}^{{[4]}}$ with
$\,y'''=(z_c^{-1}g,\,n')$,
$\,n=[n'+c]\,$ and $\,\gamma_3=z_b^{-1}z_c$,
the isomorphism (\ref{123}) identifies \qq
[g,\zeta,ijl]\otimes[g,\zeta',iln]\ \cong\
\chi_{_{k\lambda_{l'[n-b]}}}(w_b
\,w_a^{-1}w_{[b-a]}^{-1})\,\,\cr\cr \cdot\
[g,\zeta,ijn]\otimes[z_a^{-1}g,
\zeta',j'[l-a][n-a]]\,.\qqq The flat bundle
$\,P^{^\gamma}$ will be taken trivial and the isomorphisms
$\,\iota_{_{\gamma_1,\gamma_2}}$ of \,(\ref{iogg}) will be defined by
\qq
(g,\zeta)\otimes(z_a^{-1}g,\zeta')\ \longrightarrow\
\,(g,\zeta\zeta')\otimes[g,u_{ijl},ijl]\label{8}\qqq for $\,g\in
O_{ijl}$ and $\,u_{ijl}\in U(1)$. According to the definition of the
classes $[g,u_{ijl},ijl]$, see (\ref{8}), we must have
\qq
u_{i_2j_2l_2}\ =\ \chi_{_{k\lambda_{l'_1l'_2}}}(w_b\,w_a^{-1}
w_{[b-a]}^{-1})\,\,u_{i_1j_1l_1}\label{trpi}
\qqq
for the same $\,a\,$ and $\,b\,$ that we suppressed in the notation
for $\,u_{ijl}$. Equation (\ref{asa}) is a special case of
the above relation. Property (\ref{assoi}) reduces to
(\ref{assop}) which is consistent with the transformation properties
(\ref{trpi}). It is then enough to consider $\,u_{0ab}\equiv
u_{a[b-a]}$ in which case (\ref{assop}) reduces to (\ref{chc0}).
\vskip 0.2cm

\nappendix{D}
\vskip 0.3cm

\noindent We shall prove here that the amplitude $\CA(\phi)$ of
equation (\ref{amplg}), when interpreted as a number following the
procedure described in Sect.\,\,6, coincides with the expression
(\ref{ampl}). First, note that
$\int\limits_c\phi_{c}^{\,*}{B}=\int\limits_c\phi^* {B}_{i_c}$.
Next, observe that the holonomies in ${L}$ are \qq
\CH(\phi_{cb})\,=\,\exp\Big[i\int\limits_b
\phi^*{A}_{i_ci_b}\Big]\,\mathop{\otimes}\limits_{v\in b}
s_{i_ci_b}(y_c,y_b)\,. \non \qqq We have then to compute
the numbers assigned for every vertex $v$ to \qq
s_{i_{b_1}i_{c_1}}\hspace{-0.06cm}(y_{b_1},y_{c_1}) \otimes
s_{i_{c_1}i_{b_2}}\hspace{-0.06cm}
(y_{c_1},y_{b_2})\otimes\,\cdots\,\otimes s_{i_{c_n}i_{b_1}}
\hspace{-0.06cm}(y_{c_n},y_{b_1})\,, \non
\qqq see Fig.\,\,1. For $i_v$ such that $v\in O_{i_v}$ and
$\,y_v=\sigma_{i_v}(\phi(v))$, we shall insert at every second
place in the last chain the tensor $\,s_{i_{b_r}i_v}
\hspace{-0.06cm}(y_{b_r},y_v)\otimes s_{i_vi_{b_r}}
\hspace{-0.06cm}(y_v,y_{b_r})\,$ mapped by $\mu$ to
$1\in{L}_{(y_{b_r}, y_{b_r})}$. This permits to split the chain to
the blocks \qq
s_{i_vi_{b_{r}}}\hspace{-0.06cm}(y_v,y_{b_{r}})\otimes
s_{i_{b_{r}}i_{c_r}}\hspace{-0.06cm}(y_{b_{r}}, y_{c_r})\otimes
s_{i_{c_r}i_{b_{r+1}}}\hspace{-0.06cm}(y_{c_r},
y_{b_{r+1}})\otimes s_{i_{b_{r+1}}i_v}\hspace{-0.06cm}
(y_{b_{r+1}},y_v)\,. \non
\qqq
The latter give rise under $\mu$ to the factors $\,g_{i_vi_{b_{r}}
i_{c_r}}(\phi(v))\,g^{-1}_{i_vi_{b_{r+1}}i_{c_r}}
(\phi(v))\,$ which build up the product appearing in (\ref{ampl}).
\vskip 0.2cm
\eject

\nappendix{E} \vskip 0.3cm

\noindent Let us show that the isomorphism $\,\iota'\,$ of
(\ref{triv}) between $\,L'\otimes p_1^*{N'}^{^{-1}}\hspace{-0.1cm}
\otimes p_2^*N'\,$ and the trivial bundle
$\,{Z'}^{{[2]}}\times\NC\,$ intertwines the groupoid
multiplication if and only if the relation (\ref{frel}) holds. Let
$\,y_1=(g,i)$, $\,y_2=(z_a^{-1}g,j')$, and
$\,y_3=(z_b^{-1}g,l')\,$ for $\,g=\gamma\,\ee^{2\pi i\tau}
\gamma^{-1}$. \,Consider the elements \qq f_1\ =\
[\gamma,1]_{{k\lambda_{ij}}}\otimes\,[\gamma,1]^{-1}_{{k
(\tau-\lambda_{i})}}\otimes\,[\gamma
w_a^{-1},1]_{_{k(\sigma_a(\tau)- \lambda_{j'})}} \label{L1} \qqq
in the fiber $\,\,\Big(L'\otimes p_1^*{N'}^{^{-1}}\hspace{-0.1cm}
\otimes p_2^*N'\Big)_{{(y_1,y_2)}}\,$ and \qq f_2\ =\ [\gamma
w_a^{-1},1]_{{k\lambda_{j'[l'+b-a]}}}\otimes\, [\gamma
w_a^{-1},1]^{-1}_{{k(\sigma_a(\tau)-\lambda_{j'})}}\hspace{1cm}\cr\cr
\otimes\,[\gamma w_a^{-1}w_{[b-a]}^{-1},1]_{{k(\sigma_b(\tau)
-\lambda_{l'})}} \label{L2} \qqq in the fiber $\,\,\Big(L'\otimes
p_1^*{N'}^{^{-1}}\hspace{-0.1cm} \otimes p_2^*N'\Big)_{{
(y_2,y_3)}}$. \,The product of those two elements, \qq
\mu'(f_1\otimes f_2)\ =\ [\gamma, u_{ijl}]_{k\lambda_{il}}\otimes
\,[\gamma,1]^{-1}_{k(\tau-\lambda_i)}\otimes\,[\gamma
w_a^{-1}w_{[b-a]}^{-1}, 1]_{k(\sigma_b(\tau)-\lambda_{l'})}\,,
\qqq see (\ref{muij}), where the last tensor factor may be
rewritten as \qq [\gamma w_b^{-1},\chi_{k(\sigma_b(\tau)
-\lambda_{l'})}(w_b\,w_a^{-1}w_{[b-a]}^{-1})]_{k(\sigma_b(\tau)
-\lambda_{l'})}\,. \qqq From the definition of the isomorphism
$\,\iota'\,$ we have \qq &&\hbox to 2.5cm{$\iota'(f_1)$\hfill}\ =\
(y_1,y_2,\, \,v_{\tau,a})\,,\alabel{io}{a}\cr &&\hbox to
2.5cm{$\iota'(f_2)$\hfill}\ =\ (y_2,y_3,\,\,
v_{\sigma_a(\tau),[b-a]})\,. \aeqno{b}\cr &&\hbox to
2.5cm{$\iota'(\mu'(f_1\otimes f_2))$\hfill}\ =\ (y_1,y_3,\,
\,v_{\tau,b}\,u_{ijl}\,
\chi_{k(\sigma_b(\tau)-\lambda_{l'})}(w_b\,w_a^{-1}w_{[b-a]}^{-1})\,)\,.
\hspace{1cm}\aeqno{c}\cr \qqq \vskip-0.4cm \noindent  The product
of the first two elements of the trivial bundle over
$\,{Z'}^{[2]}\,$ is equal to the third one if and only if \qq
v_{\tau,a}\,v_{\sigma_a(\tau),[b-a]}\ =\ v_{\tau,b}\,u_{ijl}\,\,
\chi_{k(\sigma_b(\tau)-\lambda_{l'})}(w_b\,
w_a^{-1}w_{[b-a]}^{-1})\,\,\cr =\
v_{\tau,b}\,\,\chi_{k\sigma_b(\tau)}(w_b\,w_a^{-1}
w_{[b-a]}^{-1})\,\,u_{a[b-a]}\,, \qqq where the last equality
follows from (\ref{asa}). Upon the shift $\,b\mapsto[a+b]\,$ this
reduces to (\ref{frel}).
\vskip 0.2cm

\nappendix{F} \vskip 0.3cm

\noindent Here we prove the cocycle identity (\ref{cocy}). The
left hand side is equal to \qq
&\chi_{k\sigma_{[a+b+c]}(\tau)}(w_{[b+c]}w_b^{-1}w_a^{-1})\,\,
\chi^{-1}_{k\sigma_{[a+b+c]}(\tau)}(w_{[a+b+c]}w_{[a+b]}^{-1}w_c^{-1})&\cr\cr
&\cdot\ \chi_{k\sigma_{[a+b+c]}(\tau)}(w_{[a+b+c]}w_a^{-1}
w_{[b+c]}^{-1})\,\,\chi^{-1}_{k\sigma_{[a+b]}(\tau)}(w_{[a+b]}
w_a^{-1}w_b^{-1})&\cr\cr &\cdot\
u_{bc}\,u^{-1}_{[a+b]c}\,u_{a[b+c]}\,u^{-1}_{bc}\,.& \qqq With the
use of identity \qq \chi_{k\sigma_a(\tau)}(w_at\,w_a^{-1})\ =\
\chi_{k\tau}(t)\,\, \chi^{-1}_{k\lambda_a}(t)\label{itu}\qqq
holding for $\,t\,$ in the Cartan subgroup and of relation
(\ref{chc0}), this may be rewritten as \qq
&\chi_{k\sigma_{[a+b+c]}(\tau)}(w_{[b+c]}w_b^{-1}w_c^{-1})\,\,
\chi^{-1}_{k\sigma_{[a+b+c]}(\tau)}(w_{[a+b+c]}w_{[a+b]}^{-1}w_c^{-1})\cr\cr
&\cdot\ \chi_{k\sigma_{[a+b+c]}(\tau)}(w_{[a+b+c]}w_a^{-1}
w_{[b+c]}^{-1})\,\,\chi^{-1}_{k\sigma_{[a+b]}(\tau)}(w_c\,w_{[a+b]}
w_a^{-1}w_b^{-1}w_c^{-1})& \qqq which is equal to $\,1\,$ by the
multiplicativity of the characters.
\vskip 0.2cm

\nappendix{G} \vskip 0.3cm

\noindent Let us prove that  $\,v_{\tau,a}=v_{\tau,a}^{\lambda_0}\,$
given by (\ref{solu}) provides a special solution of (\ref{frel})
for $N'$ even and $N''$ odd. For $\,\tau=\sigma_d(\tau_0)$, we
have \qq
&&v_{\sigma_a(\tau),b}\,\,v_{\tau,[a+b]}^{-1}\,v_{\tau,a}\ =\
\psi({[a+d],b})^{-1}\,\,\psi({d,[a+b]})\,\,\psi({d,a})^{-1}\,\,
v_{\tau_0,b}\,\,v_{\tau_0,[a+b]}^{-1}\,v_{\tau_0,a}\cr &&=\
\psi({a,b})^{-1}
\,\,\chi_{k\lambda_d}^{-1}(u_{[a+b]}u_a^{-1}u_b^{-1})\,\,
v_{\tau_0,b}\,\,v_{\tau_0,[a+b]}^{-1}\,v_{\tau_0,a}\,, \qqq where
we have used the relations (\ref{etv}). Since, by (\ref{itu}), \qq
V_{\sigma_d(\tau_0),ab}\ =\
V_{\tau_0,ab}\,\,\chi^{-1}_{k\lambda_d}
(u_{[a+b]}u_a^{-1}u_b^{-1})\,, \qqq the identity (\ref{frel}) will
follow if we show that \qq
v_{\tau_0,b}\,\,v_{\tau_0,[a+b]}^{-1}\,v_{\tau_0,a}\ =\
V_{\tau_0,ab}\, \,\psi({a,b})\,.\label{ttp} \qqq The left hand
side is \qq
\chi_{k\tau_0}^{-1}(u_b)\,\,\chi_{k\tau_0}(u_{[a+b]})\,\,
\chi_{k\tau_0}^{-1}(u_a)\,\,\chi_{k\lambda_b}(u_b)\,\,\chi^{-1}_{k
\lambda_{[a+b]}}(u_{[a+b]})\,\,\chi_{k\lambda_a}(u_a)\cr \cdot\
\cases{\,\hbox to 3cm{$ \,1$\hfill}\cr \,\hbox to
3cm{$(-1)^{{b^2-[a+b]^2+a^2\over 2n''}}$\hfill}} \cr\cr =\
\chi_{k\tau_0}(u_{[a+b]}u_a^{-1}u_b^{-1})\,\,\chi_{k\lambda_{[a+b]}}
^{-1}(u_{[a+b]}u_a^{-1}u_b^{-1})\,\,\chi_{k\lambda_a}^{-1}(u_b)\,\,
\chi_{k\lambda_b}^{-1}(u_a)\cr\cdot\ \cases{\,\hbox to
1.5cm{$\,1$\hfill}\cr \,\hbox to 1.5cm{$(-1)^{-{ab\over
n''}}$\hfill}} \qqq and it indeed coincides with the right hand
side, as may be seen from the relation \qq V_{\tau,ab}\ =\
\chi_{k\tau}(w_a^{-1}w_b^{-1}w_{[a+b]})\,\,\chi^{-1}_{k\lambda_{[a+b]}}
(w_a^{-1}w_b^{-1}w_{[a+b]})\,\,u_{ab}\qqq following from the
definition (\ref{vtab}) and (\ref{itu}).
\vskip 0.2cm
\eject

\nappendix{H} \vskip 0.3cm

Here we show that the modification (\ref{mcc}) with an
appropriately chosen character  $\,\rho_{\lambda_0}(a)\,$ of
$\,\CZ\,$ depending on $\,\lambda_0=k\tau_0\,$ with
$\,\tau_0\in[\tau]\,$ trivializes the cocycle
$\,\phi_{\lambda_0}(b,a)\,$ of (\ref{eqr1}). We only have to
consider the case of $\,N'\,$ even and $\,N''\,$ odd since for
the other cases $\,\phi_{\lambda_0}(b,a)=1$. \,First, for
$\,z_a\in\CZ_\tau$, \,i.e.\,\,for $\,a=a''n''$, \qq \psi(b,a)&=&
(-1)^{-{a''(n'-a'')n''bk\over n'}}\cdot\cases{\,\hbox to
1.6cm{$1$\hfill}\ \ {\rm for}\ \ {k\over n'}\ \ {\rm even}\,,\cr
\,\hbox to 1.6cm{$(-1)^{-a''b}$\hfill}\ \ {\rm for}\ \ {k\over
n'}\ \ {\rm odd}\,,}\cr\cr &=&\cases{\,\hbox to 2.3cm{$1$\hfill}\
\ {\rm for}\ \ {k\over n'}\ \ {\rm even}\,,\cr \,\hbox to
2.3cm{$(-1)^{a''b(n''-1)}$\hfill}\ \ {\rm for}\ \ {k\over n'}\ \
{\rm odd}\,,}\qqq see (\ref{81},b). \,Let \qq
\rho^0_{\lambda_0}(a)\ =\ (-1)^{
\,a\sum\limits_{i'=0}^{n''-1}i'n_{i'}}\,,\qquad
\rho^1_{\lambda_0}(a)\ =\ (-1)^{\,{a\over
n''}\sum\limits_{i'=0}^{n''-1} i'n_{i'}}\,,\label{rhoa} \qqq with
$\,\rho^1\,$ defined for \,$k\over n'\,$ odd. A direct check shows
that for $\,a=a''n''$, \qq
\chi_{\lambda_0}^{-1}(u_a)\,\rho_{\lambda_0}^0(a) \qqq does not
depend on the choice of $\,\tau_0\in[\tau]$. \,It follows that,
for $\,{k\over n'}\,$ even, $\,\phi_{\lambda_0}\,$ is trivialized
by (\ref{mcc}) if we take $\,\rho=\rho^0$. \,Finally, for
$\,{k\over n'}\,$ odd, $\,\sum\limits_{i'=0}^{n''-1}i'n_{i'}\,$
preserves or changes its parity under the shift
$\,\lambda_0\mapsto{}^b \hspace{-0.06cm}\lambda_0\,$ for $\,b\,$
even or odd, respectively, so that \qq
{\rho^0_{{}^b\hspace{-0.06cm}\lambda_0}(a)\,\,\rho^1_{\lambda_0}(a)
\over\rho^0_{\lambda_0}(a)\,\,\rho^1_{{}^b\hspace{-0.06cm}\lambda_0}(a)}
\ =\ (-1)^{a''b(n''-1)} \qqq and, as the result,
$\,\phi_{\lambda_0}(b,a)\,$ is trivialized by (\ref{mcc}) with
$\,\rho=\rho^1$. \vskip 0.3cm

On the other hand, relations (\ref{941}) and (\ref{ioc}), together
with the proportionality of the matrix elements
$\,S_{\lambda'}^\lambda\,$ and
$\,\check{S}_{\check{\lambda}'}^{\check{\lambda}}\,$ imply that
\qq \phi_{\lambda_0}(b,a)\ =\
\ee^{2\pi i\,b'(Q_J(\lambda)+a'X)\,-\,2\pi i\, b
\,\check{Q}_{\check{J}}(\check{\lambda})}\,\label{all} \qqq with
$\,\lambda\,$ another weight such that
$\,{}^a\hspace{-0.06cm}\lambda =\lambda$. \,In particular,
$\,\phi_{\lambda_0}(b,a)\,$ appearing in (\ref{941}) is
$\,\lambda_0$-independent. \,With the use of
expressions (\ref{Q}), together with (\ref{45}) and (\ref{46}),
one checks that \qq \ee^{2\pi i\,b'Q_J(\lambda)\,-\,2\pi i\, b
\,\check{Q}_{\check{J}}(\check{\lambda})}&& \cr\cr =\
(-1)^{bk({\check{n}}'-1)\over{\check{n}}'}&=&\cases{\,\hbox to
1.5cm{$(-1)^b$\hfill}\ {\rm for} \ \,N'\,\ {\rm even},\ \,N''\,\
{\rm odd},\ \,{k\over{\check{n}}'}\ \, {\rm odd}\,,\cr\,\hbox to
1.5cm{$1$\hfill}\ {\rm otherwise}\,.}\qqq On the right hand side,
the condition that $\,{k\over{\check{n}}'}\,$ be odd may be
replaced by the requirement that $\,{k\over n'}\,$ be odd if we at
the same time we replace $\,(-1)^{b}\,$ by $\,(-1)^{a''b}\,$ for
$\,a=a''n''$. \,On the other hand, \qq \ee^{2\pi i\,a'b'X}\,=\,
(-1)^{-{a''b\,k(N-1)\over n'}}\,=\,\cases{\,\hbox to
1.5cm{$(-1)^{-a''b}$\hfill}\ {\rm for} \ \,N'\,\ {\rm even},\
\,N''\,\ {\rm odd},\ \,{k\over{n}'}\ \, {\rm odd}\,,\cr \,\hbox to
1.5cm{$1$\hfill}\ {\rm otherwise}\,.}\ \ \qqq It follows that
$\,\phi_{\lambda_0}(b,a)$, \,as given by (\ref{all}), is equal to
$\,1$.
\vskip 0.2cm

\nappendix{I} \vskip 0.3cm

\noindent We construct here the canonical element
$\,\Phi(\tilde\varphi,\varphi)\,$ of (\ref{nzem}), where
$\,\varphi:[0.\pi]\rightarrow G\,$ with
$\,\varphi(0)\in\CC_{\tau_0}$, $\,\varphi(\pi)\in\CC_{\tau_1}\,$
and $\,\tilde\varphi=z_a^{-1}\varphi\,$ for $\,z_a\in\CZ$.  \,Let
us choose, for a sufficiently fine split (partition) of
$\,[0,\pi]\,$ into subintervals $\,b$, \,arbitrary lifts
$\,\phi_{_b}\,$ and $\,\tilde\phi_{_b}\,$ of $\,\varphi|_b\,$ and
$\,\tilde\varphi|_b\,$ to $\,Y=\sqcup\CO_i$. \,In other words, \qq
\phi_{_b}\,=\,(\varphi|_b\,,\,i_{_b})\,,\qquad
\tilde\phi_{_b}\,=\,(\tilde\varphi|_b\,,\,\tilde i_{_b}) \qqq for
some choice of indices such that
$\,\varphi(b)\subset O_{_{i_{_b}}}\,$ and
$\,\tilde\varphi(b)\subset O_{_{\tilde i_{_b}}}$. \,Let
$\,{\psi}_{_b}\,$ denote the mapping from
$\,b\subset[0,\pi]\,$ to $\,{Y'}^{[2]}\,$ defined by \qq
{\psi}_{_b}(x)\ =\ (\tilde\phi_{_b}(x)\,,\,\phi_{_b}(x))\,.\qqq
\,We also choose lifts $\,y_{v}\,$ and $\,\tilde y_{v}\,$ to
$\,Y\,$ of $\,\varphi(v)\,$ and $\,\tilde\varphi(v)$,
\,respectively, for vertices $\,v\,$ of the partition of
$\,[0,\pi]$. \,Let us set \qq &&\Phi(\tilde\varphi,\varphi)\cr\cr
&&\quad=\ {\iota'_0}^{-1}(\tilde
y_{_{\{0\}}},y_{_{\{0\}}},1)\,\otimes
\,{\iota'_1}^{-1}(y_{_{\{\pi\}}},\tilde y_{_{\{\pi\}}},1)
\,\otimes\,\Big(\mathop{\otimes}\limits_{b\subset[0,\pi]}\CH_{_{L'}}
({\psi}_{_b})\Big)\,,\qquad\quad\label{Phi} \qqq where
$\,\iota_s'\,$ are the bundle isomorphisms given by (\ref{triv}). \,We
shall show how $\,\Phi(\tilde\varphi,\varphi)\,$ may be considered
in a canonical way as an element of the line
$\,(\CL_{_{\tilde\CD_0\tilde\CD_1}})^{^{-1}}_{_{\tilde\varphi}}\otimes\,
(\CL_{_{\CD_0\CD_1}})_{_{\varphi}}$. \,As it stands, \qq
&&\Phi(\tilde\varphi,\varphi)\ \ \in \ \ L'_{_{(\tilde
y_{_{\{0\}}},\, y_{_{\{0\}}})}}\otimes\,(N'_0)^{\,-1}_{_{\tilde
y_{_{\{0\}}}}}\otimes \,(N'_0)_{_{y_{_{\{0\}}}}}\cr\cr
&&\hspace{1.5cm}\otimes\,L'_{_{(y_{_{\{\pi\}}},\tilde
y_{_{\{\pi\}}})}}
\otimes\,(N'_1)^{\,-1}_{_{y_{_{\{\pi\}}}}}\otimes
\,(N'_1)_{_{\tilde
y_{_{\{\pi\}}}}}\otimes\,\Big(\mathop{\otimes}\limits_{v \in
b\subset[0,\pi]}L'_{_{(\tilde y_{b},\,y_{b})}}\Big)
\,.\hspace{0,6cm} \qqq where $\,y_{b}=\phi_{_b}(v)\,$ and $\,
\tilde y_{b}=\tilde\phi_{_b}(v)$. \,With the use of the groupoid
multiplication, the last factor is canonically isomorphic to \qq
\mathop{\otimes}\limits_{v\in b\subset[0,\pi]}\Big(L'_{_{(\tilde
y_b, \,\tilde y_v)}}\otimes\,L'_{_{(\tilde y_v,\,y_v)}}\otimes\,
L'_{_{(y_v,\,y_b)}}\Big)\,. \qqq But the line \qq L'_{_{(\tilde
y_{_{\{0\}}},\,y_{_{\{0\}}})}}\otimes\,
\Big(\mathop{\otimes}\limits_{v\in b\subset[0,\pi]}L'_{_{(\tilde
y_v,\,y_v)}} \Big)\,\otimes\,{L'}_{_{(y_{_{\{\pi\}}},\,\tilde
y_{_{\{\pi\}}})}} \qqq is canonically trivial since the factors
appear in dual pairs. We infer that, in a canonical way, \qq
\Phi(\tilde\varphi,\varphi)\ \ &\in\ & (N'_0)^{\,-1}_{_{\tilde
y_{_{\{0\}}}}}\otimes\,\Big(\mathop{\otimes} \limits_{v\in
b\subset[0,\pi]}L'_{_{(\tilde y_b,\,\tilde y_v)}}\Big) \,\otimes
(N'_1)_{_{\tilde y_{_{\{\pi\}}}}} \,\,\cr\cr
&\otimes&(N'_0)_{_{y_{_{\{0\}}}}}\otimes\,\Big(\mathop{\otimes}
\limits_{v\in b\subset[0,\pi]}L'_{_{(y_v,\,y_b)}}\Big)\,\otimes
(N'_1)^{\,-1}_{_{y_{_{\{\pi\}}}}}\,. \qqq Recalling that, by
construction, the lines bundles $\,N'_s\,$ over the subsets
$\,\pi^{-1}(\CC_{\tau_s})\subset Y\,$ coincide with the bundles
$\,N_s\,$ and using the definitions (\ref{lp}) and (\ref{ldd}), we
infer that the last line is canonically isomorphic with the line 
$\,(\CL_{_{\tilde\CD_0\tilde\CD_1}})^{^{-1}}_{_{\tilde\varphi}}
\otimes\,(\CL_{_{\CD_0\CD_1}})_{_{\varphi}}$. \vskip 0.3cm

The fact that the isomorphisms $\,\iota'_s\,$ preserve the groupoid 
multiplication and the associativity of the groupoid multiplication 
in $\,L'\,$ result in the canonical identification (\ref{mp}).
We leave the details to the reader. \,Finally, formula (\ref{aa'}) 
is a consequence of the fact that, from the point of view of group 
$\,G'$, \,multiplication by $\,\Phi\,$ gives the canonical isomorphism 
used to identify two different realizations of the same fiber of 
the line bundle $\,\CL'_{_{\CD'_0\CD'_1}}$.

\end{document}